\documentclass[11pt]{article}
\usepackage{graphicx}
\usepackage{amsmath}
\usepackage{multirow}
\usepackage{amssymb}
\usepackage{amsthm}
\usepackage{booktabs}
\usepackage{bm}
\usepackage{diagbox}
\usepackage{natbib} 
\usepackage{moreverb}
\usepackage{geometry}
\usepackage{bbm}

\usepackage{setspace}
\newcommand{\1}{\mathbbm{1}}

\newcommand{\cF}{{\mathcal{F}}}

\newcommand{\cP}{{\mathcal{P}}}

\newcommand{\cZ}{{\mathcal{Z}}}

\newcommand{\indep}{\perp\!\!\!\perp}

\newtheorem{theorem}{Theorem}
\newtheorem{proposition}{Proposition}
\newtheorem{corollary}{Corollary}
\newtheorem{lemma}{Lemma}

\newtheorem{assumption}{Assumption}
\newtheorem*{conditions*}{Sufficient Conditions}

\newtheorem{definition}{Definition}
\theoremstyle{remark}
\newtheorem{remark}{Remark}

\title{Tilted sensitivity analysis in matched observational studies}
\author{Colin B. Fogarty \thanks{Department of Statistics, University of Michigan, Ann Arbor, MI.}}
\date{}
\begin{document}
\thispagestyle{empty}
\maketitle
\abstract{We present a new procedure for conducting a sensitivity analysis in matched observational studies. For any candidate test statistic, the approach defines tilted modifications dependent upon the proposed strength of unmeasured confounding. The framework subsumes both (i) existing approaches to sensitivity analysis for sign-score statistics; and (ii) sensitivity analyses using conditional inverse probability weighting, wherein one weights the observed test statistic based upon the worst-case assignment probabilities for a proposed strength of hidden bias. Unlike the prevailing approach to sensitivity analysis after matching, there is a closed form expression for the limiting worst-case distribution when matching with multiple controls. Moreover, the approach admits a closed form for its design sensitivity, a measure used to compare competing test statistics and research designs, for matching with multiple controls, whereas the conventional approach generally only does so for pair matching. The tilted sensitivity analysis improves design sensitivity under a host of generative models. The proposal may also be adaptively combined with the conventional approach to attain a design sensitivity no smaller than the maximum of the individual design sensitivities. Data illustrations indicate that tilting can provide meaningful improvements in the reported robustness of matched observational studies.}
\newpage
\normalsize
\section{Introduction}
In an observational study employing matching, treated and control individuals with similar observed covariates are placed into matched sets through the solution of an optimization problem \citep{ han04, zub12, bru24}. Despite their similarity with respect to available covariates, individuals in the same matched set may differ in terms of their probability of receiving the treatment due to unobserved factors. A sensitivity analysis in an observational study examines the extent to which adjustment for observables alone may have failed to produce genuine evidence for a treatment effect due to the potential presence of hidden bias. 

\citet{ros87} introduced a model for sensitivity analysis suitable for matched observational studies, which bounds the odds ratio of the probabilities of receiving the treatment for any two individuals in the same matched set. While the model for biased treatment assignments of \citet{ros87} is applicable regardless of the structure of the matched sets, methods and supporting theory for conducting a sensitivity analysis under this model vary substantially depending upon whether or not one has a paired observational study (a study with exactly one treated and one control unit in each matched set). For paired observational studies, when testing the sharp null hypothesis of no treatment effect one can provide bounds on the $p$-value through the construction of a stochastically dominating random variable. These bounds are valid in finite samples and provide an interpretable characterization of the worst-case pattern of hidden bias: individuals with larger responses are given a higher probability of receiving the treatment. For other forms of matching, similar results are attainable for a class of test statistics referred to as sign-score statistics in \citet{ros88}. While this class includes useful members, it does not contain commonly deployed statistics such as the difference in means with continuous outcomes. 

Outside of sign-score statistics, new complications arise. When computing the worst-case $p$-value under the sharp null there are $\prod_{i=1}^I(n_i-1)$ candidate patterns of hidden bias to consider, where $n_i$ is the number of individuals in the $i$th matched set and $I$ is the number of matched sets \citep{ros90}. The exponential scaling as a function of the number of matched sets renders exact calculation of the worst-case pattern of hidden bias infeasible in practice when matching with multiple controls. A major breakthrough in facilitating sensitivity analysis in matched designs beyond pair matching was attained by \citet{gas00}, who showed that one may find asymptotically valid bounds on the $p$-value through the solution of $I$ optimization problems, each with only $n_i-1$ decision variables. The procedure, known as the \textit{separable algorithm}, is reviewed in \S \ref{sec:separable}. Unlike in the paired case, the worst-case pattern of unmeasured confounding returned by the separable algorithm is not available in closed form. Furthermore, the worst-case pattern of hidden bias can change as the allowed degree of hidden bias within the sensitivity model changes. 

The differences in sensitivity analysis between paired studies and the broader class of matched designs have significant implications for theoretical comparisons of competing test statistics. One particular manifestation is in the calculation of design sensitivity \citep{ros04}, a metric describing a test statistic's ability to distinguish non-negligible treatment effects from biased treatment assignments. For paired studies, closed-form expressions for the design sensitivity are readily available, facilitating analytic comparisons between test statistics and furnishing broader intuition about the drivers of robustness. In more general matched designs, with few exceptions one must rely upon Monte Carlo simulation to compare design sensitivities between competing test statistics, hindering the development of the deeper insights that are available in the paired setting. 

This paper introduces a new procedure for sensitivity analysis in matched observational studies. For any candidate test statistic, we introduce in \S \ref{sec:tilt} a \textit{tilted} modification based upon the degree of hidden bias assumed to exist at any given stage of the sensitivity analysis. The modification is constructed in such a way that even beyond paired designs, the worst-case pattern of unmeasured confounding is both available in closed form and does not vary as the allowed degree of hidden bias increases. Rather than allowing the output of the separable algorithm to vary as the allowed strength of confounding increases, one instead modifies the test statistic itself in such a way that the bound on the worst-case expectation remains constant. The resulting inference is equivalent to that produced by the usual approach in paired designs, but differs from the conventional approach to sensitivity analysis outside of paired studies. This and related modifications were previously considered in \citet{fog23} for testing weak null hypotheses with the difference-in-means statistic. In general, the tilting modification does not confer validity for testing the weak null without additional restrictions. Here we instead show that when testing sharp nulls, tilting provides an alternative extension of sensitivity analysis to matching with multiple controls which eliminates many of the technical hurdles arising from the prevailing approach. Despite the matched context, we also highlight connections between the tilted modifications and approaches to sensitivity analysis involving inverse probability weighting \citep{zha19, dor22}. In contrast with the conventional approach, we show in \S \ref{sec:ds} that the tilted sensitivity analysis furnishes test statistics with a closed form for their design sensitivity for any matched structure. The tilted sensitivity analysis provides an improvement in design sensitivity relative to the conventional approach for a wide range of generative models, but does not uniformly dominate the conventional approach. In \S \ref{sec:adapt}, we note that the tilted and conventional approaches can be adaptively combined to attain a design sensitivity no smaller than the maximum of tilted and conventional design sensitivities. In \S \ref{sec:data}, we highlight the benefits conferred by tilting in multiple real-data applications.

\section{Background and review}
\subsection{Sensitivity analysis after matching}
 There are $I$ matched sets formed on the basis of pretreatment covariates. For notational ease we assume that the $i$th of $I$ matched sets contains 1 treated individual and $n_i-1$ controls, but the proposals herein extend immediately to matched structures returned by full matching \citep{han04}; see \citet[Problem 4.12]{obs} for details of the extension. Let $Z_{ij}$ be the treatment indicator for the $j$th individual in the $i$th matched set, taking the value 1 if treated and 0 otherwise. Let $y_{1ij}$ and $y_{0ij}$ be the potential outcomes under treatment and control, let $x_{ij}$ be the vector of observed covariates, and let $0\leq u_{ij}\leq 1$ be an unobserved covariate for the $ij$th individual. The observed response for individual $ij$ is $Y_{ij} = Z_{ij}y_{1ij} + (1-Z_{ij})y_{0ij}$; implicit in this representation is that the Stable Unit-Treatment Value Assumption (SUTVA) holds \citep{rub74}. Let $\mathbf{Z} = (Z_{11},...,Z_{In_I})^T$ be the vector of treatment assignments across all matched sets, let $\mathbf{Z}_i = (Z_{i1},...,Z_{in_i})^T$ be the assignments in matched set $i$, and let the analogous notation hold for other quantities such as $\mathbf{u}_i = (u_{i1},...,u_{in_i})^T$. Let $\mathbf{Y}_\mathbf{Z}$ be the lexicographically ordered vector of observed responses under assignment $\mathbf{Z}$.  Let $\mathcal{F} = \{y_{0ij}, y_{1ij}, x_{ij}, u_{ij}\}$ be the features of the observed study population, let $\Omega = \{\mathbf{z} : \sum_{j=1}^{n_{i}}z_{ij} = 1, \;\; i=1,...,I\}$ be the set of treatment assignments satisfying the matched design, and let $\mathcal{Z} = \{\mathbf{Z}\in \Omega\}$.

Letting $\pi_{ij} = \text{pr}(Z_{ij}\mid \mathcal{F})$, the sensitivity model introduced in \citet{ros87} proposes the following logit model for the assignment probabilities:
\vspace{-5pt}
\begin{align}\label{eq:sensmodel}
\log\left(\frac{\pi_{ij}}{1-\pi_{ij}}\right) = \kappa_i + \log(\Gamma) u_{ij},
\end{align}
where $\Gamma \geq 1$ is a sensitivity parameter controlling the impact of unmeasured confounding on the assignment probabilities. Observe that at $\Gamma=1$, all individuals in the same matched set have the same assignment probabilities. 

Conditioning on $\mathcal{Z}$ to return attention to the matched structure, we have 
\vspace{-5pt}
\begin{align}\label{eq:rho}
\varrho_{ij}:=\text{pr}(Z_{ij}=1\mid \mathcal{F}, \mathcal{Z}) = \frac{\exp\{\log(\Gamma) u_{ij}\}}{\sum_{\ell=1}^{n_i}\exp\{\log(\Gamma) u_{i\ell}\}};\\
\text{pr}_{\mathbf{u}}(\mathbf{Z} = \mathbf{z}\mid \mathcal{F}, \mathcal{Z}) = \frac{\exp\{\log(\Gamma) \mathbf{z}^T \mathbf{u}\}}{\sum_{\mathbf{b}\in \Omega}\exp\{\log(\Gamma) \mathbf{b}^T \mathbf{u}\} }\nonumber, 
\end{align}
such that conditioning on $\mathcal{Z}$ removes dependence on the nuisance parameters $\kappa_i$. At $\Gamma=1$, conditioning on $\mathcal{Z}$ recovers the usual randomization distribution in a block-randomized experiment, while $\Gamma > 1$ allows for departures from this idealization. Balance checks and tests can assess whether matching has succeeded in recovering this idealized design in the absence of hidden bias \citep[\S 6]{han08balance, ITOS}. Note that (\ref{eq:rho}) does not model variation in the matched structure across treatment allocations; see \citet{pim24} for a discussion of how variation in matched structure can impact inference.

 Consider a test statistic of the form $T = \mathbf{Z}^T\mathbf{q}= \sum_{i=1}^I\sum_{j=1}^{n_i}Z_{ij}q_{ij}$, where $\mathbf{q} = q(\mathbf{Y}_\mathbf{Z})$ is a function of the observed responses under assignment vector $\mathbf{Z}$, $\mathbf{Y}_\mathbf{Z}$; such statistics are called sum statistics, and include the treated-minus-control difference in means and Wilcoxon's rank sum test among many \citep{obs}. Assuming that (\ref{eq:sensmodel}) holds at $\Gamma$, the right-tail probability for $T$ is \begin{align}\label{eq:tailprob}
\text{pr}(T \geq a \mid \mathcal{F}, \mathcal{Z}) &= \frac{\sum_{\mathbf{b} \in \Omega} \1\{\mathbf{b}^Tq(\mathbf{Y}_\mathbf{b})\geq a\}\exp\{\log(\Gamma)\mathbf{b}^T\mathbf{u}\}}{\sum_{\mathbf{b}\in \Omega}\exp\{\log(\Gamma) \mathbf{b}^T \mathbf{u}\}}.
\end{align}

This distribution is unknown to the practitioner, due to its dependence on observed outcomes $\mathbf{Y}_\mathbf{b}$ for assignments $\mathbf{b}\neq \mathbf{Z}$ and on the unobserved covariates $\mathbf{u}$. With a few exceptions \citep{ros02att, fog23}, the literature on sensitivity analysis after matching typically considers tests of sharp null hypotheses under the finite population model, hypotheses that impute the missing values of the potential outcomes for each individual. For ease of exposition, we consider the sharp null hypothesis of no effect at all, $H_{F}: y_{0ij} = y_{1ij}\;\; (i=1,...,I;j=1,...,n_i)$. Under this null, $q(\mathbf{Y}_{\mathbf{b}}) = q(\mathbf{Y}_\mathbf{z})$ for any $\mathbf{b}, \mathbf{z}\in \Omega$, as the observed responses $\mathbf{Y}_\mathbf{Z}$ provide the hypothetical responses $\mathbf{Y}_\mathbf{b}$ for all possible randomizations. At $\Gamma>1$, the right-tail probability in (\ref{eq:tailprob}) remains unknown due to its dependence on the unmeasured confounders $u_{ij}$. In a sensitivity analysis, one computes upper bounds on the tail probability (\ref{eq:tailprob}) at any given $\Gamma>1$. One then finds the largest value of $\Gamma$ for which $H_{F}$ can be rejected. This changepoint $\Gamma$, the \textit{sensitivity value} \citep{zha19senval}, quantifies the robustness of a study's findings to unmeasured confounding.

For various extensions of the sensitivity model (\ref{eq:sensmodel}) to accommodate features ranging from stochastic potential outcomes to heterogeneous bias across matched sets, see \citet{fog19extend, su24, zha24, che25} and \citet{wu25}.

\subsection{The separable algorithm}\label{sec:separable}
For a given $\Gamma$, asymptotic bounds on (\ref{eq:tailprob}) under the sharp null hypothesis can be attained using the separable algorithm introduced in \cite{gas00}. The algorithm finds the vector(s) of unmeasured confounders yielding the worst-case expectation under the sharp null. If multiple vectors provide the same worst-case expectation, the one providing the largest variance is selected. One then conducts inference using a normal random variable with this expectation and variance. 

Let $T_i = \sum_{j=1}^{n_i}Z_{ij}q_{ij}$, such that $T = \sum_{i=1}^IT_i$. Rearrange the values $q_{ij}$ in each matched set $i$ such that $q_{i1} \geq q_{i2} \geq ... \geq q_{i_{n_i}}$. Using results in \cite{ros90}, the worst-case expectation for $T_i$ under the sharp null, optimized over all possible $\mathbf{u}_i\in [0,1]^{n_i}$, can be restricted to a search over $n_i-1$ candidates:\vspace{-5pt}\begin{align}\mu_{\Gamma i} &= \max_{ a = 1,..., n_i-1}\;\frac{\Gamma\sum_{j=1}^{a}q_{ij} + \sum_{j = a+1}^{n_i} q_{ij}}{\Gamma a + (n_i-a)}.\label{eq:AS}
\end{align} Let $\mathcal{A}_i$ be the set of values for $a$ attaining the maximal expectation $\mu_{\Gamma i}$, and define the largest variance for $T_i$ for patterns of hidden bias within $\mathcal{A}_i$ as \\$\nu^2_{\Gamma i} := \max_{a \in \mathcal{A}_i} \left\{\Gamma \sum_{j=1}^{a}q_{ij}^2 + \sum_{j= a+1}^{n_i} q_{ij}^2\right\}/\left\{\Gamma a + (n_i-a)\right\} - (\mu_{\Gamma i})^2$. Under mild conditions, the tail probability for $T$ in (\ref{eq:tailprob}) is asymptotically upper bounded by the tail probability of a normal random variable with expectation $\sum_{i=1}^I\mu_{\Gamma i}$ and variance $\sum_{i=1}^I\nu^2_{\Gamma i}$. This approximation requires solving $I$ tractable optimization problems, each requiring the enumeration of $n_i-1$ candidate solutions.

The following proposition, proven in the web-based supporting materials, provides a fixed-point representation for (\ref{eq:AS}) which will be useful in what follows.

\begin{proposition} \label{prop:MGamma}
Let $\bar{q}_i = n_i^{-1}\sum_{j=1}^{n_i}q_{ij}$. Then, $\mu_{\Gamma i} = \bar{q}_i + \left\{(\Gamma-1)/(1+\Gamma)\right\}M_{\Gamma i}$, where 
\begin{align}
M_{\Gamma i} &:= \mathtt{SOLVE}\left\{c:\; \frac{1}{n_i}\sum_{j=1}^{n_i}\left|q_{ij} - \bar{q}_i - \left(\frac{\Gamma-1}{1+\Gamma}\right)c\right| = c\right\} \label{eq:MGamma},
\end{align} and $\mathtt{SOLVE}(\cdot)$ denotes the unique solution to the displayed equation.\end{proposition} 

\section{Tilted sensitivity analysis}\label{sec:tilt}

%
We now introduce a modified approach to conducting a sensitivity analysis. Suppose that the sharp null hypothesis $H_F$ holds. For each matched set $i$, define $T_{\Gamma i}$, referred to as the \textit{tilted} version of $T_i$ henceforth, as   \begin{align}
T_{\Gamma i} &:= T_i-\bar{q}_i - \left(\frac{\Gamma-1}{1+\Gamma}\right)|T_i-\bar{q}_i| \label{eq:AGamma},
\end{align} which has the equivalent representation \begin{align}
T_{\Gamma i}=\frac{\{2\Gamma/(1+\Gamma)\}(T_i-\bar{q}_i)}{\exp\{\log(\Gamma) \1\{T_i >  \bar{q}_i\}\}}
= \begin{cases}\{2/(1+\Gamma)\}(T_i-\bar{q}_i) & T_i > \bar{q}_i\\
\{2\Gamma/(1+\Gamma)\}(T_i-\bar{q}_i) & T_i \leq \bar{q}_i\end{cases}. \label{eq:AGammaweight}\end{align} 

Without the factor $(2\Gamma)/(1+\Gamma)$ in (\ref{eq:AGammaweight}), observe that $f_\Gamma(x) := x/\exp\{\log(\Gamma)\1\{x\geq 0\}\}$ is the 45 degree line through the origin for $x \leq 0$, with slope equal to 1. For $x>0$, the slope changes from 1 to 1/$\Gamma$, tilting towards the horizontal axis.

Let $w_{\Gamma i} \geq 0 $, $i=1,...,I$, be nonnegative multipliers which are fixed under the sharp null, and consider as a test statistic $T_\Gamma(\mathbf{w_\Gamma}) = \sum_{i=1}^I w_{\Gamma i}T_{\Gamma i}$. \citet{ros14} describes the benefits of weighting for improving the power of a sensitivity analysis. Different choices of weights will also provide connections between sensitivity analyses using tilted statistics and both (i) using sign-score statistics \citep{ros88}; and (ii) using inverse probability weighting.

We now describe how to conduct a sensitivity analysis after tilting when (\ref{eq:sensmodel}) is assumed to hold at $\Gamma$. For each matched set $i$, define the variance $\tilde{\nu}^2_{\Gamma i}$ as\begin{align}\label{eq:nutilt}
\tilde{\nu}_{\Gamma i}^2 &:= \frac{4\Gamma^2}{(1+\Gamma)^2}\left(\frac{1}{\sum_{j=1}^{n_i}\exp\{\log(\Gamma)\1\{q_{ij} > \bar{q}_i\}\}}\right) \sum_{j=1}^{n_i}\frac{(q_{ij}-\bar{q}_i)^2}{\exp\{\log(\Gamma)\1\{q_{ij} > \bar{q}_i\}\}}.
\end{align}Further define $c_{\Gamma,\alpha}(\mathbf{w_\Gamma}) = \Phi^{-1}(1-\alpha)\sqrt{\sum_{i=1}^Iw_{\Gamma i}^2\tilde{\nu}_{\Gamma i}^2}$, where $\Phi^{-1}(\cdot)$ is the quantile function for the standard normal distribution. The following theorem, proven in the web-based supporting material, justifies conducting a sensitivity analysis for each $\Gamma$ by rejecting the sharp null hypothesis when $T_{\Gamma}(\mathbf{w_\Gamma}) \geq c_{\Gamma,\alpha}(\mathbf{w_\Gamma})$. That is, $c_{\Gamma,\alpha}(\mathbf{w_\Gamma})$ is an asymptotically valid critical value when  (\ref{eq:sensmodel}) holds at $\Gamma$.

\begin{theorem}\label{thm:AGamma}
Suppose that the sharp null hypothesis holds and that the sensitivity model (\ref{eq:sensmodel}) holds at $\Gamma$. Suppose that the stratum sizes are bounded, $n_i \leq \tilde{N}$ for all $i$, and suppose further that the constants $q_{ij}$ and weights $w_{\Gamma i}$ satisfy 
\begin{align*} \frac{\max_{i=1,...,I}\;\; w_{\Gamma i}^2(q_{i1}-q_{in_i})^2}{\sum_{i=1}^Iw_{\Gamma i}^2(q_{i1}-q_{in_i})^2} \rightarrow 0\end{align*} as $I\rightarrow \infty$, where $q_{ij}$ are again sorted so that $q_{i1}\geq q_{i2}\geq...\geq q_{in_i}$. Let $\vartheta_{\Gamma i}$ and $\sigma^2_{\Gamma i}$ denote the true expectation and variance for $T_{\Gamma i}$. Let $\mathcal{B}_I = \{i: \sigma^2_{\Gamma i} > \tilde{\nu}^2_{\Gamma i}\}$, and suppose that
$\sqrt{I^{-1}\sum_{i\in\mathcal{B}_I} w_{\Gamma i}^2(\sigma^2_{\Gamma i} - \tilde{\nu}^2_{\Gamma i})} = o\left(I^{-1/2}|\sum_{i=1}^I w_{\Gamma i}\vartheta_{\Gamma i}|\right)$. Then, for all $\epsilon > 0$, there exists an $I^*$ such that $I > I^*$ implies that for any $\alpha < 1/2$,\begin{align*}
\text{pr}\{T_{\Gamma}(\mathbf{w_\Gamma}) \geq c_{\Gamma,\alpha}(\mathbf{w_\Gamma})\mid \mathcal{F}, \mathcal{Z}\} \leq \alpha + \epsilon.
\end{align*}
\end{theorem}

Theorem \ref{thm:AGamma} implies that $1-\Phi\left(T_\Gamma(\mathbf{w_\Gamma})/\sqrt{\sum_{i=1}^Iw_{\Gamma i}^2\tilde{\nu}_{\Gamma i}^2}\right)$ provides an asymptotic upper bound on the $p$-value for testing the sharp null when (\ref{eq:sensmodel}) holds at $\Gamma$. The condition on $\max w_{\Gamma i}^2(q_{i1}-q_{in_i})^2/\sum w_{\Gamma i}^2(q_{i1}-q_{in_i})^2$ provides a central limit theorem by Theorem 1 of \citet[\S 6.1.2]{haj99}. A condition analogous to $\sqrt{I^{-1}\sum_{i\in\mathcal{B}_I} w_{\Gamma i}^2(\sigma^2_{\Gamma i} - \tilde{\nu}^2_{\Gamma i})} = o\left(I^{-1/2}|\sum_{i=1}^I w_{\Gamma i}\vartheta_{\Gamma i}|\right)$ is needed to justify the separable algorithm of \citet{gas00} for \textit{any} test statistic, and is not unique to tilting. For intuition, note that if $I^{-1}\sum_{i=1}^I w_{\Gamma i}\vartheta_{\Gamma i}$ and $I^{-1}\sum_{i=1}^Iw_{\Gamma i}^2\sigma^2_{\Gamma i}$ have finite nonzero limits, then \\$\sqrt{I^{-1}\sum_{i\in\mathcal{B}_I} w_{\Gamma i}^2(\sigma^2_{\Gamma i} - \tilde{\nu}^2_{\Gamma i})} = O(1)$ and $I^{-1/2}|\sum_{i=1}^I w_{\Gamma i}\vartheta_{\Gamma i}| = O(I^{1/2})$, hence satisfying the condition. For a pathological example where this fails, see the web-based supporting material.

The following lemma, proven in the web-based supporting materials, allows one to find a closed-form expression for the output of the separable algorithm described in \S \ref{sec:separable} even when matching with multiple controls, which is generally not possible without deploying tilting. The remainder of the proof of Theorem \ref{thm:AGamma} follows in a straightforward way from the proof of Proposition 1 of \citet{gas00}, and Theorem 1 of \citet[\S 6.1.2]{haj99}. 

\begin{lemma} \label{lemma:AGamma}
Suppose that (\ref{eq:sensmodel}) holds at $\Gamma$. Then,  $E(T_{\Gamma i}\mid \mathcal{F}, \mathcal{Z}) \leq 0$ for all $i=1,...,I$, with $E(T_{\Gamma i}\mid \mathcal{F}, \mathcal{Z}) = 0$ when $u_{ij}= \1\{q_{ij} > \bar{q}_i\}$. When $u_{ij}=\1\{q_{ij}>\bar{q}_i\}$, the variance of $T_{\Gamma i}$ equals $\tilde{\nu}^2_{\Gamma i}$ in (\ref{eq:nutilt}). If there exists another pattern of unmeasured confounding $\mathbf{u}'_{i} \neq \mathbf{u}_i$ for which $E(T_{\Gamma i}\mid \mathcal{F}, \mathcal{Z}) = 0$, the variance for $T_{\Gamma i}$ under pattern $\mathbf{u}'_{i}$ will be no larger than $\tilde{\nu}^2_{\Gamma i}$.
\end{lemma}

\begin{remark}\label{remark:separable} A result analogous to Theorem \ref{thm:AGamma} would hold for the conventional sensitivity analysis when using a statistic $T(\mathbf{w_\Gamma}) = \sum_{i=1}^Iw_{\Gamma i}T_i$, with the critical value $c_{\Gamma,\alpha}(\mathbf{w_\Gamma})$ replaced by $\sum_{i=1}^Iw_{\Gamma i}\mu_{\Gamma i} + \Phi^{-1}(1-\alpha)\sqrt{\sum_{i=1}^Iw^2_{\Gamma i}v^2_{\Gamma i}}$. Unlike for $T_{i}$, the worst-case expectation for $T_{\Gamma i}$ when (\ref{eq:sensmodel}) is assumed to hold at $\Gamma$ may be calculated \textit{without} resorting to the separable algorithm described in \S \ref{sec:separable}:  the worst-case deviate returned by the separable algorithm has a closed-form solution when using $T_\Gamma(\mathbf{w_\Gamma})$ as a test statistic, while it generally does not when using $T(\mathbf{w_\Gamma})$. In the web-based supporting materials, we further show that tilting is the \textit{unique} (up to scalar positive multiples) coordinate-wise, non-degenerate continuous non-decreasing transformation of $q_{ij}-\bar{q}_i$ that yields a closed-form expression for the output of the separable algorithm for all possible $(q_{i1}-\bar{q}_i,...,q_{in_i}-\bar{q}_i)^T$.
\end{remark}
\begin{remark} The pattern of unmeasured confounding used by tilting has an intuitive form: for any value of $\Gamma$, individuals $ij$ above the matched-set average of $q_{ij}$, $\bar{q}_i$, are given $u_{ij}=1$, making them relatively more likely to be treated, while individuals with $q_{ij}$ at or below the stratum-specific mean are given $u_{ij}=0$, making them relatively less likely to be treated. To see why this maximizes the expectation, use (\ref{eq:AGammaweight}) and define $q_{\Gamma ij} = \{(2\Gamma)/(1+\Gamma)(q_{ij}-\bar{q}_i)\}/\exp\{\log(\Gamma)\1\{q_{ij} > \bar{q}_i\}\}$. Consider the numerator in (\ref{eq:AS}) applied to $q_{\Gamma ij}$. The largest numerator is attained by giving the largest weight to positive $q_{\Gamma ij}$, and the smallest weight to negative $q_{\Gamma ij}$. Setting $a^* = \sum_{j=1}^{n_i}\1\{q_{\Gamma ij} > 0\} \Leftrightarrow \sum_{j=1}^{n_i}\1\{q_{ij} > \bar{q}_i\}$ implies positive $q_{\Gamma ij}$ receive a multiplier $\Gamma$ and negative $q_{\Gamma ij}$ receive a multiplier 1. Noting the division by $\exp\{\log(\Gamma) \1\{q_{ij}>\bar{q}_i\}\}$ in $q_{\Gamma ij}$, this yields a worst-case numerator of 0 since $\sum_{j=1}^{n_i}(q_{ij}-\bar{q_i}) = 0.$ Because the worst-case numerator is 0 and the denominator is strictly positive, 0 is also the worst-case expectation in (4).
\end{remark}


\subsection{Tilting, pair matching, and sign-score statistics}\label{sec:signscore}

We first describe connections between sensitivity analyses using the tilted test statistics $T_{\Gamma}(\mathbf{w_\Gamma})$ and existing results for sensitivity analyses for an important subset of statistics commonly used in matched observational studies. Suppose $T_i$ takes on at most two distinct values for $i=1,...,I$, admitting the representation
\begin{align}\label{eq:ss}T_i &= a_{i2} + (a_{i1}-a_{i2})\sum_{j=1}^{n_i}Z_{ij}\1\{q_{ij} = a_{i1}\} \end{align} with $a_{i1}\geq a_{i2}$ and $(a_{i1},a_{i2})^T$ allowed to differ from set to set. From (\ref{eq:ss}), we see that $\sum_{i=1}^I T_i$ is equivalent to an important class of statistics known as sign-score statistics \citep[\S\S 4.3-4.4]{ros88, obs}. Examples include any test statistic in a matched pairs design, along with several common statistics with multiple controls such as the Mantel-Haenszel test statistic with binary outcomes. For such test statistics, one can construct a stochastically dominating random variable $B_i$ when (\ref{eq:sensmodel}) holds at $\Gamma$ by assigning $u_{ij}=1$ to observations with $q_{ij} = a_{i1}$, and $u_{ij}=0$ to observations with $q_{ij}=a_{i2}$; see Proposition 14 of \citet[\S 4.4.1]{obs} for details. It follows that $\mu_{\Gamma i}$ in (\ref{eq:AS}), the worst-case expectation that would be returned by the separable algorithm, equals the expectation of the stochastically bounding random variable $B_i$. The following shows an equivalence between $T_i-\mu_{\Gamma i}$ and $w_{\Gamma i}T_{\Gamma i}$ for a particular choice of the weights, $w^{(ss)}_{\Gamma i}$.

\begin{proposition}\label{prop:signscore}
Suppose that $T_i$ admits the representation (\ref{eq:ss}). Then,
\begin{align}\label{eq:signscore}
T_i - \mu_{\Gamma i} &= \underbrace{\left(\frac{\Gamma+1}{2}\right)\left(\frac{n_i}{n_i+(\Gamma-1)\sum_{j=1}^{n_i}\1\{q_{ij}> \bar{q}_i\}}\right)}_{=:w^{(SS)}_{\Gamma i}}T_{\Gamma i}.
\end{align}
\end{proposition}

First consider the implications of Proposition \ref{prop:signscore} for matched pair designs $(n_i=2)$. The representation for $T_{\Gamma i}$ in (\ref{eq:AGamma}) is commonly encountered when conducting sensitivity analyses with pairs; see, for instance, the numerator of the second display in Equation 6 of \citet{ros07}. In paired designs, $|T_i - \bar{q}_i|$ is constant across all randomizations when testing the sharp null, and the worst-case expectation when conducting a sensitivity analysis at $\Gamma$ is $\mu_{\Gamma i} = \bar{q}_i + \{(\Gamma-1)/(1+\Gamma)\}|T_i-\bar{q}_i|$. For paired designs, $w_{\Gamma i}^{(SS)} = 1$, and $T_{\Gamma i} = T_i - \mu_{\Gamma i}$ is thus the observed value for the test statistic $T_i$ in stratum $i$ minus the worst-case expectation when (\ref{eq:sensmodel}) holds at $\Gamma$. Proposition \ref{prop:signscore} shows that in the case of pair matching, the tilted sensitivity analysis reproduces the standard approach to sensitivity analysis without modification. Beyond paired observational studies, Proposition \ref{prop:signscore} illustrates a more general connection between the tilted sensitivity analysis and the conventional sensitivity analysis when restricting attention to sum statistics of the form (\ref{eq:ss}). Notably, statistics of the form (\ref{eq:ss}) are the statistics for which one can bypass the separable algorithm. Within this class, finite-sample valid sensitivity analysis using tilting can proceed using the stochastically dominating random variable to compute $p$-values. With the choice of weights $w_{\Gamma i}^{(ss)}$ in (\ref{eq:signscore}), the tilted sensitivity analysis exactly recovers the conventional sensitivity analysis for sign-score statistics.

\subsection{Tilting and conditional inverse probability weighting}\label{sec:IPW}
The equivalent form for $T_{\Gamma i}$ in (\ref{eq:AGammaweight}), together with Theorem \ref{thm:AGamma}, inspires an alternative motivation for the tilted statistics through connections with inverse probability weighting (IPW). Suppose that (\ref{eq:sensmodel}) holds at $\Gamma>1$, that the sharp null is true, and that the true conditional assignment probabilities are $\varrho_{ij} = \exp\{\log(\Gamma) u_{ij}\}/\sum_{\ell=1}^{n_i}\exp\{\log(\Gamma) u_{i\ell}\}$. Then, the inverse probability weighted $n_i^{-1}\sum_{j=1}^{n_i}Z_{ij}(q_{ij}-\bar{q}_i)/\varrho_{ij}$ has expectation zero; at $\Gamma=1$, this returns $\sum_{j=1}^{n_i}Z_{ij}(q_{ij}-\bar{q}_i)$. When conducting a sensitivity analysis, we do not have access to the true assignment probabilities $\varrho_{ij}$ because $u_{ij}$ is unknown. Consequently, the realization of the random variable $n_i^{-1}\sum_{j=1}^{n_i}Z_{ij}(q_{ij}-\bar{q}_i)/\varrho_{ij}$ is unknown. Consider instead forming an IPW statistic using any feasible vector of conditional probabilities $p_{ij}$ when (\ref{eq:sensmodel}) holds at $\Gamma$. If $p_{ij}$ were actually the true assignment probabilities the IPW statistic would have expectation 0; in general, because $p_{ij}$ is feasible at $\Gamma$,
\begin{align}
\max_{0 \leq u_{ij} \leq 1} E\left\{ n_i^{-1}\sum_{j=1}^{n_i}Z_{ij}(q_{ij}-\bar{q}_i)/p_{ij}\mid \mathcal{F}, \mathcal{Z}\right\} \geq 0. \label{eq:maxIPW}
\end{align} 

Consider now finding the minimax vector of candidate IPW denominators $(p_{i1},...,p_{in_i})^T$ for the above problem, i.e. a vector that minimize the maximum value of the expectation in (\ref{eq:maxIPW}) over the set of possible vectors of conditional assignment probabilities when (\ref{eq:sensmodel}) holds at $\Gamma$. One solution is obtained by setting $p_{ij} = \tilde{\varrho}_{ij}$, with 
\begin{align*}
\tilde{\varrho}_{ij} &= \frac{\exp\{\log(\Gamma)\1\{q_{ij}> \bar{q}_i\}\}}{\sum_{\ell=1}^{n_i}\exp\{\log(\Gamma)\1\{q_{i\ell}> \bar{q}_i\}\}} = \frac{\exp\{\log(\Gamma)\1\{q_{ij}> \bar{q}_i\}\}}{n_i + (\Gamma-1)\sum_{\ell=1}^{n_i}\1\{q_{i\ell}> \bar{q}_i\}},
\end{align*} 
The corresponding IPW statistic is\begin{align}
IPW_{\Gamma i} = \frac{1}{n_i}\sum_{j=1}^{n_i}\frac{Z_{ij}(q_{ij}-\bar{q}_i)}{\tilde{\varrho}_{ij}} = \underbrace{\left(\frac{\Gamma+1}{2\Gamma}\right)\left\{\frac{n_i + (\Gamma-1)\sum_{j=1}^{n_i}\1\{q_{ij}> \bar{q}_i\}}{n_i}\right\}}_{=:w^{(IPW)}_{\Gamma i}}T_{\Gamma i},\label{eq:IPW}
\end{align}
where the last equality uses (\ref{eq:AGammaweight}). Noting that $w_{\Gamma i}^{(IPW)}$ in (\ref{eq:IPW}) is a constant under the sharp null, Lemma \ref{lemma:AGamma} implies that under the sharp null, when (\ref{eq:sensmodel}) holds at $\Gamma$, $E(IPW_{\Gamma i}\mid \mathcal{F}, \mathcal{Z}) \leq 0$, with equality when $u_{ij} = \1\{q_{ij} > \bar{q}_i\}$. Weighting by $\tilde{\varrho}_{ij}$ yields a random variable whose expectation is bounded above by 0 when (\ref{eq:sensmodel}) holds at $\Gamma$, with equality when $\tilde{\varrho}_{ij}$ are the actual assignment probabilities.

Equation (\ref{eq:IPW}) illustrates strong connections between tilted statistics $T_{\Gamma i}$ and the minimax-centered IPW statistic $IPW_{\Gamma i}$: they differ only in the form of the matched-set weights $w_{\Gamma i}$, with the unweighted tilted statistic ignoring the normalizing constant $n_i + (\Gamma-1)\sum_{j=1}^{n_i}\1\{q_{ij}>\bar{q}_i\}$ for the worst-case assignment probabilities. When $T_i$ is a sign-score statistic, equations (\ref{eq:signscore}) and (\ref{eq:IPW}) further demonstrate striking yet previously unexplored similarities between sensitivity analyses after matching using sign-score statistics and sensitivity analyses using the minimax centered inverse probability weighting within each matched set: the methods are both weighted versions of the tilted sensitivity analysis, differing in the choice of weights $w_{\Gamma i}$. See the web-based supporting materials for an illustration of these connections using the Mantel-Haenszel statistic with binary outcomes. Tilting provides a unifying framework for considering these seemingly disparate approaches to conducting a sensitivity analysis after matching.

\section{Comparing the tilted and conventional sensitivity analyses through design sensitivity} \label{sec:ds}
\subsection{Generative model for comparison}\label{sec:gen}
While Proposition \ref{prop:signscore} reveals connections between tilted statistics and conventional sensitivity analyses using sign-score statistics, in general the sensitivity analyses based upon tilted and untilted statistics are fundamentally different. Without restricting to sign-score statistics, the tilted and conventional approaches are no longer equal up to the choice of weights $w_{\Gamma i}$. In particular, the unmeasured confounders yielding the worst-case expectation for $T_{\Gamma i}$ and for $T_i$ can differ. While the tilted sensitivity analysis is more straightforward to conduct because it avoids the separable algorithm, one would be hesitant to apply the tilted sensitivity analysis to the broader class of sum statistics if it demonstrated deficiencies in terms of statistical performance relative to the conventional approach. In this section, we highlight circumstances under which the tilted sensitivity analysis proves advantageous in terms of a metric known as \textit{design sensitivity} \citep{ros04}.

For theoretical comparisons among competing approaches to sensitivity analysis, it is common to assume (i) a superpopulation generative model for the responses in each matched set; and (ii) that we are within the favorable setting where the treatment has an effect and there is no unmeasured confounding \citep{ros04, ros13}. For $i=1,...,I$, $j=1,...,n_i$, potential outcomes and treatment assignments are generated as \begin{align}\label{eq:gen} &y_{1ij} = \alpha_i + \tau + \varepsilon_{1ij};\;\;
y_{0ij} = \alpha_i + \varepsilon_{0ij},\end{align}where $\alpha_i$ are matched set fixed effects, $\tau \geq 0$ is the mean treatment effect, and $(\varepsilon_{0ij}, \varepsilon_{1ij})^T$ are $iid$ draws from a common mean zero, exchangeable bivariate error distribution with marginal distribution $F_\epsilon$ for $i=1,...,I$. Treatment assignments are generated satisfying
$\text{pr}(\mathbf{Z} = \mathbf{z} \mid \mathcal{F}, \mathcal{Z}) = |\Omega|^{-1}$, which implies $Z_{ij}\indep (\varepsilon_{0ij}, \varepsilon_{1ij})$ since the $\varepsilon_{zij}$ are contained in $\mathcal{F}$. The observed responses remain $Y_{ij} = Z_{ij}y_{1ij} + (1-Z_{ij})y_{0ij}$, but are now random due to both randomized treatment assignment and randomness across draws for $(\varepsilon_{0ij}, \varepsilon_{1ij})$. The observed errors are $\epsilon_{ij} = Z_{ij}\varepsilon_{1ij} + (1-Z_{ij})\varepsilon_{0ij}$, and are also $iid$ with marginal distribution $F_\epsilon$. The test statistics under consideration remain of the form $\mathbf{Z}^T\mathbf{q} = \sum_{i=1}^I\sum_{j=1}^{n_i}Z_{ij}q_{ij} = \sum_{i=1}^I T_i $, with the vector $q = q(\mathbf{Y}_\mathbf{Z})$ also varying across draws from the generative model.

We consider the tilted sensitivity analysis with $w_{\Gamma i} = 1$ for all $i$; results extend naturally to weighted variants. $T_{\Gamma i}$ and $T_{i}-\mu_{\Gamma i}$, the observed statistic minus its worst-case expectation under the tilted and conventional approaches, when (\ref{eq:sensmodel}) is assumed to hold at $\Gamma$ may be expressed using (\ref{eq:MGamma}) and (\ref{eq:AGamma}) as 
\begin{align*}
T_{\Gamma i} &= T_i - \bar{q}_i - \left(\frac{\Gamma-1}{1+\Gamma}\right)|T_i-\bar{q}_i|;\\
T_i - \mu_{\Gamma i} &= T_i - \bar{q}_i - \left(\frac{\Gamma-1}{1+\Gamma}\right)M_{\Gamma i},
\end{align*}
where $M_{\Gamma i} = \texttt{SOLVE}\{c: c = n_i^{-1}\sum_{j=1}^{n_i}|q_{ij}-\bar{q}_i - \left\{(\Gamma-1)/(1+\Gamma)\right\}c|\}$ as before. The right-hand sides of the above displays differ only in the random variable by which $\{(\Gamma-1)/(1+\Gamma)\}$ is multiplied: $|T_i-\bar{q}_i|$ for the tilted sensitivity analysis, and $M_{\Gamma i}$ for the conventional approach. As suggested by the notation, the random variable $M_{\Gamma i}$ changes with the value of $\Gamma$ at which the sensitivity analysis is conducted, while  $|T_i-\bar{q}_i|$ does not. Differences in the distributions of $M_{\Gamma i}$ and $|T_i-\bar{q}_i|$ will play a fundamental role in determining the comparative performance of the tilted and conventional sensitivity analyses. 

Throughout this section, we make the following high-level assumptions about the limiting behavior of $T_i$, $T_{\Gamma i}$, and $M_{\Gamma i}$. Primitive conditions will depend upon the particular function $q(\mathbf{Y}_{\mathbf{Z}})$ used to form a given test statistic.
\begin{assumption}[Convergence in probability of first and second sample moments]\label{as:plim}
For all $\Gamma \geq 1$, $I^{-1}\sum_{i=1}^I (T_i - \bar{q}_i)$, $I^{-1}\sum_{i=1}^I |T_i - \bar{q}_i|$, $I^{-1}\sum_{i=1}^I M_{\Gamma i}$, $I^{-1}\sum_{i=1}^I\nu^2_{\Gamma i}$, and $I^{-1}\sum_{i=1}^I\tilde{\nu}^2_{\Gamma i}$ converge in probability to constant limits. 
\end{assumption}
\begin{assumption}[Asymptotic normality]\label{as:clt}
For all $\Gamma \geq 1$,
$I^{-1}\sum_{i=1}^I(T_{\Gamma i} - T_{\Gamma i}') = o_p(I^{-1/2})$ and $I^{-1}\sum_{i=1}^I(T_{i} - T_{i}') = o_p(I^{-1/2})$ where $\{T_{\Gamma i'}\}$ and $\{T_i'\}$ are sequences of independent random variables. $I^{-1/2}\sum_{i=1}^I\{T_{i}' - E(T_{i}')\}/\sqrt{I^{-1}\sum_{i=1}^I \text{Var}(T_{i}')}$ and $I^{-1/2}\sum_{i=1}^I\{T_{\Gamma i}' - E(T_{\Gamma i}')\}/\sqrt{I^{-1}\sum_{i=1}^I \text{Var}(T_{\Gamma i}')}$ converge jointly in distribution to a multivariate normal.
\end{assumption}

Moving forward, let $\theta = \text{plim}\; I^{-1}\sum_{i=1}^I(T_i-\bar{q}_i)$, let $\eta = \text{plim}\; I^{-1}\sum_{i=1}^I|T_i-\bar{q}_i|$, and let $m_{\Gamma} = \text{plim}\;I^{-1}\sum_{i=1}^IM_{\Gamma i}$, where $\text{plim}$ denotes the limit in probability.

\subsection{Design sensitivity for the tilted sensitivity analysis}

\citet{ros04} introduced design sensitivity to compare the limiting performance of test statistics in a sensitivity analysis. The design sensitivity is derived under the favorable setting for an observational study of a positive treatment effect and no hidden bias. Whether this favorable setting actually holds is unknown to the practitioner, so even under the favorable setting a researcher would like to assess the robustness of their findings to hidden bias through conducting a sensitivity analysis. Under the generative model described in \S \ref{sec:gen} with $\tau > 0$ in (\ref{eq:gen}) and under Assumptions \ref{as:plim} and \ref{as:clt}, there exists a number $\tilde{\Gamma}$, the design sensitivity, such that a sensitivity analysis under (\ref{eq:sensmodel}) will reject the null of no effect with probability tending to 1 at $\Gamma < \tilde{\Gamma}$, and with probability tending to zero for $\Gamma > \tilde{\Gamma}$. Test statistics conferring larger values for $\tilde{\Gamma}$ are preferred. 

The design sensitivity for the conventional sensitivity analysis is\begin{align}\label{eq:dsconv}
\tilde{\Gamma}_{conv} &= \texttt{SOLVE}\left\{\tilde{\Gamma}: \theta = \left(\frac{\tilde{\Gamma}-1}{1+\tilde{\Gamma}}\right)m_{\tilde{\Gamma}}\right\}.
\end{align}In words, it is the value $\tilde{\Gamma}_{conv}$ such that the limiting value for the observed test statistic equals the limiting value of the worst-case expectation (assuming the sharp null) returned by the separable algorithm. Unfortunately, for  $n_i-1\geq	 2$ controls the design sensitivity $\tilde{\Gamma}_{conv}$ does not admit a closed form for the general class of sum statistics precisely because the separable algorithm does not provide a closed form for $M_{\Gamma i}$. In practice, it is instead calculated through Monte Carlo simulation and numeric root finding. This has hindered theoretical comparisons of different choices of test statistic using design sensitivity as a guide when matching with multiple controls.

For the tilted sensitivity analysis, no such difficulty arises. By using a test statistic with a known upper bound on the worst-case expectation under the null when (\ref{eq:sensmodel}) is assumed to hold at $\Gamma$, the tilted sensitivity analysis \textit{does} admit a closed-form solution for its design sensitivity, even when matching with multiple controls. The following theorem provides the closed form.
\begin{theorem}\label{thm:closed}
Under Assumptions \ref{as:plim} and \ref{as:clt} and for $0 < \theta < \eta$, the design sensitivity for the tilted sensitivity analysis is \begin{align}\label{eq:tiltdesign}
\tilde{\Gamma}_{tilt} &=  \texttt{SOLVE}\left\{\tilde{\Gamma}: \theta = \left(\frac{\tilde{\Gamma}-1}{1+\tilde{\Gamma}}\right)\eta\right\} = \frac{\eta + \theta}{\eta - \theta}.
\end{align}
\end{theorem}

The form of the design sensitivity, available for matching with multiple controls, aligns with existing closed-form solutions for design sensitivity in the conventional approach for the case of matched pairs \citep{ros13}. For tilting, this implies that available intuition for strategies increasing design sensitivity in paired designs port over naturally to matching with multiple controls. Essential to the proof of Theorem \ref{thm:closed} is that the output of the separable algorithm when applied to the tilted sensitivity analysis is available in closed form by Lemma \ref{lemma:AGamma}. The design sensitivity $\tilde{\Gamma}_{tilt}$ is increasing in the ratio $\theta/\eta$, illustrating how test statistics with larger design sensitivity have expectations that are large relative to their absolute expectations under the favorable setting; this helps explain why strategies de-emphasizing smaller realized values of the test statistic such as inner trimming can improve design sensitivity, as these values contribute relatively more to the absolute value of the statistic than they do to the statistic itself.

\subsection{Illustration: $M$-statistics and the number of controls}
A closed-form solution for the design sensitivity enables analytic comparisons between competing research designs and test statistics. To illustrate, we consider the impact of the number of matched controls on design sensitivity for $m$-statistics, studied for the conventional approach by Monte Carlo simulation in \citet{ros13}. $M$-statistics are statistics asymptotically equivalent to 
\begin{align}\label{eq:mstat}
T &= \sum_{i=1}^I\sum_{j=1}^{n_i}Z_{ij}\sum_{\ell=1}^{n_i}\psi(Y_{ij}-Y_{i\ell}),
\end{align} 
where $\psi$ is a non-stochastic function assumed odd, monotone non-decreasing and uniformly continuous. The following theorem establishes after tilting, that design sensitivity is non-decreasing in the number of matched controls $J$ when using $m$-statistics, and is strictly increasing under additional support conditions.

\begin{theorem}\label{thm:mstat}
Suppose that $n_i = (J+1)$ for all $i=1,...,I$ with $J\geq 1$. Let $\theta_J = E\{\sum_{j=1}^{n_i} Z_{ij}\sum_{\ell=1}^{n_i}\psi(Y_{ij}-Y_{i\ell})\}$, and let $\eta_J =  E|\sum_{j=1}^{n_i}Z_{ij}\sum_{\ell=1}^{n_i}\psi(Y_{ij}-Y_{i\ell})\}|$ with $0 < \theta_J < \eta_J$. Under the generative model in \S \ref{sec:gen}, the design sensitivity for the tilted sensitivity analysis using $m$-statistics is $\tilde{\Gamma}_{tilt, J}= (\eta_J + \theta_J)/(\eta_J - \theta_J)$. Moreover, $\tilde{\Gamma}_{tilt, J}$ is non-decreasing in $J$, and is strictly increasing in $J$ under mild conditions on the distribution $F_\epsilon$. 
\end{theorem}

\subsection{Analytic comparison of design sensitivities}\label{sec:compare}

Comparing (\ref{eq:dsconv}) and (\ref{eq:tiltdesign}) provides the following simple condition for determining whether or not the tilted sensitivity analysis outperforms the conventional approach in terms of design sensitivity.

\begin{proposition}\label{prop:compare}
Suppose that Assumptions \ref{as:plim} and \ref{as:clt} hold, and let $\tilde{\Gamma}_{tilt}$ and $\tilde{\Gamma}_{conv}$ be the design sensitivity for the tilted and conventional approaches respectively. Assume the design sensitivities are uniquely defined. Then, the design sensitivity for the tilted sensitivity analysis is greater than that of the conventional sensitivity analysis if and only if $m_{\tilde{\Gamma}_{tilt}} > \eta$, or equivalently, if and only if $m_{\tilde{\Gamma}_{conv}} > \eta$.
\end{proposition}

The lack of a closed form for $m_{\Gamma}$ makes it challenging to precisely characterize circumstances under which the design sensitivity of the tilted sensitivity analysis will be larger (better) than that of the conventional approach for any particular $\tau > 0$. For a subclass of $m$-statistics studied in \citet{ros13}, we instead present a comparison along these lines when the treatment effect is small, in the sense that one can establish the existence of a $\tilde{\tau} > 0$ such that one approach is superior to the other for $0 < \tau < \tilde{\tau}$. This assesses a statistic's ability to distinguish a small treatment effect from a small degree of unmeasured confounding. 

Consider the following class of trimmed $m$-statistics
\begin{align}\label{eq:trim}
\psi_{\iota, h}(y) &= \left(\frac{h}{h-\iota}\right)\text{sgn}(y)\max\{0, \min\{|y|,h\}-\iota\}\;\;\;\; (0\leq \iota < h).
\end{align}This function evaluates to 0 for all $|y|<\iota$, increases linearly for $\iota \leq |y| \leq h$, and evaluates to $\pm h$ for $y \geq h$ or $y\leq -h$ respectively; $\iota$ and $h$ are referred to as the degrees of inner and outer trimming. We further define
\begin{align}\label{eq:special}
\psi_{lin}(y) = \lim_{h \rightarrow \infty} \psi_{0,h}(y) = y;\;\;\;\;
\end{align} Evaluating (\ref{eq:mstat}), we see that $\psi_{lin}$ returns $n_i-1$ times the treated-minus-control difference in means for the $i$th contribution to the observed test statistic $T_i$. 

Suppose $n_i = J+1$ for all $i=1,...,I$. Consider again the generative model of \S \ref{sec:gen} for $\tau > 0$, and note that the observed responses satisfy $Y_{ij} = \alpha_i+\epsilon_{ij}+Z_{ij}\tau$ with $\epsilon_{ij}\overset{iid}{\sim}F_\epsilon$. Define $q_{ij}(\tau) = \sum_{\ell=1}^{J+1}\psi\left\{\epsilon_{ij}-\epsilon_{i\ell} + (Z_{ij}-Z_{i\ell})\tau\right\}$ and $T_i(\tau) = \sum_{j=1}^{J+1}Z_{ij}q_{ij}(\tau)$; these are the values for $q_{ij}$ and the observed statistic $T_i$ expressed as a function of the treatment effect $\tau$.  Define $\bar{A}_{i} = (J+1)^{-1}\sum_{j=1}^{J+1}|q_{ij}(0)|$ and $\bar{S}_{i} = (J+1)^{-1}\sum_{j=1}^{J+1}\text{sgn}(q_{ij}(0))$ as the average of the absolute values and signs of $q_{ij}(0)$ respectively; $q_{ij}(0)$ represents what the values of $q_{ij}$ would have been if there were actually no treatment effect.

For any $G$ such that $\epsilon_{i1},...,\epsilon_{i(J+1)}\overset{iid}{\sim} G$, let $\beta_{\psi, G} = \left[\frac{\partial}{\partial \tau} \;E_G\{T_i(\tau)\}\Bigr|_{\tau=0^+}\right]/E_{G}(\bar{A}_{i})$.  Define the skewness functional $K(\psi, F_\epsilon) = \{B(\psi, F_\epsilon) - B(\psi, F_{-\epsilon})\}/2$, where 
\begin{align*}
B(\psi,G) &=   \frac{\partial}{\partial \tau}\left\{\frac{1}{J+1}\sum_{j=1}^{J+1}E_G|q_{ij}(\tau) - \tau \beta_{\psi, G}\bar{A}_i | -  E_G|T_i(\tau)|\right\}\Bigr|_{\tau=0^+}\end{align*} and $F_{-\epsilon}$ is the distribution of $-\epsilon_{ij}$ when $\epsilon_{ij} \sim F_{\epsilon}$. By construction $K(\psi, F_\epsilon) =  0$ when $F_\epsilon$ is centrally symmetric, and it further satisfies $K(\psi, F_\epsilon) = -K(\psi, F_{-\epsilon})$, both typical properties of skewness measures. In the web-based supplementary material we further argue that it tends to be positive for right-skewed distributions and negative for left-skewed distributions.

The following theorem describes sufficient conditions on the skewness measure $K(\psi, F_\epsilon)$ for comparing the tilted and conventional sensitivity analysis for different $\psi$. Recall that $F_\epsilon$ centrally symmetric implies $K(\psi, F_\epsilon)=0$.
\begin{theorem}\label{thm:designsens}
Set $J=n_i-1 \geq 2$ and consider comparing the tilted and conventional sensitivity analyses. Call approach $A$ superior to approach $B$ in the small-$\tau$ regime if there exists a $\tilde{\tau} > 0$ such that for all $0 < \tau < \tilde{\tau}$, the design sensitivity for approach $A$ is greater than that of approach $B$. Under suitable regularity conditions on $F_\epsilon$, we have the following conclusions:
\begin{itemize}
 \item[(i)] Consider $\psi_{lin}$. If $K(\psi_{lin}, F_\epsilon) > 0$, the tilted approach is superior in the small-$\tau$ regime. For $K(\psi_{lin}, F_\epsilon) < 0$, the conventional approach is superior in the small-$\tau$ regime. For $K(\psi_{lin}, F_\epsilon) = 0$, the tilted approach is superior if
\begin{align}\label{eq:slice}
\frac{f_0E(\bar{A}_i^2\mid T_i(0)=0) + E(\bar{A}_i\bar{S}_i^2) - E(\bar{A}_i)\{E(\bar{S}_i)\}^2}{\{E(\bar{A}_i)\}^2} > f_0\left(\frac{J-1}{J}\right),
\end{align}
where $f_0$ is the density of $q_{ij}(0)$ evaluated at 0. If the inequality in (\ref{eq:slice}) is reversed when $K(\psi_{lin}, F_\epsilon) = 0$, the conventional approach is superior in the small-$\tau$ regime. For Gaussian $F_\epsilon$ the inequality in (\ref{eq:slice}) holds, and the tilted approach is superior in the small-$\tau$ regime.

\item[(ii)] Consider $\psi_{\iota, h}$ for $0 < \iota < h < \infty$. If $K(\psi_{\iota, h}, F_\epsilon) \geq 0$, the tilted approach is superior in the small-$\tau$ regime. 
\item[(iii)] Consider $\psi_{0, h}$ with $0<h<\infty$ and let $J$ be even. If $K(\psi_{0,h}, F_\epsilon) \geq 0$, the tilted approach is superior in the small-$\tau$ regime.
\item[(iv)] Consider $\psi_{0, h}$ with $0<h<\infty$ and let $J$ be odd. If $K(\psi_{0,h}, F_\epsilon) > 0$, the tilted approach is superior in the small-$\tau$ regime. If $K(\psi_{0,h}, F_\epsilon) < 0$, the conventional approach is superior in the small-$\tau$ regime. If $K(\psi_{0,h}, F_\epsilon)=0$, a distribution-dependent second order coefficient determines whether the tilted or conventional approach is superior in the small-$\tau$ regime. 
\end{itemize}
\end{theorem}

See the web-based supporting materials for a proof along with regularity conditions. The main difficulty in the proof comes from establishing a tractable expansion for $m_{\Gamma} = E(M_{\Gamma i})$ for $\tau$ small given that no closed form exists for $M_{\Gamma i}$. The coefficients of the resulting expansions highlight the role of the shape of the distribution of $F_\epsilon$ in determining which approach is superior. Skewness produces first-order contributions for the expansions. Across all $\psi$, right-skewness is favorable for the tilted sensitivity analysis, while left-skewness is often favorable for the conventional sensitivity analysis. Both the parity dependence for $\psi_{0,h}$ and the contrasting conclusions for $\psi_{0,h}$ versus $\psi_{\iota, h}$ for $\iota > 0$ stem from the behavior of the distribution of $q_{ij}(0)$ at 0. For $\epsilon_{ij}$ continuously distributed and for $\psi_{\iota,h}$ with $\iota > 0$, the distribution of $q_{ij}(0)$ has an atom at 0 for all $J$ under mild conditions on the support of $q_{ij}(0)$. For $\psi_{0,h}$ atoms at 0 arise when $J$ is even but do not arise for $J$ odd. Atoms at 0 provide a first-order contribution which pushes the minimal value of $K(\psi, F_\epsilon)$ needed for the tilted approach to be superior below 0, in turn guaranteeing superiority of tilting under central symmetry. For the conventional approach to be superior in cases (ii) and (iii), $F_\epsilon$ must be sufficiently left-skewed to overcome the positive contribution from atoms at 0.

When atoms are absent at 0, the comparison at $K(\psi, F_\epsilon)=0$, which includes all centrally symmetric $F_\epsilon$, requires a second-order expansion in $\tau$.  For the linear statistic $\psi_{lin}$ there are no atoms at 0 for any $J$ when $\epsilon_{ij}$ is continuously distributed. The inequality (\ref{eq:slice}) arises from these second-order expansions for $\psi_{lin}$. The dependence in the left-hand side of (\ref{eq:slice}) on the conditional distribution given $T_i(0)=0$ complicates analytic comparisons even under an assumption that $F_\epsilon$ is centrally symmetric. The unique properties of conditional normal distributions enable the proof that tilted sensitivity analysis is superior for Gaussians when using $\psi_{lin}$. For $\psi_{0,h}$ and $J$ odd, the threshold for determining superiority of the tilted approach at $K(\psi_{0,h}, F_\epsilon)=0$ is the sum of two terms, one akin to (\ref{eq:slice}) and another non-negative term from the kinks in $\psi_{0,h}$ at $\pm h$. See the web-based supporting materials for a discussion of these kink contributions.

\subsection{Monte Carlo comparison of design sensitivities}
\label{sec:sim}
The expansion-based results in Theorem \ref{thm:designsens} are valid for $m$-statistics of the form (\ref{eq:trim}) for $\tau$ in a range $0<\tau<\tilde{\tau}$; however, the value $\tilde{\tau}$ depends upon both the test statistic and the error distribution, making it less clear how well the results characterize behavior for moderately sized treatment effects. We now use Monte Carlo simulation to compare the design sensitivities of the tilted and conventional approaches under the generative model described in \S\ref{sec:gen}.  We vary the number of matched controls, the distribution $F_\epsilon$ in (\ref{eq:gen}), and the choice of test statistic. We consider matching with $J=2,3$ and $5$ controls and compare performance with three different test statistics. We use the difference in means statistic;  the aligned rank statistic of \citet{hod63} with alignment by stratum means; and the trimmed $m$-statistic with $\iota=0, h=2.5\hat{s}$, where $\hat{s}$ is the sample median of the $IJ(J+1)/2$ values $|Y_{ij}-Y_{i\ell}|$, $j< \ell$. This $m$-statistic is described in detail in \citet{ros13}.  We consider distributions for the errors in (\ref{eq:gen}) that are Normal, $t_3$, $\text{Exp}(1)-1$, and $1-\text{Exp}(1)$, where Exp($\lambda$) is an exponential distribution with rate parameter $\lambda$. The Normal and $t_3$ distributions are symmetric, while $\text{Exp}(1)-1$, and $1-\text{Exp}(1)$ are right-skewed and left-skewed respectively. We set $\tau/\sigma = 1/2$ where $\tau$ is defined as in (\ref{eq:gen}) and $\sigma^2 = \text{Var}(\varepsilon_{1ij} - \varepsilon_{0i\ell})$, $j\neq \ell$; this accounts for differences in the variances of the generative models, measuring treatment effects in units of the standard deviations of the potential observed differences for individuals in the same matched set. For each setting, we compute the design sensitivities of the tilted and conventional approaches through $I=100,000$ Monte Carlo simulations.

Table \ref{tab:designsens} shows the results. We see that as $J$ increases, the design sensitivities of both the tilted and conventional approaches increase; see the web-based supporting information for a discussion of why this holds for tilted $m$-statistics. We see that tilting provides an improvement in design sensitivity over the conventional approach for the symmetric and right-skewed error distributions for all three test statistics and all choices of $J$ except for the aligned rank statistic with $J=5$ and a $t_3$ generative model, where the conventional approach slightly outperforms the tilted variant. The conventional approach has superior design sensitivity for the left-skewed generative distribution, 1-Exp(1). 

In the web-based supporting materials, we include an additional simulation study which plots the difference between $m_{\tilde{\Gamma}_{tilt}}$ and $\eta$ as a function of $\tau/\sigma$ across a range of generative models. This plot illustrates that the comparisons in design sensitivity persist across a range of effect sizes. We also introduce two heuristics, one more interpretable and one more accurate, for predicting which of the tilted or conventional approaches will perform better in terms of both design sensitivity and sensitivity values when applied to real data. The heuristics apply across different test statistics and effect sizes, and compare features of the observed distribution of the within-set scores $q_{ij}-\bar{q}_i$ to the treated contributions $T_i-\bar{q}_i$. The more accurate heuristic correctly predicted the winning method for each setting in Table \ref{tab:designsens}. Together, these clarify the role that observable data features play in determining which procedure will perform better. 

\begin{table}\caption{Design sensitivities when matching with $J=2$, $J=3$ and $J=5$ controls for the tilted and conventional sensitivity analysis. Three test statistics are displayed across the columns. Across the rows, there are four error distributions $F_\epsilon(\cdot)$: Normal, $t_{3}$, Exp(1)-1, and 1-Exp(1), where Exp(1) is the exponential distribution with rate parameter 1. The signal to noise ratio is $\tau/\sigma = 1/2$ for all settings.}
\begin{center}
\begin{tabular}{l l c c c c c c c c c}
& & \multicolumn{3}{c}{Difference in Means} & \multicolumn{3}{c}{Huber Outer Trim} & \multicolumn{3}{c}{Aligned Rank}\\
& &$J=2$ & $J=3$ & $J=5$ & $J=2$ & $J=3$ & $J=5$& $J=2$ & $J=3$ & $J=5$\\
\hline
\multirow{2}{*}{Norm.} &Conv.& 4.08 & 4.35 & 4.86 & 3.88&4.20&4.64&3.69&4.06&4.43 \\
&Tilt. &  4.32 & 4.62 & 5.10 & 4.09 &4.45&4.84&3.87&4.17&4.50\\
\multirow{2}{*}{$t_3$}&Conv. &  4.97 & 5.38 & 5.89 & 5.19 & 5.80&6.54&5.13&5.76&6.70\\
&Tilt.& 5.22 & 5.70 & 6.31 & 5.52 &6.25&6.92&5.37&5.95&6.60\\
\multirow{2}{*}{E(1)-1}&Conv.& 4.14 & 4.22 & 4.40 & 4.16&4.54&4.86&4.34 & 5.07&5.94\\
&Tilt.&  5.37 & 6.37 & 8.06 & 5.45&7.13&9.24&5.62&7.54&11.0\\
\multirow{2}{*}{1-E(1)} &Conv. &  5.09 & 6.15 & 8.01 &5.23&6.17&7.96&4.96&5.91&7.08 \\
&Tilt.&  4.21 & 4.36 & 4.58 & 4.25&4.44&4.67&4.19&4.43&4.56\\
\end{tabular}
\end{center}\label{tab:designsens}
\end{table}

\section{Adaptive combination}\label{sec:adapt}

Theorem \ref{thm:designsens} and Table \ref{tab:designsens} reveal that the tilted sensitivity analysis does not uniformly improve upon the conventional approach in terms of design sensitivity; rather, the improvement is tied to the underlying distribution of $q_{ij}-\bar{q}_i$. An incorrect \textit{a priori} choice of approach could unnecessarily decrease the reported robustness to hidden bias. Fortunately, one need not choose ahead of time between the tilted and conventional approaches. Rather, the approaches may be combined to attain a design sensitivity no smaller than the maximum of the tilted and conventional design sensitivities by employing the approach for multivariate one-sided testing introduced in \citet{coh20mult}. Moreover, with this combination method one can attain a strictly larger design sensitivity than the maximum of the tilted and conventional design sensitivities. That is, one could do strictly worse in terms of design sensitivity by prespecifying a particular method before looking at the data than would be possible with a data-adaptive choice. In short, the method adaptively chooses a weighted combination of the tilted and conventional test statistics, allowing the method to adaptively combine the two approaches based upon whatever weighting performs best in the context of the sensitivity analysis. We defer a description of the combination method to the web-based supporting materials, where we also present simulation studies comparing the power of a sensitivity analysis using the tilted, conventional, and adaptive approaches in finite samples. These illustrate the minimal loss in power from adaptivity relative to whichever of the tilted and conventional approaches performed better in a given generative model. Furthermore, the simulations show the potential for large gains relative to the worse of the tilted and conventional approaches. See also \citet{hen21} for an alternative approach to adaptive combination of test statistics for improved design sensitivity.

\section{Data examples}\label{sec:data}
Design sensitivity provides but one of many metrics by which methods for sensitivity analysis may be judged. While the favorable setting of no bias and a treatment effect enables theoretical comparisons, whether benefits under this setting reflect benefits in practice is not a foregone conclusion. In real-world examples, it is unlikely that  (\ref{eq:sensmodel}) actually holds at $\Gamma=1$ (no unmeasured confounding). In this section, we compare the performance of the tilted, conventional, and adaptive approaches in several real-data examples. For each matched comparison, we consider a sensitivity analysis for the sharp null hypothesis of no effect at $\alpha=0.05$. For each method, we report the \textit{sensitivity value} \citep{zha19senval}, the largest value of $\Gamma$ for which the sensitivity analysis rejects the null hypothesis. Larger sensitivity values imply that the findings are more robust to hidden bias. 

We consider the following data sets, all of which are either available within packages on \texttt{CRAN} or are included in tabular form within the cited papers.
\begin{itemize}
\item[(SL)] Smoking and lead levels in the blood. $I=150$. One treated, five controls in each matched set \citep{ros13}.
\item[(FM)] Fish consumption and mercury levels in the blood. $I=397$. One treated, two controls in each matched set \citep{ros14}.
\item[(DC)]  Dropping out of high school and cognitive achievement. $I=12$. One treated, two controls in each matched set \citep{gas00}.
\item[(AB)]  Alcohol and blood pressure. $I=206$. One treated, two controls in each matched set \citep[\S 1.5]{ITOS}.
\item[(LH)]  Light drinking and HDL cholesterol levels. $I=200$. One treated, three controls in each matched set \citep[\S 1.4]{ITOS}.
\end{itemize}
We consider four test statistics. As in the simulations in \S \ref{sec:compare}, we consider the difference in means statistic, the $m$-statistic $\psi_{hu, 2.5}$, and the aligned-rank test of \citet{hod63}. In addition, we consider the weighted rank test denoted \texttt{u868} in \citet{ros24}, highlighted therein for its impressive performance in sensitivity analysis. 

\begin{table}
\begin{center}
\begin{tabular}{l || c c c | c c c | c c c | c c c}
& \multicolumn{3}{c}{Difference in Means} & \multicolumn{3}{c}{Huber Outer Trim} & \multicolumn{3}{c}{Aligned Rank} & \multicolumn{3}{c}{Rank Sum, \texttt{u868}}\\
& C & T & A & C & T & A & C & T & A & C & T & A\\
SL & 1.49 & 1.53 & 1.52 & 2.07 & 2.18 & 2.15 & 2.00 & 2.10 & 2.07 & 1.50 & 1.49 & 1.49\\
FM& 15.9 & 20.8 & 20.4 & 14.0 & 19.9 & 19.4 & 15.3&21.2&20.6 & 18.1 & 37.0 & 32.1 \\
DC & 1.32 & 1.34 & 1.32 & 1.30 & 1.34 & 1.31 & 1.36 & 1.38 & 1.35 & 1.63 & 1.64 & 1.57 \\
AB& 2.18 & 2.20 & 2.18 & 2.17 & 2.18 & 2.16 & 2.11 & 2.11 & 2.10 & 2.02 & 2.05 & 2.02\\
LH& 3.74 & 3.64 & 3.69 & 3.38 & 3.30 & 3.34 & 3.30 & 3.15 & 3.25 & 4.47 & 4.11 & 4.86\\
\end{tabular}

\caption{\label{tab:change} The largest value of $\Gamma$ such that the null hypothesis is rejected at $\alpha=0.05$ using the conventional (C), tilted (T), and adaptive (A) approaches with various test statistics.} 

\end{center}
\end{table}

Table \ref{tab:change} shows the results. We see that for the data sets FM, DC, and AB, the tilted sensitivity analysis outperformed the conventional approach for all choices of test statistic. For LH, the conventional sensitivity analysis outperforms the tilted approach for all test statistics. For SL, the tilted sensitivity analysis outperformed the conventional approach for all test statistics except for the \texttt{u868} rank test from \citet{ros24}. The  heuristics developed in the web-based supporting material which compare the observed within-set spread of the centered scores $q_{ij}-\bar{q}_i$ to that of the realized treated contribution $T_i-\bar{q}_i$ largely mirrored the sensitivity value comparisons in Table 2. The refined heuristic predicted the better performing method for every data-set/statistic combination except SL/\texttt{u868}, where the conventional value only slightly exceeded the tilted value, 1.50 versus 1.49.  

The adaptive approach generally lags behind the better of the two approaches, but can perform materially better than the laggard; this is particularly striking in the findings from data set FM, where choosing the conventional sensitivity analysis would have yielded materially lower sensitivity values than the tilted or adaptive approaches. Observe that for the \texttt{u868} rank-sum test in data set LH, the adaptive procedure outperforms both the conventional \textit{and} the tilted sensitivity analyses. This occurs because, for $4.47 < \Gamma < 4.86$, there is no single pattern of hidden bias under which the tilted and conventional test statistics simultaneously fail to reject the null. In particular, the tilted test statistic rejects the null hypothesis when assessed at the worst-case unmeasured confounder used by the conventional approach, and the conventional statistic rejects the null when assessed at the worst-case unmeasured confounder used by tilting. The adaptive approach correctly enforces the constraint that there must be a single vector of assignment probabilities governing the distribution of both test statistics. 

 
\section{Discussion}
The tilted sensitivity analysis has appealing practical and theoretical properties: it may be implemented in closed form without use of the separable algorithm of \citet{gas00}; it has a closed-form expression for its design sensitivity even when matching with multiple controls; it improves design sensitivity for a range of generative models; and it improves reported robustness to hidden bias in many real-data examples as illustrated in Table \ref{tab:change}. Furthermore, one need not choose between the tilted and conventional approaches to sensitivity analysis with matched controls: the adaptive approach outlined in \S \ref{sec:adapt} provides insurance against a poor choice. While the exposition herein has focused on tests of sharp null hypotheses for the general class of sum statistics, for the difference in means statistic \citet{fog23} employed tilting in an investigation of tests of the weak null of no effect on average. There it was shown that the untilted difference in means fails to provide a valid sensitivity analysis for the average treatment effect, and that left-skewed generative models can provide large gaps between the nominal and actual Type I error rates. It is of interest that left-skewed generative models are precisely the models where the conventional approach outperformed the tilted variant, as demonstrated in Table \ref{tab:designsens}. Part of the deficit in power for the tilted approach in this setting may be attributed to the fact that it also provides valid inference for the average treatment effect under additional restrictions.


When using $m$-statistics with inner trimming, we have established that the tilted approach outperforms the conventional approach in the small-$\tau$ regime for centrally symmetric errors, and for the difference in means we have done so for Gaussian errors. We do not currently have a characterization of a broader subclass of centrally symmetric laws under which tilting will perform better for the difference in means. The class of centrally convex unimodal laws \citep[Definition 2.5]{dha88} is too large: one can construct mixtures of centered Gaussians for which the conventional approach is superior. These mixture laws are centrally convex unimodal, but are not log-concave. An open question is whether when $F_\epsilon$ log-concave, one can establish superiority of the tilted approach in the small-$\tau$ regime for the difference in means.

\section*{Acknowledgments}
This work is supported in part by funds from the National Science Foundation (NSF DMS-2413484).

\appendix
\begin{center}\Large{Appendix}\end{center}

\section{Comparing design sensitivities and sensitivity values}
\subsection{Comparing design sensitivity as a function of effect size}
Proposition \ref{prop:compare} provides a straightforward means of comparing the design sensitivities of the tilted and conventional approaches by way of Monte Carlo simulation. It is particularly convenient to leverage Proposition \ref{prop:compare} at $\Gamma = \tilde{\Gamma}_{tilt}$, given the closed-form expression for the tilted sensitivity analysis available from Theorem \ref{thm:closed} as a function of $\eta$ and $\theta$. We now compare the tilted and conventional sensitivity analyses under the generative model described in \S\ref{sec:gen}, varying the number of matched controls and the distribution $F_\epsilon$. We consider using a trimmed $m$-statistic with $\iota = 0$, $h=2.5\hat{s}$, where $\hat{s}$ is the sample median of the $IJ(J+1)/2$ values $|Y_{ij}-Y_{i\ell}|$, $j < \ell$. This particular choice of $\psi$ in (\ref{eq:mstat}) is an $m$-statistic inspired by the loss function of \citet{hub81}; see \citet{ros07} for more details. We consider choices for the marginal distributions for $(\varepsilon_{0ij},\varepsilon_{1ij})$ that are (i) standard normal; (ii) $t_3$; (iii) standard double exponential; (iv) $\text{Exp}(1)-1$; and (v) $1-\text{Exp}(1)$, where $\text{Exp}(\lambda)$ is an exponential distribution with rate $\lambda$. All distributions have mean zero. Distributions (i)-(iii) are symmetric about zero, distribution (iv) is right-skewed, and distribution (v) is left-skewed. For each choice of generative distribution and number of matched controls, we evaluate Monte Carlo estimates of $m_{\tilde{\Gamma}_{tilt}} - \eta$ for a range of values of $\tau > 0$ based upon $I=100,000$ matched sets. 

\begin{figure}
\centering \includegraphics[scale=.8]{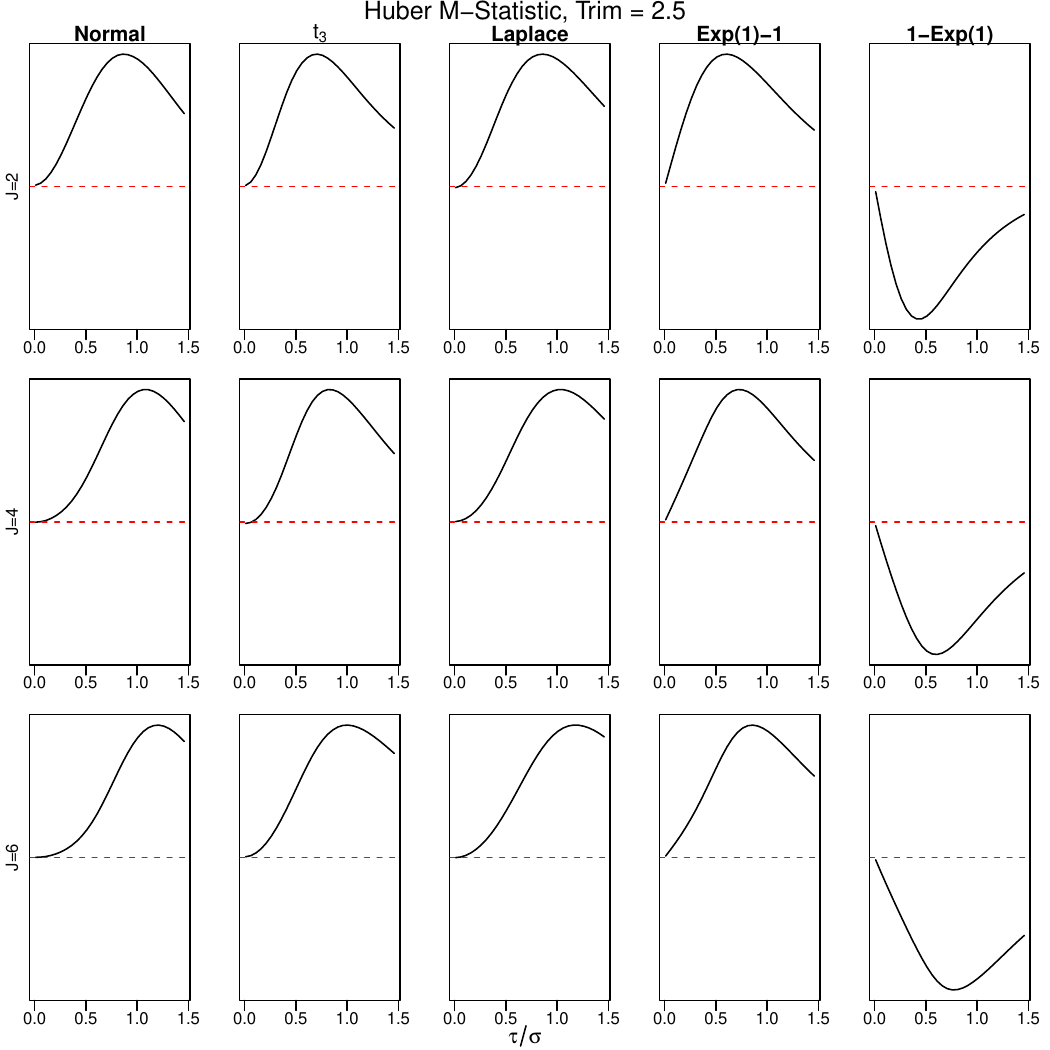}
\caption{Comparing the design sensitivities of the tilted and conventional approaches for the trimmed $m$-statistic with $\iota=0$, $h=2.5\hat{s}$. The plots show Monte Carlo estimates of $m_{\tilde{\Gamma}_{tilt}} - \eta$ as a function of $\tau/\sigma$, along with a horizontal dotted line at $y=0$ for a given choice of the marginal distributions for $(\varepsilon_{0ij}, \varepsilon_{1ij})$ (varying across columns) and number of matched controls (varying across rows) in the generative model in \S \ref{sec:gen}. By Proposition \ref{prop:compare}, if for any given generative model and number of matched controls $J$ the function lies above the dotted line at a given value of $\tau/\sigma$, the tilted sensitivity analysis has a better design sensitivity than the conventional approach for that value of $\tau/\sigma$; otherwise, the conventional approach has the larger design sensitivity.}\label{fig:compare} 
\end{figure}
Figure \ref{fig:compare} shows the Monte Carlo estimates of $m_{\tilde{\Gamma}_{tilt}} - \eta$ for the trimmed Huber $m$-statistic with different numbers of controls $J$ and generative distributions as a function of $\tau/\sigma$, where $\sigma^2 = \text{Var}(\varepsilon_{1ij} - \varepsilon_{0i\ell})$, $j\neq \ell$; this accounts for differences in the variances of the generative models, measuring treatment effects in units of the standard deviations of the potential treated minus control differences for individuals in the same matched set. We see that for the normal, $t_3$, double exponential, and $\text{Exp}(1)-1$ distributions, the y coordinates of the plots are above zero for all displayed values of $\tau/\sigma$ and all displayed numbers of controls. By Proposition \ref{prop:compare}, this implies that the tilted sensitivity analysis has a larger design sensitivity than the conventional approach for each of these scenarios and for all values of $\tau/\sigma$. When the distribution is instead $1-\text{Exp}(1)$, a left-skewed distribution, we see that the conventional sensitivity analysis outperforms the tilted approach for all numbers of controls displayed.
\subsection{Heuristics for comparing the tilted and conventional approaches}
Here we develop heuristics for comparing both the design sensitivities and sensitivity values of the tilted and conventional approaches, meant to apply broadly across choices of test statistics, generative models, and effect sizes. Theorem \ref{thm:designsens} provides a rigorous analogue of these heuristics for small treatment effects and trimmed $m$-statistics under the favorable setting of no hidden bias. Let $T_i = \sum_{j=1}^{n_i}Z_{ij}q_{ij}$ and consider the conventional and tilted statistics with stratum-wise weights $w = (w_1,...,w_I)^T$ satisfying $w_i \geq 0$ and normalized so that $I^{-1}\sum_{i=1}^I w_i=1$:
\begin{align*}
T(w)&= \sum_{i=1}^I w_iT_i;\\ T_\Gamma(w) &= \sum_{i=1}^I w_iT_{\Gamma i} = \sum_{i=1}^Iw_i\left[T_i - \bar{q}_i - \{(\Gamma-1)/(1+\Gamma)\}|T_i-\bar{q}_i|\right].
\end{align*}

The forthcoming heuristics arise from the fixed-point representation for the solution to the separable algorithm in Proposition \ref{prop:MGamma}. Define the fixed-point map $A_{\Gamma i}(t) = n_i^{-1}\sum_{j=1}^{n_i}|q_{ij}-\bar{q}_i-\kappa_\Gamma t|$ with $\kappa_{\Gamma} = (\Gamma-1)/(\Gamma+1)$. Then, from Proposition \ref{prop:MGamma}, \begin{align*}
M_{\Gamma i} &= \frac{1}{n_i}\sum_{j=1}^{n_i}|q_{ij}-\bar{q}_i - \kappa_\Gamma M_{\Gamma i}|\\
&= \texttt{SOLVE}\{t : t = A_{\Gamma i}(t)\}
\end{align*}

As $\kappa_\Gamma < 1$ for $\Gamma < \infty$, the fixed-point equation $A_{\Gamma i}(\cdot)$ is a contraction mapping: for any $s$ and $s'$, using the reverse triangle inequality, observe
\begin{align*}
|A_{\Gamma i}(s) - A_{\Gamma i}(s')| &= \left| \frac{1}{n_i}\sum_{j=1}^{n_i}|q_{ij}-\bar{q}_i - \kappa_\Gamma s| - \frac{1}{n_i}\sum_{j=1}^{n_i}|q_{ij}-\bar{q}_i - \kappa_\Gamma s'|\right|\\
& \leq \kappa_\Gamma |s-s'|.
\end{align*} This implies that for any initial $s^{(0)}\neq M_{\Gamma i}$, the output $s^{(1)} = A_{\Gamma i}(s^{(0)})$ from one round of fixed-point iteration is strictly closer to $M_{\Gamma i}$ than $s^{(0)}$ is:\begin{align*}
|M_{\Gamma i} - s^{(1)}| &= |A_{\Gamma i}(M_{\Gamma i}) - A_{\Gamma i}(s^{(0)})|\\
&\leq \kappa_{\Gamma}|M_{\Gamma i} - s^{(0)}| < |M_{\Gamma i} - s^{(0)}|. 
\end{align*} More generally, for $\tilde{\ell} > \ell \geq 0$ and for iterates of the form $s^{(\ell+1)} = A_{\Gamma i}(s^{(\ell)})$, $|M_{\Gamma i} - s^{(\tilde{\ell})}| \leq  \kappa_{\Gamma}^{\tilde{\ell}-\ell}|M_{\Gamma i} - s^{(\ell)}|$.

Define $\hat{\theta} = I^{-1}\sum_{i=1}^I w_i(T_i-\bar{q}_i)$ and $\hat{\eta} = I^{-1}\sum_{i=1}^Iw_i|T_i-\bar{q}_i|$, and let $\theta$ and $\eta$ denote the probability limits of $\hat{\theta}$ and $\hat{\eta}$ respectively. For the tilted statistic, so long as the required probability limits and limiting sensitivity values exist, the same proof used in Theorem \ref{thm:closed} shows that the limiting sensitivity value is $\tilde{\Gamma}_{tilt} = (\eta+\theta)/(\eta-\theta)$ even outside of the favorable setting, yielding $\kappa_{\tilde{\Gamma}_{tilt}} = \theta/\eta$. Moreover, the same proof as Proposition \ref{prop:compare} shows that under the regularity conditions needed for the existence of the limiting sensitivity value and for $0<\theta<\eta$, one can determine which method has the larger limiting sensitivity value by comparing $m_{\tilde{\Gamma}_{tilt}} = \text{plim}\; I^{-1}\sum_{i=1}^Iw_iM_{\tilde{\Gamma}_{tilt}i}$ to $\eta$: $m_{\tilde{\Gamma}_{tilt}} > \eta$ implies the tilted approach is superior, while $m_{\tilde{\Gamma}_{tilt}} < \eta$ implies the conventional approach is superior. 


This inspires the following finite-sample heuristic. First initialize $\hat{s}^{(0)} = \hat{\eta}$. Then, consider a single round of fixed-point iteration using the fixed-point maps $A_{\Gamma i}(\hat{s}^{(0)})$ at $\Gamma=\hat{\Gamma}$, where $\hat{\Gamma} = (\hat{\eta}+\hat{\theta})/(\hat{\eta}-\hat{\theta})$. Observe that $\hat{\kappa}= \hat{\theta}/\hat{\eta}$, so that $\hat{\kappa}\hat{s}^{(0)} = \hat{\theta}$. The fixed-point iteration yields $\hat{s}^{(1)}_i = A_{\hat{\Gamma}i}(\hat{s}^{(0)}) = n_i^{-1}\sum_{j=1}^{n_i}|q_{ij}-\bar{q}_i-\hat{\theta}|$. As $A_{\hat{\Gamma} i}(t) -t$ is strictly decreasing with a unique zero at $M_{\hat{\Gamma} i}$, on a set-by-set basis $\hat{s}_i^{(1)} - \hat{\eta}$ will be positive (negative)  precisely when $M_{\hat{\Gamma} i} - \hat{\eta}$ is positive (negative). 

Averaging the matched-set differences, we arrive at the heuristic
\begin{align*}
\hat{H}^{(1)} &= I^{-1}\sum_{i=1}^I\frac{w_i}{n_i}\sum_{j=1}^{n_i}|q_{ij}-\bar{q}_i-\hat{\theta}| - I^{-1}\sum_{i=1}^Iw_i|T_i-\bar{q}_i|. 
\end{align*}
The heuristic predicts that the tilted approach will be superior for $\hat{H}^{(1)} > 0$, and that the conventional approach will be superior for $\hat{H}^{(1)} < 0$. The population-level limiting analogue instead forms $H^{(1)} = \text{plim}_{I\rightarrow \infty} \hat{H}^{(1)}$.

A more accurate but less interpretable heuristic conducts an additional round of fixed-point iteration, plugging $\hat{s}^{(1)}_i$ into the fixed-point equation, returning $\hat{s}^{(2)}_i = A_{\hat{\Gamma} i}(\hat{s}^{(1)}_i)$, and comparing the weighted average of $\hat{s}^{(2)}_i$ to $\hat{\eta}$:
\begin{align*}
\hat{H}^{(2)} &= I^{-1}\sum_{i=1}^Iw_i A_{\hat{\Gamma}i}(\hat{s}^{(1)}_i)- I^{-1}\sum_{i=1}^Iw_i|T_i-\bar{q}_i|  \\
&=   I^{-1}\sum_{i=1}^I\frac{w_i}{n_i}\sum_{j=1}^{n_i}\left|q_{ij}-\bar{q}_i - \frac{\hat{\theta}/\hat{\eta}}{n_i}\sum_{\ell=1}^{n_i}|q_{i\ell}-\bar{q}_i - \hat{\theta}|\right| - I^{-1}\sum_{i=1}^Iw_i|T_i-\bar{q}_i|. 
\end{align*}The population analogue $H^{(2)}$ again takes the probability limit of $\hat{H}^{(2)}$.

\subsubsection{The role of distribution shape}
The heuristic $H^{(1)}$ highlights the role of distribution shape in determining whether the tilted or conventional approach will be superior. As an illustration, consider the generative model of \S \ref{sec:gen} at $w_i=1$ and with $n_i = J+1 > 2$ for all $i$. Assume without loss of generality that the first individual in each matched set received the treatment, and consider the difference-in-means statistic. For this statistic, we have $\theta = E(T_i-\bar{q}_i) = \tau$. Let $\bm{\zeta} \in \mathbb{R}^{J+1}$ with $\zeta_1=\theta$, $\zeta_j = -\theta/(J)$ for $j = 2,...,J+1$, and let $\epsilon_{ij} = Z_{ij}\varepsilon_{1ij} + (1-Z_{ij})\varepsilon_{0ij}$. Then, for the difference in means test, $q_{ij} = \tilde{q}_{ij} + \zeta_{j}$, where $\tilde{q}_{ij} = \epsilon_{ij} - \sum_{\ell \neq j}{\epsilon}_{i\ell}/(J)$. Therefore, $\eta = E|T_i-\bar{q}_i| = E|\tilde{q}_{ij} + \theta|$, while $n_i^{-1}\sum_{j=1}^{J+1}E|q_{ij}-\bar{q}_i - \theta| = \{1/(J+1)\}E|\tilde{q}_{ij}| + \{J/(J+1)\}E|\tilde{q}_{ij}-\{(J+1)/J\}\theta|$.

Let $h(a) = E|\tilde{q}_{ij} + a|$. The limiting heuristic $H^{(1)}$ is
\begin{align*}
H^{(1)} &= \left(\frac{1}{J+1}\right)h(0) + \left(\frac{J}{J+1}\right)h\left(-\frac{J+1}{J}\theta\right) - h(\theta)\\
&= \left(\frac{1}{J+1}\right)\{h(0) - h(\theta)\} + \left(\frac{J}{J+1}\right)\left\{h\left(-\frac{J+1}{J}\theta\right) - h(\theta)\right\}
\end{align*}

Let $F(a)$ be the CDF of $\tilde{q}_{ij}$ and assume $F(a)$ is continuous; then, $h'(a) = 1 - 2F(-a)$. Letting $\lambda_J = 1/(J+1)$ and $c_J = 1/(1-\lambda_J) = (J+1)/(J) > 1$, we have
\begin{align*}
H^{(1)} &= -\lambda_J\int_{0}^\theta h'(s)\;ds  -(1-\lambda_J)\int_{-c_J \theta}^{\theta} h'(s)\;ds\\
&= -\int_{0}^\theta h'(s)\;ds -(1-\lambda_J)\int_{-c_J \theta}^{0} h'(s)\;ds\\
&= -\int_{0}^\theta h'(s)\;ds -c_J(1-\lambda_J)\int_{-\theta}^{0} h'(c_Js)\;ds\\
&= -\int_{0}^\theta h'(s)\;ds - \int_{0}^{\theta} h'(-c_Js)\;ds\\
&= 2\int_0^\theta\left\{F(-s) + F(c_Js) - 1\right\}\; ds = 2\int_0^\theta\left\{\text{pr}(\tilde{q}_{ij} \leq -s) - \text{pr}(\tilde{q}_{ij} > c_J s)\right\}\; ds
\end{align*}
First consider evaluating the integrand at $s=0$. We then see that the integrand will be positive if $F(0) > 1/2$, and negative otherwise. Recalling that $E(\tilde{q}_{ij}) = 0$, the integrand at $s=0$ will be positive if the median lies below the mean, and negative if the median lies above the mean. These are properties commonly associated with right- and left-skewed data, respectively. More generally, when averaged over the range $0 \leq s \leq \theta$ the integrand tends to be positive for right-skewed data (right-skewed distributions with expectation zero have many moderate negative observations, and fewer positive observations that are larger in magnitude), and tends to be negative for left-skewed data (left-skewed distributions with expectation zero have many moderate positive observations, and fewer negative observations that are larger in magnitude). The heuristic tends to favor tilting for right-skewed distributions and the conventional approach for left-skewed distributions.

For centrally symmetric $\tilde{q}_{ij}$ with continuous CDF $F(a)$, $F(-s) = 1 - F(s)$ and $F(c_J s) - F(s) \geq 0$ for any $s \geq 0$ since $c_J  > 1$. Therefore, for centrally symmetric distributions,
 \begin{align*}
 H^{(1)} &= 2\int_0^\theta\left\{F(-s) + F(c_Js) - 1\right\}\; ds\\
 &= 2\int_0^\theta\left\{-F(s)+ F(c_Js)\right\} \geq 0.
 \end{align*}

Therefore, the heuristic favors tilting for centrally symmetric $\tilde{q}_{ij}$.

\subsubsection{Assessing heuristic accuracy}
To illustrate the accuracy of the heuristics, we consider both the Monte Carlo simulation in \S \ref{sec:sim} which produced Table \ref{tab:designsens} and the data analysis in \S \ref{sec:data} summarized in Table \ref{tab:change}. Applying the heuristics to the design-sensitivity simulations tests their ability to predict design-sensitivity comparisons, while applying them to the real data analyzed in \S \ref{sec:data} tests their utility in finite samples and outside of the favorable setting of no hidden bias. 

For the simulations in \S \ref{sec:sim}, we applied the heuristics with $w_i=1$ for all $i=1,...,I$. The heuristic $H^{(1)}$ successfully predicted which method would perform best in all settings except for the setting with $t_3$ errors at $J=5$ using the aligned rank statistic: here it predicted the tilted approach would have the better design sensitivity, but in reality the conventional approach had a higher design sensitivity (6.7 vs 6.6). The more accurate heuristic $H^{(2)}$ corrected this error, and successfully predicted all of the displayed design sensitivity comparisons in Table \ref{tab:designsens}.

For the data sets analyzed in \S \ref{sec:data}, we set $w_i=1$ for $i=1,...,I$ when using the difference in means, the $m$-statistic using outer trimming, and the aligned rank statistic. The \texttt{u868} rank statistic is a weighted sum of blockwise rank statistics, and we set $w_i$ equal to the corresponding weights, normalized so that $I^{-1}\sum_{i=1}^I w_i = 1$. The finite-sample heuristic $\hat{H}^{(1)}$ correctly predicted which method would have the larger sensitivity value in all data-set/test-statistic combinations except for the difference in means and outer-trimmed $m$-statistic in the light-drinking and HDL cholesterol data set (\texttt{LH}) and the \texttt{u868} rank statistic in the smoking and lead data set (\texttt{SL}). The refined heuristic $\hat{H}^{(2)}$ corrected two of the three errors: it made correct predictions for all data sets and test statistics except for \texttt{u868} in data set \texttt{SL}. There, the conventional approach had the slightly larger sensitivity value (1.50 vs 1.49), but the heuristic predicted that tilting would perform better.

\section{Adaptive combination of the tilted and conventional approaches}
\subsection{Adaptive weighting through a two-person game}
We now describe how the approach of \citet{coh20mult} for multivariate one-sided testing may be used to adaptively combine the tilted and conventional sensitivity analyses. As will be asserted in Proposition \ref{prop:adapt}, the resulting procedure has design sensitivity that is no smaller than the \textit{maximum} of the design sensitivities of the tilted and conventional approaches. For a fixed $\Gamma\geq 1$ let $\mathbf{a}_{ij} = (a_{ij1}, a_{ij2})^T$ with $a_{ij1} = q_{ij}$, $a_{ij2} = w_{\Gamma i}[q_{ij} - \bar{q}_i - \{(\Gamma-1)/(1+\Gamma)\}|q_{ij} - \bar{q}_i|]$. Consider the vector of test statistics $\mathbf{A}_\Gamma= (T, T_\Gamma(w_\Gamma))^T$, which contains the conventional test statistic in its first element and the tilted test statistic with weights $w_{\Gamma}$ in its second element. For any fixed value of the assignment probabilities $\bm{\varrho} = (\varrho_{11}, \varrho_{12},...,\varrho_{In_I})^T$, the expectation and covariance for $\mathbf{A}_{\Gamma}$ are, for $k,\ell = 1,2$,
    \begin{align*}
        \mu_{\Gamma k}({\varrho}) = \sum_{i = 1}^{I}\sum_{j = 1}^{n_{i}}a_{ijk}\varrho_{ij};\;\; 
        \Sigma_{\Gamma k \ell}({\varrho}) =  \sum_{i = 1}^{I} \left\{\sum_{j = 1}^{n_{i}}a_{ijk}a_{ij\ell}\varrho_{ij} - \left(\sum_{j = 1}^{n_{i}}a_{ijk}\varrho_{ij}\right)\left(\sum_{j = 1}^{n_{i}}a_{ij\ell}\varrho_{ij}\right)   \right\}.
    \end{align*} 
    
Let $\cP_\Gamma$ be the set of possible assignment probabilities when (\ref{eq:sensmodel}) holds at $\Gamma$. Then, for any $\Gamma \geq 1$ in (\ref{eq:sensmodel}), define $B_{\Gamma}$ as \begin{align}\label{eq:modify}
B_{\Gamma} = \underset{\bm{\varrho}\in\cP_\Gamma}{\min}\;\; \underset{\bm{\lambda} \in \Lambda_{+}}{\sup}\;\; \max\left[0,\frac{\bm{\lambda}^T\{\mathbf{A}_\Gamma-\bm{\mu}_\Gamma(\bm{\varrho})\}}{\{\bm{\lambda}^T\Sigma_\Gamma(\bm{\varrho})\bm{\lambda}\}^{1/2}}\right]^2,
\end{align}where $\Lambda_{+} = \{(\lambda_1, \lambda_2)^T: \lambda_1\geq 0, \lambda_2 \geq 0, (\lambda_1,\lambda_2)^T \neq (0,0)^T\}$ is the nonnegative orthant excluding the zero vector. Observe that for any fixed $\lambda$, (\ref{eq:modify}) corresponds to a sensitivity analysis based upon the linear combination $\lambda_1T + \lambda_2 T_{\Gamma}(w_\Gamma)$ with a greater-than alternative when (\ref{eq:sensmodel}) holds at $\Gamma$. Assuming bivariate asymptotic normality for $A_\Gamma$, for $\alpha \leq 0.5$ the worst-case deviate for a fixed $\bm{\lambda}$ could be compared to the square of the $1-\alpha$ standard normal quantile to determine whether or not the sensitivity analysis rejects at $\Gamma$. Optimizing $\bm{\lambda}$ over $\Lambda_{+}$ further allows for adaptively chosen linear combinations of the tilted and conventional sensitivity analyses that may outperform either approach on its own. The critical value used to perform inference, $c_{\Gamma,\Lambda_{+}}(\alpha)$, must account for the additional optimization over $\Lambda$, and may no longer be chosen based on a standard normal quantile. Instead, $c_{\Gamma,\Lambda_{+}}(\alpha)$ is a quantile from a chi-bar-square distribution; see \citet{coh20mult} for details. The proposed method for combining the conventional and tilted approaches rejects the sharp null when
$B_{\Gamma} \geq c_{\Gamma, \Lambda_{+}}(\alpha).$ As $c_{\Gamma, \Lambda_{+}}(\alpha) \geq \{\Phi^{-1}(1-\alpha)\}^2$, adaptively optimizing the test statistic over all nonnegative linear combinations requires a larger critical value. So long as $\mathbf{A}_\Gamma$ has a bivariate normal limiting distribution under the sharp null, the adaptive sensitivity analysis asymptotically controls the Type I error rate at $\alpha$ if (\ref{eq:sensmodel}) holds at $\Gamma$. Under the generative model considered in \S \ref{sec:gen}, the procedure can further provide uniform improvements in design sensitivity, as described in the following proposition.

\begin{proposition}\label{prop:adapt} Let $\tilde{\Gamma}_{tilt}$ and $\tilde{\Gamma}_{conv}$ be the design sensitivities for the tilted and conventional sensitivity analyses respectively when using test statistics $T_{\Gamma}(w_{\Gamma})$ and $T$. Consider the procedure that rejects the sharp null when $B_{\Gamma} \geq c_{\Gamma,\Lambda_{+}}(\alpha)$; call its design sensitivity $\tilde{\Gamma}_{adapt}$. Then, under Assumptions \ref{as:plim} and \ref{as:clt},
$\tilde{\Gamma}_{adapt} \geq \max\{\tilde{\Gamma}_{tilt}, \tilde{\Gamma}_{conv}\}$.
\end{proposition} 
 
Proposition \ref{prop:adapt} is an immediate consequence of Theorem 2 of \citet{coh20mult}, and the proof is omitted. It states that in terms of design sensitivity, one need not choose \textit{a priori} between the tilted and conventional sensitivity analyses. Rather, the two approaches may be combined with a resulting design sensitivity that is at least as good as the better of the two individual approaches.

\subsection{The power of a sensitivity analysis}

Proposition \ref{prop:adapt} suggests that in terms of design sensitivity alone, the adaptive procedure should be preferred: it is never worse than either the tilted or conventional sensitivity analysis, and it can be strictly better than both of them when $n_i > 2$. Recall however that design sensitivity calculations focus solely on the limit as $I\rightarrow \infty$ and ignore small-sample considerations. In moderate sample sizes, the differences in the sizes of the critical values used to perform inference in the adaptive approach (based upon the $\bar{\chi}^2$ distribution) versus either the conventional or tilted approaches (based upon the standard normal) will impact the power of these methods, such that the adaptive approach may not provide an improvement in finite samples.

To investigate this further, we consider power simulations under the generative model in \S \ref{sec:gen} for moderate sample sizes $I$. In each of 10,000 Monte Carlo simulations, we simulate data sets with $I=200$ and $J=3$, so that each matched set has $n_i=4$, with one treated and three control observations. We use the generative model described in \S \ref{sec:gen} to generate potential responses under treatment and control. We consider three marginal distributions for $\varepsilon_{ijz}$ in (\ref{eq:gen}): (i) $t_3$; (ii) $\text{Exp}(1)-1$; and (iii) $1-\text{Exp}(1)$, where $\text{Exp}(\lambda)$ denotes an exponential distribution with rate $\lambda$. The effect-to-noise ratio is set at $\tau/\sigma=1/2$ for all three distributions, with $\sigma = \sqrt{\text{Var}(\varepsilon_{1ij}-\varepsilon_{0i\ell})}$, $j\neq \ell$, as before, and we proceed using the difference in means as the test statistic. From Table \ref{tab:designsens}, we see that the tilted sensitivity analysis has higher design sensitivity than the conventional approach in generative model (i) and (ii) (5.70 versus 5.38 and 6.37 versus 4.22 respectively), while the conventional approach has higher design sensitivity than the tilted sensitivity analysis in generative model (iii) (6.15 versus 4.36).  By Proposition \ref{prop:adapt}, the adaptive procedure has a design sensitivity no smaller than the larger of the two individual design sensitivities. Therefore, the design sensitivity for the adaptive procedure will be lower bounded by 5.70, 6.37, and 6.15 in settings (i)-(iii).

\begin{figure}
\centering \includegraphics[scale=.6]{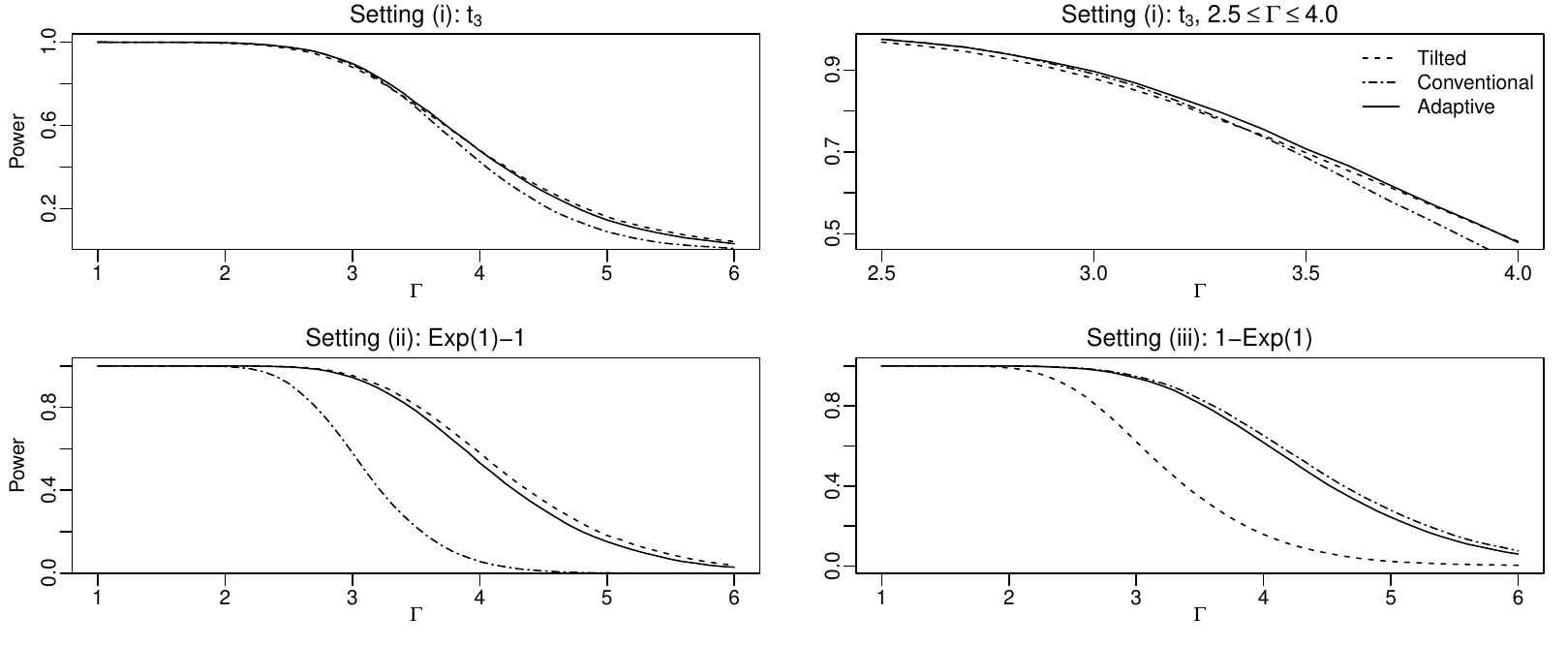}
\caption{The power of a sensitivity analysis for the tilted (dashed line), conventional (dot-dashed line), and adaptive (solid line) procedures with $I=200$, $J=3$, $\tau/\sigma = 1/2$, with the marginal distribution for $\varepsilon_{ijz}$ in (\ref{eq:gen}) chosen to be (i) $t_{df=3}$; (ii) $\text{Exp}(1)-1$, and (iii) $1-\text{Exp}(1)$, where $\text{Exp}(1)$ is an exponential distribution with rate 1. The plot in the top-right zooms in on the power plots for setting (i) between $\Gamma=2.5$ and $\Gamma=4$ to better visualize behavior within that range. Curves are Monte Carlo estimates calculated using 10,000 data sets.}\label{fig:power}
\end{figure}

For each choice of generative model, we produce Monte Carlo estimates across our 10,000 simulations for the power of the tilted, conventional, and adaptive sensitivity analyses for $\Gamma$ ranging from 1 to 6. Figure \ref{fig:power} shows the results. The plots in the top half of the figure are from setting (i), the $t_{df=3}$ generative model, and reveal that while the tilted sensitivity analysis has higher design sensitivity than the conventional approach, it does not provide uniformly higher power than the conventional approach for all $\Gamma$ in finite samples. The zoomed-in view between $\Gamma=2.5$ and $\Gamma=4.0$ in the top-right shows this more clearly. For $\Gamma < 3.3$, the conventional approach improves upon the power of the tilted sensitivity analysis, while for $\Gamma \geq 3.3$ the tilted sensitivity analysis has higher power. The figure illustrates additional benefits of the adaptive procedure: despite using a larger critical value, the adaptive procedure attains power larger than that of both the tilted sensitivity analysis and the conventional sensitivity analysis for $\Gamma \in [1.8, 3.9]$. Moreover, for $\Gamma > 3.9$ the adaptive approach hews closely to the power curve of the tilted sensitivity analysis (which attains the largest power in that range). The largest gap in the power between the adaptive and tilted sensitivity analyses in that range is 0.017, reflecting a small price for adaptivity.

The plots from setting (ii), bottom-left, and setting (iii), bottom-right correspond to generative models for which there is a large gap between the design sensitivities of the tilted and conventional approaches. The tilted sensitivity analysis has larger design sensitivity in setting (ii), and the conventional approach has larger design sensitivity in setting (iii). The power profiles reflect a similar behavior: in terms of power, the tilted sensitivity analysis dominates the conventional approach in the plot from setting (ii), but is dominated by the conventional approach in the plot from setting (iii). The discrepancies are particularly large for $\Gamma$ ranging from 3 to 5 in those plots. The power of the conventional approach can lag behind that of the tilted sensitivity analysis by as much as 0.60 in the setting (ii) plot, and the tilted approach's power is smaller than that of the conventional approach by as much as 0.51 in the setting (iii) plot.  As anticipated, the adaptive procedure closely tracks the power profile of the superior method in each plot by emphasizing the superior procedure in its weighted combination, providing insurance against choosing an underperforming method. The cost of this insurance is the gap between the power of the adaptive approach and the superior method for each model. The maximum value for this gap is 0.05 in the second plot, and 0.04 in the third. 

\section{Proofs and examples}
We now provide proofs and illustrations for all results except for Theorem \ref{thm:designsens}, whose proof will be presented in a standalone section.
\subsection{Proof of Proposition \ref{prop:MGamma}}

As described in \S \ref{sec:separable}, the worst-case expectation $\mu_{\Gamma i}$ returned by the separable algorithm is

\begin{align*}
\mu_{\Gamma i} &= \max_{ a = 1,..., n_i-1}\;\frac{\Gamma\sum_{j=1}^{a}q_{ij} + \sum_{j = a+1}^{n_i} q_{ij}}{\Gamma a + (n_i-a)},
\end{align*}
where $q_{i1}\geq...\geq q_{in_i}$

We claim that $\mu_{\Gamma i} = \bar{q}_i + \left\{(\Gamma-1)/(1+\Gamma)\right\}M_{\Gamma i}$, where 
\begin{align*}
M_{\Gamma i}
&= \mathtt{SOLVE}\left\{c:\; \frac{1}{n_i}\sum_{j=1}^{n_i}\left|q_{ij} - \bar{q}_i - \left(\frac{\Gamma-1}{1+\Gamma}\right)c\right| = c\right\}.
\end{align*}

To show this, first note
\begin{align*}
\mu_{\Gamma i} &= \max_{ a = 1,..., n_i-1}\;\frac{\Gamma\sum_{j=1}^{a}q_{ij} + \sum_{j = a+1}^{n_i} q_{ij}}{\Gamma a + (n_i-a)}\\
&= \bar{q}_i + \max_{ a = 1,..., n_i-1}\;\frac{\Gamma\sum_{j=1}^{a}(q_{ij}-\bar{q}_i) + \sum_{j = a+1}^{n_i} (q_{ij}-\bar{q}_i)}{\Gamma a + (n_i-a)}.\\
&= \bar{q}_i+\max_{ a = 1,..., n_i-1}\;\frac{(\Gamma-1)\sum_{j=1}^{a}(q_{ij}-\bar{q}_i)}{(\Gamma-1)a  - (\Gamma-1)n_i/2 + (\Gamma+1)n_i/2}.\\
&= \bar{q}_i + \left(\frac{\Gamma-1}{1+\Gamma}\right)\max_{ a = 1,..., n_i-1}\;\frac{\sum_{j=1}^{a}(q_{ij}-\bar{q}_i)}{n_i/2 + \frac{\Gamma-1}{1+\Gamma}(a-n_i/2)},\\
\end{align*}
where the third line uses that $\sum_{j=1}^{n_i}(q_{ij}-\bar{q}_i)=0$, so $\sum_{i=1}^a(q_{ij}-\bar{q}_i) = -\sum_{i=a+1}^{n_i}(q_{ij}-\bar{q}_i)$.

We now show that 
\begin{align*}
M_{\Gamma i} := \max_{ a = 1,..., n_i-1}\;\frac{\sum_{j=1}^{a}(q_{ij}-\bar{q}_i)}{n_i/2 + \frac{\Gamma-1}{1+\Gamma}(a-n_i/2)} &= \mathtt{SOLVE}\left\{c:\; \frac{1}{n_i}\sum_{j=1}^{n_i}\left|q_{ij} - \bar{q}_i - \left(\frac{\Gamma-1}{1+\Gamma}\right)c\right| = c\right\}.
\end{align*}

Let $N(a) = \sum_{j=1}^{a}(q_{ij}-\bar{q}_i)$ and $D(a) = n_i/2 + \frac{\Gamma-1}{1+\Gamma}(a-n_i/2)$ be the numerator and denominator of the objective function on the left hand side. Then, $M_{\Gamma i} = \max_{ a = 1,..., n_i-1} N(a)/D(a)$, which is a fractional optimization problem. For notational ease, expand the maximization to be over $a=0,...,n_i$; as $N(a) = 0$ at either endpoint, neither $a=0$ nor $a=n_i$ will be a maximizer of the fractional program. For any $c$, consider instead the parametric problem 
\begin{align}\label{eq:Fgamma}
H_{\Gamma i}(c) = \max_{ a = 0,..., n_i} N(a) - cD(a).
\end{align}
By \citet{din67}, we have that $H_{\Gamma i}(M_{\Gamma i}) = 0$ and that the zero is unique. That is, the zero of the parametric problem as a function of $c$ is attained when setting $c$ equal to the maximizer of the fractional problem. Moreover, we see that $H_{\Gamma i}(c) > 0$ if and only if $M_{\Gamma i} > c$.

Consider a fixed value for $c$ in $H_{\Gamma i}(c)$:
\begin{align*}H_{\Gamma i}(c) = \max_{ a = 0,..., n_i} \sum_{j=1}^{a}\left\{q_{ij}-\bar{q}_i - \left(\frac{\Gamma-1}{1+\Gamma}\right)c\right\} + \left(\frac{\Gamma-1}{1+\Gamma}\right)(n_i/2)c - (n_i/2)c.
\end{align*}

The $\arg\max$ of the above is simply $a^* = \sum_{j=1}^{n_i}\1\left\{q_{ij}-\bar{q}_i - \left(\frac{\Gamma-1}{1+\Gamma}\right)c > 0\right\}$, resulting in only the nonnegative values of $q_{ij}-\bar{q}_i - \left(\frac{\Gamma-1}{1+\Gamma}\right)c$ being summed. As $\sum_{j=1}^{n_i}\left\{q_{ij}-\bar{q}_i - \left(\frac{\Gamma-1}{1+\Gamma}\right)c\right\} = -n_i\left(\frac{\Gamma-1}{1+\Gamma}\right)c$ since $\sum_{j=1}^{n_i}(q_{ij}-\bar{q}_i)=0$ and noting $-n_i\left(\frac{\Gamma-1}{1+\Gamma}\right)c = \sum_{j=1}^{n_i}\left\{q_{ij}-\bar{q}_i - \left(\frac{\Gamma-1}{1+\Gamma}\right)c\right\} = \sum_{j=1}^{a^*}\left|q_{ij}-\bar{q}_i - \left(\frac{\Gamma-1}{1+\Gamma}\right)c\right| - \sum_{j=a^*+1}^{n_i}\left|q_{ij}-\bar{q}_i - \left(\frac{\Gamma-1}{1+\Gamma}\right)c\right|$,  we have
\begin{align*}
\sum_{j=1}^{a^*}\left\{q_{ij}-\bar{q}_i - \left(\frac{\Gamma-1}{1+\Gamma}\right)c\right\} &= \sum_{j=1}^{a^*}\left|q_{ij}-\bar{q}_i - \left(\frac{\Gamma-1}{1+\Gamma}\right)c\right|\\
&= -n_i\left(\frac{\Gamma-1}{1+\Gamma}\right)c + \sum_{j=a^*+1}^{n_i}\left|q_{ij}-\bar{q}_i - \left(\frac{\Gamma-1}{1+\Gamma}\right)c\right|\\
&= -(n_i/2)\left(\frac{\Gamma-1}{1+\Gamma}\right)c + (1/2)\sum_{j=1}^{n_i}\left|q_{ij}-\bar{q}_i - \left(\frac{\Gamma-1}{1+\Gamma}\right)c\right| 
\end{align*}
Hence,
\begin{align} \label{eq:Fgamma2} H_{\Gamma i}(c) &= (1/2)\sum_{j=1}^{n_i}\left|q_{ij}-\bar{q}_i - \left(\frac{\Gamma-1}{1+\Gamma}\right)c\right| - (n_i/2)c\end{align}
As $M_{\Gamma i}$ is the zero of the above, it must satisfy $n_i^{-1}\sum_{j=1}^{n_i}\left|q_{ij}-\bar{q}_i - \left(\frac{\Gamma-1}{1+\Gamma}\right)M_{\Gamma i}\right| = M_{\Gamma i}$ as desired. 

\subsection{Theorem 1 and the separable algorithm}

\subsubsection{Proof of Lemma \ref{lemma:AGamma}}

The bound on the expectation was proved as Lemma 1 in \citet{fog23}, but is included here for completeness. For any vector of unmeasured confounders $\mathbf{u}_i = (u_{i1},...,u_{in_i})$, if (\ref{eq:sensmodel}) holds at $\Gamma > 1$ we have
\begin{align}\label{eq:numAi}
E_{\mathbf{u}_i}(T_{\Gamma i}\mid \cF, \cZ) &= \frac{\sum_{j=1}^{n_i}\exp\{\log(\Gamma) u_{ij}\}[q_{ij} - \bar{q}_i - \left\{(\Gamma-1)/(1+\Gamma)\right\}|q_{ij}-\bar{q}_i|]}{\sum_{j=1}^{n_i}\exp\{\log(\Gamma) u_{ij}\}}
\end{align}
Observe that
\begin{align*}
q_{ij} - \bar{q}_i -  \left(\frac{\Gamma-1}{1+\Gamma}\right)|q_{ij} - \bar{q}_i|&=\begin{cases} 2(q_{ij} - \bar{q}_i)/(1+\Gamma) & q_{ij}> \bar{q}_i\\
2\Gamma (q_{ij}-\bar{q}_i)/(1+\Gamma) & q_{ij} \leq \bar{q}_i\end{cases}.
\end{align*}
A maximizer for (\ref{eq:numAi}) is attained by setting $u_{ij} =  \1\{q_{ij} > \bar{q}_i\}$, resulting in the largest possible multiplier, $\exp\{\log(\Gamma) u_{ij}\} = \Gamma$, for positive values of $q_{ij}-\bar{q}_i$, and the smallest possible value of the multiplier, $\exp\{\log(\Gamma) u_{ij}\} = 1$ to nonpositive values of $q_{ij}-\bar{q}_i$. Under this choice of $u_{ij}$, the numerator of (\ref{eq:numAi}) becomes $2\Gamma/(1+\Gamma)\sum_{j=1}^{n_i}(q_{ij}-\bar{q}_i) = 0$  as $\sum_{j=1}^{n_i}(q_{ij}-\bar{q}_i) = 0$.  The confounder $u_{ij} =  \1\{q_{ij} \geq \bar{q}_i\}$ thus also maximizes (\ref{eq:numAi}), as any other choice ${u}'_{ij}$ will yield an expectation that is less than or equal to zero.

For $\Gamma > 1$ positive values for $q_{ij}-\bar{q}_i$ must receive $u_{ij}=1$, and negative values must receive $u_{ij}=0$; otherwise, the expectation will be below zero. If there are units $i\ell$ such that $q_{i\ell}=\bar{q}_i$,  $u_{i\ell} = \1\{q_{i\ell} > \bar{q}_i\}$ is not the unique maximizer of the expectation. In this case, replacing $u_{i\ell} = \1\{q_{i\ell} > \bar{q}_i\}$ with any $u_{i\ell} \in [0,1]$ would also maximize the expectation at 0. In the separable algorithm, if there are multiple values for the unmeasured confounder attaining the maximal expectation for $T_{\Gamma i}$, the one which maximizes the variance is chosen. To complete the proof of the lemma, we must thus show that of the values for the unmeasured confounder attaining the largest expectation, $u_{ij} = \1\{q_{ij} > \bar{q}_i\}$ produces the largest possible variance for $T_{\Gamma i}$.

Among $\mathbf{u}_i$ which attain the expectation upper bound of 0, the variance for $T_{\Gamma i}$ is  

\begin{align*}
\text{var}_{\mathbf{u}_{i}}(T_{\Gamma i}\mid \mathcal{F}, \mathcal{Z}) &:= \frac{4\Gamma^2}{(1+\Gamma)^2}\left(\frac{1}{\sum_{j=1}^{n_i}\exp\{\log(\Gamma)u_{ij}\}}\right) \sum_{j=1}^{n_i}\frac{(q_{ij}-\bar{q}_i)^2}{\exp\{\log(\Gamma)\1\{q_{ij} > \bar{q}_i\}\}}.\end{align*}

The values of $u_{ij}$ are fixed for all $ij$ such that $q_{ij}\neq \bar{q}_i$. For $\Gamma>1$ we see that the variance is maximized by setting $u_{i\ell}=0$ for all $\ell$ such that $q_{i\ell}=\bar{q}_i$. Among $\mathbf{u}_i$ maximizing the expectation, the variance maximizer thus sets $u_{ij} = \1\{q_{ij} > \bar{q}_i\}$. 
\subsubsection{Proof of uniqueness of tilting in Remark  \ref{remark:separable}}
Here we prove a sense in which tilting is unique (up to positive scalar multiples) among procedures yielding a closed form for the worst-case expectation. 

\begin{proposition} \label{prop:unique}Consider the set of all coordinate-wise transformations of $q_{ij}-\bar{q}_i$ of the form $w_{\Gamma i} h_{\Gamma}(q_{ij}-\bar{q}_i)$ for non-degenerate, continuous and non-decreasing function $h_{\Gamma}(\cdot)$ and set-specific multipliers $w_{\Gamma i} > 0$. Suppose that for any $n_i\geq 2$ and any candidate vector $(q_{i1}-\bar{q}_i,...,q_{in_i}-\bar{q}_i)^T$, $\max_{u \in [0,1]^{n_i}} E_u\{\sum Z_{ij}w_{\Gamma i}h_{\Gamma}(q_{ij}-\bar{q}_i)\} = 0$ when the sensitivity model (\ref{eq:sensmodel}) holds at $\Gamma$. Then, for some constant $C_{\Gamma} > 0$, $h_{\Gamma}(q_{ij}-\bar{q}_i) = C_{\Gamma}(q_{ij}-\bar{q}_i)/\exp\{\log(\Gamma)\1\{q_{ij}-\bar{q}_i > 0\}\}.$
\end{proposition}
\begin{proof}
Consider worst-case expectation returned by the separable algorithm in (\ref{eq:AS}). Using an equivalent expression and ignoring the constant $w_{\Gamma i}$, the requirement that the worst-case expectation is 0 for each $i$ can be stated as
\begin{align*}
\max_{1\leq g_{ij} \leq \Gamma} \frac{\sum_{j=1}^{n_i} g_{ij} h_{\Gamma}(q_{ij}-\bar{q}_i)}{\sum_{j=1}^{n_i} g_{ij}} &= 0.
\end{align*}
As the denominator is strictly positive, the fact that maximal expectation is 0 implies that
\begin{align*}
\max_{1\leq g_{ij} \leq \Gamma}\;\; \sum_{j=1}^{n_i} g_{ij} h_{\Gamma}(q_{ij}-\bar{q}_i)  &= 0.
\end{align*}
Let $\phi_\Gamma(y) = y$ if $y \leq 0$, and $\phi_\Gamma(y) = \Gamma y$ if $y > 0$. Then,
\begin{align*}
\max_{1\leq g_{ij} \leq \Gamma}\;\; \sum_{j=1}^{n_i} g_{ij} h_{\Gamma}(q_{ij}-\bar{q}_i)  &= \sum_{j=1}^{n_i} \phi_\Gamma\{ h_{\Gamma}(q_{ij}-\bar{q}_i)\} = 0.
\end{align*}
Define the composition $b_\Gamma(y) = \phi_\Gamma(h_\Gamma(y))$, and consider first $n_i=2$. If $q_{i1}-\bar{q}_i = t$, we have $-t = (q_{i2}-\bar{q}_i)$. Then, for any $t$,
\begin{align*}
b_\Gamma(t) + b_\Gamma(-t) = 0 \Leftrightarrow -b_\Gamma(t) = b_\Gamma(-t)
\end{align*}
Now suppose $n_i=3$ and let $(q_{i1}-\bar{q}_i, q_{i2}-\bar{q}_i, q_{i3}-\bar{q}_i) = (s, t, -s-t)$. Then, for any $s,t$:
\begin{align*}
b_\Gamma(s) + b_\Gamma(t) + b_\Gamma(-(s+t)) = 0.
\end{align*}
As $-b_\Gamma(s+t) = b_\Gamma(-(s+t))$ from the derivation at $n_i=2$, we have
\begin{align*}
b_\Gamma(s+t) = b_\Gamma(s)+b_\Gamma(t).
\end{align*}
Hence $b_\Gamma(\cdot)$ is additive. Since $h_\Gamma(\cdot)$ is non-decreasing by assumption and $\phi_\Gamma$ is non-decreasing, $b_\Gamma(\cdot)$ is also non-decreasing. Hence $b_\Gamma(\cdot)$ must be linear, i.e $b_\Gamma(y) = C_{\Gamma} y$, where $C_\Gamma > 0$ since we have assumed $h_\Gamma(\cdot)$ is non-degenerate and we have established that $b_\Gamma$ is non-decreasing.

Inverting $\phi_\Gamma(y)$ to form $h_\Gamma(y) = \phi^{-1}_\Gamma(b_\Gamma(y))$, we obtain $h_\Gamma(y) = C_\Gamma y/\Gamma$ for $y > 0$, and $h_\Gamma(y) = C_\Gamma y$ for $y \leq 0$. That is,
\begin{align*}
h_\Gamma(y) &= \frac{C_\Gamma y}{\exp\{\log(\Gamma)\1\{y > 0\}\}}.
\end{align*}

This completes the proof. The particular choice $C_\Gamma = (2\Gamma)/(1+\Gamma)$ returns the tilting transformation in (\ref{eq:AGammaweight}).

\end{proof}

\subsubsection{Proof of Theorem \ref{thm:AGamma}}
Suppose that the sharp null hypothesis holds and that the sensitivity model (\ref{eq:sensmodel}) holds at $\Gamma$. Under the conditions on $q_{ij}$ and $w_{\Gamma i}$ provided in the Theorem's statement, Theorem 1 of \citet{haj99} applies to $T_\Gamma(w_{\Gamma})$. Let $\vartheta_{\Gamma i}$ and $\sigma^2_{\Gamma i}$ be the true expectation and variance of $T_{\Gamma i}$. We thus have that for every $\epsilon$ there exists an $I_1$ such that for any $k$, $I\geq I_1$ implies
\begin{align*}
\text{pr}\left(\frac{T_{\Gamma}(w_{\Gamma}) - \sum_{i=1}^Iw_{\Gamma i}\vartheta_{\Gamma i}}{\sqrt{\sum_{i=1}^Iw^2_{\Gamma i}\sigma^2_{\Gamma i}}} \geq k \right) \leq 1-\Phi(k) + \epsilon
\end{align*}

As $\vartheta_{\Gamma i}$ and $\sigma^2_{\Gamma i}$ are unknown, the separable algorithm attempts to upper bound the above tail probability when (\ref{eq:sensmodel}) holds at $\Gamma$. By Lemma \ref{lemma:AGamma}, $\vartheta_{\Gamma i} \leq 0$ for all $i$. Moreover, for matched sets $i$ where $\vartheta_{\Gamma i} = 0$, we have that $\tilde{\nu}^2_{\Gamma i} \geq \sigma^2_{\Gamma i}$, where $\tilde{\nu}^2_{\Gamma i}$ is the candidate variance returned by the separable algorithm. 

For any $k > 0$, it suffices to show that there exists an $I_2$ such that $I > I_2$ implies 

\begin{align*}
\text{pr}\left(T_\Gamma(w_\Gamma) - k\sqrt{\sum_{i=1}^I w^2_{\Gamma i} \tilde{\nu}^2_{\Gamma i}} \geq 0\right) &\leq \text{pr}\left(T_\Gamma(w_\Gamma) - \sum_{i=1}^Iw_{\Gamma i}\vartheta_{\Gamma i} - k\sqrt{\sum_{i=1}^Iw^2_{\Gamma i}{\sigma}^2_{\Gamma i}} \geq 0\right)
\end{align*}

Equivalently, we show that there exists an $I_2$ such that $I > I_2$ implies 

\begin{align*}
k\left(\sqrt{I^{-1}\sum_{i=1}^Iw^2_{\Gamma i}\sigma^2_{\Gamma i}} - \sqrt{I^{-1}\sum_{i=1}^Iw^2_{\Gamma i}\tilde{\nu}^2_{\Gamma i}}\right) \leq -I^{-1/2}\sum_{i=1}^Iw_{\Gamma i}\vartheta_{\Gamma i}
\end{align*}

Again by Lemma \ref{lemma:AGamma} we have $\vartheta_{\Gamma i} \leq 0$ for all $i$. Because of this, when $I^{-1}\sum_{i=1}^Iw^2_{\Gamma i}\sigma^2_{\Gamma i} \leq I^{-1}\sum_{i=1}^Iw^2_{\Gamma i}\tilde{\nu}^2_{\Gamma i}$, the inequality holds. Consider instead the case that $I^{-1}\sum_{i=1}^Iw^2_{\Gamma i}\sigma^2_{\Gamma i} > I^{-1}\sum_{i=1}^Iw^2_{\Gamma i}\tilde{\nu}^2_{\Gamma i}$. Using concavity of $f(x) = \sqrt{x}$ and recalling $\mathcal{B}_I = \{i:\sigma^2_{\Gamma i} > \tilde{\nu}^2_{\Gamma i}\}$, we have

\begin{align*}
&k\left(\sqrt{I^{-1}\sum_{i=1}^Iw^2_{\Gamma i}\sigma^2_{\Gamma i}} - \sqrt{I^{-1}\sum_{i=1}^Iw^2_{\Gamma i}\tilde{\nu}^2_{\Gamma i}}\right)\\& \leq \frac{k}{2} \left(\frac{I^{-1}{\sum_{i=1}^Iw^2_{\Gamma i}(\sigma^2_{\Gamma i} - \tilde{\nu}^2_{\Gamma i})}}{\sqrt{I^{-1}\sum_{i=1}^Iw^2_{\Gamma i}\tilde{\nu}^2_{\Gamma i}}}\right)\\
&\leq \frac{k}{2} \sqrt{I^{-1}\sum_{i\in \mathcal{B}_I}w^2_{\Gamma i}(\sigma^2_{\Gamma i} -\tilde{\nu}^2_{\Gamma i})}\sqrt{\frac{I^{-1}{\sum_{i\in \mathcal{B}_I}w^2_{\Gamma i}\sigma^2_{\Gamma i}} - {I^{-1}\sum_{i\in \mathcal{B}_I}w^2_{\Gamma i}\tilde{\nu}^2_{\Gamma i}}}{I^{-1}{\sum_{i=1}^Iw^2_{\Gamma i}\tilde{\nu}^2_{\Gamma i}}}}.
\end{align*}

Under the assumption of bounded stratum sizes, there exist constants $C_{\Gamma 1}, C_{\Gamma 2}$ such that $\sigma^2_{\Gamma i} \leq C_{\Gamma 1} \sum_{j=1}^{n_i}(q_{ij}-\bar{q}_i)^2$  and $\tilde{\nu}^2_{\Gamma i}\geq C_{\Gamma 2} \sum_{j=1}^{n_i}(q_{ij}-\bar{q}_i)^2$ for all $i$. Therefore,

\begin{align*} \sqrt{\frac{I^{-1}{\sum_{i\in \mathcal{B}_I}w^2_{\Gamma i}\sigma^2_{\Gamma i}} - {I^{-1}\sum_{i\in \mathcal{B}_I}w^2_{\Gamma i}\tilde{\nu}^2_{\Gamma i}}}{I^{-1}{\sum_{i=1}^Iw^2_{\Gamma i}\tilde{\nu}^2_{\Gamma i}}}} &= O(1). \end{align*} 

We have further assumed that $\sqrt{I^{-1}\sum_{i\in \mathcal{B}_I}w^2_{\Gamma i}(\sigma^2_{\Gamma i} -\tilde{\nu}^2_{\Gamma i})} =  o(I^{-1/2}|\sum_{i=1}^Iw_{\Gamma i}\vartheta_{\Gamma i}|)$, which in turn implies $k\left(\sqrt{I^{-1}\sum_{i=1}^Iw^2_{\Gamma i}\sigma^2_{\Gamma i}} - \sqrt{I^{-1}\sum_{i=1}^Iw^2_{\Gamma i}\tilde{\nu}^2_{\Gamma i}}\right)\\ = O(1)o(I^{-1/2}|\sum_{i=1}^Iw_{\Gamma i}\vartheta_{\Gamma i}|) = o(I^{-1/2}|\sum_{i=1}^Iw_{\Gamma i}\vartheta_{\Gamma i}|)$. As $I^{-1/2}\sum_{i=1}^I w_{\Gamma i}\vartheta_{\Gamma i} \leq 0$ by Lemma \ref{lemma:AGamma}, the proof is complete.

\subsubsection{A pathological example where the separable algorithm fails}

Let $n_i=4$ for all $i = 1,...,I$, and let $q_{i1} = 1, q_{i2} = i^{-2}, q_{i3} = -i^{-2}, q_{i4}=-1$; note $\bar{q}_i = 0$ for all $i$. The tilted test statistic applies the separable algorithm to $q_{ij\Gamma} = q_{ij} - \{(\Gamma-1)/(1+\Gamma)\}|q_{ij}|$ when conducting a sensitivity analysis when (\ref{eq:sensmodel}) holds at $\Gamma$, yielding by Lemma \ref{lemma:AGamma} $u_{i1}=u_{i2}=1$, $u_{i3}=u_{i4}=0$, a worst-case expectation of 0, and a variance 
\begin{align*}\tilde{\nu}^2_{\Gamma i}&=\frac{4\Gamma^2}{(1+\Gamma)^2}\left(\frac{1}{2\Gamma+2}\right)(1+i^{-4})(1+\Gamma^{-1})
\end{align*}

Suppose that in reality $u_{i1}=1$, $u_{i2}=u_{i3}=u_{i4}=0$. Then, the true expectation and variance would be
\begin{align*}
\vartheta_{\Gamma i} &= \frac{2\Gamma}{1+\Gamma}\frac{(\Gamma^{-1} - 1)i^{-2}}{\Gamma +3}\\
\sigma^2_{\Gamma i} &= \frac{2\Gamma+2}{\Gamma+3}\tilde{\nu}^2_{\Gamma i}+ \frac{4\Gamma^2}{(1+\Gamma)^2}\left(\frac{1}{\Gamma + 3}\right)i^{-4}(\Gamma^{-2}-\Gamma^{-1}) - \vartheta_{\Gamma i}^2
\end{align*}

Evaluating the Riemann zeta function at $s=2$ and $s=4$, we have $\sum_{i=1}^{\infty} i^{-2} = \pi^2/6$, and $ \sum_{i=1}^{\infty} i^{-4} = \pi^4/90$. Therefore:
\begin{align*}
I^{-1}\sum_{i=1}^I\tilde{\nu}^2_{\Gamma i} &\rightarrow \frac{4\Gamma^2}{(1+\Gamma)^2}\left(\frac{1}{2\Gamma+2}\right)(1+\Gamma^{-1})\\
I^{-1}\sum_{i=1}^I{\sigma}^2_{\Gamma i} &\rightarrow \frac{4\Gamma^2}{(1+\Gamma)^2}\left(\frac{1}{\Gamma + 3}\right)(1+\Gamma^{-1})\\
I^{-1/2}\sum_{i=1}^I \vartheta_{\Gamma i} &\rightarrow 0.
\end{align*}

Noting that the limiting value of $I^{-1}\sum_{i=1}^I\tilde{\nu}^2_{\Gamma i} $ is strictly smaller than that of $I^{-1}\sum_{i=1}^I\sigma^2_{\Gamma i}$ for $\Gamma > 1$ by a factor $(\Gamma+3)/(2\Gamma+2)$, this implies that for any $k > 0$, there exists an $I_2$ such that $I>I_2$ implies

\begin{align*}
k\left(\sqrt{I^{-1}\sum_{i=1}^I\sigma^2_{\Gamma i}} - \sqrt{I^{-1}\sum_{i=1}^I\tilde{\nu}^2_{\Gamma i}}\right) \geq -I^{-1/2}\sum_{i=1}^I\vartheta_{\Gamma i},
\end{align*} which implies under asymptotic normality that the separable algorithm provides a Type I error rate larger than $\alpha$ for any $\alpha < 1/2$.

To explain why the separable algorithm fails here, the particular form of $q_{ij}$ considered allows one to find a pattern of hidden bias not used by the separable algorithm for which the expectation is smaller for each $i$, in an asymptotically negligible way, than the expectation deployed in the separable algorithm, but where the variance is larger in a manner that persists asymptotically. Since the separable algorithm focuses first on finding the largest expectation, it fails to identify this alternative pattern of hidden bias with a larger variance and an asymptotically equivalent expectation at $\sqrt{I}$ scaling.

\subsection{Proposition \ref{prop:signscore}, and an illustration with the Mantel-Haenszel statistic}
\subsubsection{Proof of Proposition \ref{prop:signscore}}

For statistics of the form (\ref{eq:ss}), recalling $a_{i1}\geq a_{i2}$ we have 
\begin{align*}
q_{ij} & \in \{a_{i1}, a_{i2}\}\\
\bar{q}_i &= a_{i2} + (a_{i1}-a_{i2})\frac{\sum_{j=1}^{n_i}\1\{q_{ij}=a_{i1}\}}{n_i}\\
\1\{q_{ij}>\bar{q}_i\} &= \1\{q_{ij}=a_{i1}\}\\
\mu_{\Gamma i} &= a_{i2} + \frac{\Gamma (a_{i1}-a_{i2})\sum_{j=1}^{n_i}\1\{q_{ij} > \bar{q}_i\}}{n_i + (\Gamma-1)\sum_{j=1}^{n_i}\1\{q_{ij} > \bar{q}_i\}}
\end{align*}

$T_i-\mu_{\Gamma i}$ can take on two possible values: $a_{i1}-\mu_{\Gamma i}$ or $a_{i2}-\mu_{\Gamma i}$. These may be expressed as
\begin{align}
a_{i1}-\mu_{\Gamma i} &= \frac{(a_{i1}-a_{i2})(n_i - \sum_{j=1}^{n_i}\1\{q_{ij} > \bar{q}_i\})}{n_i + (\Gamma-1)\sum_{j=1}^{n_i}\1\{q_{ij} > \bar{q}_i\}} = \left(\frac{n_i}{n_i + (\Gamma-1)\sum_{j=1}^{n_i}\1\{q_{ij} > \bar{q}_i\}}\right)(a_{i1}-\bar{q}_i) \label{eq:a1}\\
a_{i2} - \mu_{\Gamma i} &= \frac{-\Gamma (a_{i1}-a_{i2})\sum_{j=1}^{n_i}\1\{q_{ij} > \bar{q}_i\}}{n_i + (\Gamma-1)\sum_{j=1}^{n_i}\1\{q_{ij} > \bar{q}_i\}} = \left(\frac{n_i}{n_i + (\Gamma-1)\sum_{j=1}^{n_i}\1\{q_{ij} > \bar{q}_i\}}\right)\Gamma(a_{i2}-\bar{q}_i),\label{eq:a2}
\end{align}

The tilted test statistic $T_i-\bar{q}_i - \{(\Gamma-1)/(1+\Gamma)\}|T_i-\bar{q}_i|$ takes on the following two values (note $a_{i1}\geq \bar{q}_i\geq a_{i2}$):
\begin{align}
a_{i1}-\bar{q}_i - \left(\frac{\Gamma-1}{1+\Gamma}\right)|a_{i1}-\bar{q}_i| &= \left(\frac{2}{1+\Gamma}\right)(a_{i1}-\bar{q}_i); \label{eq:a1t}\\
a_{i2}-\bar{q}_i - \left(\frac{\Gamma-1}{1+\Gamma}\right)|a_{i2}-\bar{q}_i| &= \left(\frac{2}{1+\Gamma}\right)\Gamma(a_{i2}-\bar{q}_i)\label{eq:a2t}
\end{align}

Comparing (\ref{eq:a1})  to (\ref{eq:a1t}) and comparing (\ref{eq:a2}) to (\ref{eq:a2t}) completes the proof.
\subsubsection{Illustration: Mantel-Haenszel test with binary outcomes}\label{sec:MH}
To illustrate the connections made within \S\S\ref{sec:signscore}-\ref{sec:IPW}, consider the use of the Mantel-Haenszel test statistic with binary potential outcomes, $y_{zij} \in \{0,1\}$ for $z=0,1$. The Mantel-Haenszel statistic is $T = \sum_{i=1}^I\sum_{j=1}^{n_i}Z_{ij}Y_{ij}$, summing the number of events among the treated units; this corresponds to $q_{ij}=Y_{ij}$, and to $a_{i1}=1$, $a_{i2} = 0$ in (\ref{eq:ss}). Under the sharp null hypothesis, $Y_{ij} = y_{0ij}$.

For any matched set $i$, assuming (\ref{eq:sensmodel}) holds at $\Gamma$, the tilted statistic $T_{\Gamma i}$ is 
\begin{align*}
T_{\Gamma i} &= \sum_{j=1}^{n_i}Z_{ij}\left\{Y_{ij} - \bar{Y}_i- \left(\frac{\Gamma-1}{1+\Gamma}\right)|Y_{ij}-\bar{Y}_i|\right\}\\
&= \left(\frac{2\Gamma}{1+\Gamma}\right)\sum_{j=1}^{n_i}\frac{Z_{ij}(Y_{ij}-\bar{Y}_i)}{\exp\{\log(\Gamma)\1\{Y_{ij}>\bar{Y}_i\}\}}
\end{align*} 
where $\bar{Y}_i = \sum_{j=1}^{n_i}Y_{ij}/n_i$. 

In the conventional sensitivity analysis for the Mantel-Haenszel statistic, the worst-case expectation $\mu_{\Gamma i}$ is calculated by setting $u_{ij}=\1\{Y_{ij}=1\}$, such that individuals who experienced the event have the higher probability of treatment. The test statistic minus this worst-case expectation, $T_i - \mu_{\Gamma i}$, is\begin{align*}
T_i - \mu_{\Gamma i} &= \sum_{j=1}^{n_i}Z_{ij}Y_{ij} - \frac{\Gamma\sum_{j=1}^{n_i}Y_{ij}}{n_i + (\Gamma-1)\sum_{j=1}^{n_i}Y_{ij}}\\
&= \left(\frac{\Gamma n_i}{n_i + (\Gamma-1)\sum_{j=1}^{n_i}\1\{Y_{ij}>\bar{Y}_i\}}\right)\sum_{j=1}^{n_i}\frac{Z_{ij}(Y_{ij}-\bar{Y}_i)}{\exp\{\log(\Gamma)\1\{Y_{ij}>\bar{Y}_i\}\}},
\end{align*} with the second line using Proposition \ref{prop:signscore}. Finally, letting $\tilde{\varrho}_{ij}$ be the values for $\text{pr}(Z_{ij}=1 \mid \mathcal{F}, \mathcal{Z})$ when $u_{ij} = \1\{Y_{ij}> \bar{Y}_i\}$, from (\ref{eq:IPW}) the corresponding worst-case IPW statistic would instead take the form 
\begin{align*}
IPW_{\Gamma i} &=  \left(\frac{n_i + (\Gamma-1)\sum_{j=1}^{n_i}\1\{Y_{ij}>\bar{Y}_i\}}{n_i}\right)\sum_{j=1}^{n_i}\frac{Z_{ij}(Y_{ij}-\bar{Y}_i)}{\exp\{\log(\Gamma)\1\{Y_{ij}>\bar{Y}_i\}\}}.
\end{align*}

From this we see that $T_{\Gamma i}$, $T_i - \mu_{\Gamma i}$, and $IPW_{\Gamma i}$ differ only in the matched-set weights applied to the common contribution $\sum_{j=1}^{n_i}Z_{ij}(Y_{ij}-\bar{Y}_i)/\exp\{\log(\Gamma)\1\{Y_{ij}> \bar{Y}_i\}$ from each set $i$. At $\Gamma=1$ (no unmeasured confounding) the weights are equal across the three approaches, but differ at $\Gamma > 1$ when $n_i > 2$. The weights of the conventional sensitivity analysis and the IPW approach are inversely proportional, and depend upon the normalizing constant for the worst-case assignment probabilities under the sharp null, $n_i + (\Gamma-1)\sum_{j=1}^{n_i}\1\{Y_{ij}>\bar{Y}_i\}$. The tilted statistic does not make use of the normalizing constant.

\subsection{Proof of Theorem \ref{thm:closed}}

At level $\alpha$, the sensitivity analysis using Theorem \ref{thm:AGamma} rejects the null when assuming that (\ref{eq:sensmodel}) holds at $\Gamma$ when
\begin{align*}
\frac{I^{-1/2}\sum_{i=1}^I (T_i - \bar{q}_i) - I^{-1/2}\left(\frac{\Gamma-1}{1+\Gamma}\right)\sum_{i=1}^I|T_i-\bar{q}_i|}{\sqrt{I^{-1}\sum_{i=1}^I\tilde{\nu}^2_{\Gamma i}}} \geq \Phi^{-1}(1-\alpha)
\end{align*}
Under Assumption 1, the denominator converges to a constant limiting value, and $ I^{-1}\sum_{i=1}^I (T_i - \bar{q}_i) - I^{-1}\{(\Gamma-1)/(1+\Gamma)\}\sum_{i=1}^I|T_i-\bar{q}_i|$ (the numerator divided by $I^{1/2}$) also converges to a constant limiting value.  By Assumption 2, the random variable on the left hand side will follow a Gaussian distribution when \\$E\left\{I^{-1/2}\sum_{i=1}^I (T_i - \bar{q}_i) - I^{-1/2}\{(\Gamma-1)/(1+\Gamma)\}\sum_{i=1}^I|T_i-\bar{q}_i|\right\}$ tends to a constant, in which case the power of the resulting procedure will be some value $\beta \in (0,1)$. The power of the sensitivity analysis converges to 1 when  $\text{plim}\; I^{-1}\sum_{i=1}^I (T_i - \bar{q}_i) - I^{-1}\{(\Gamma-1)/(1+\Gamma)\}\sum_{i=1}^I|T_i-\bar{q}_i| = c> 0$ as the numerator tends to $\infty$ in this case, and converges to 0 when $\text{plim} \; I^{-1}\sum_{i=1}^I (T_i - \bar{q}_i) - I^{-1}\{(\Gamma-1)/(1+\Gamma)\}\sum_{i=1}^I|T_i-\bar{q}_i| = c < 0$ as the numerator tends to $-\infty$. The design sensitivity is the value $\Gamma$ such that the difference in the probability limits exactly equals 0, i.e. the $\Gamma$ that solves the equation $\text{plim}\; I^{-1}\sum_{i=1}^I (T_i - \bar{q}_i) - I^{-1}\{(\Gamma-1)/(1+\Gamma)\}\sum_{i=1}^I|T_i-\bar{q}_i| = 0$. Solving for $\Gamma$, we obtain

\begin{align*}
\tilde{\Gamma}_{tilt} &= \underset{I\rightarrow\infty}{\text{plim}}\;\; \frac{I^{-1}\sum_{i=1}^I|T_i - \bar{q}_i| +  I^{-1}\sum_{i=1}^I(T_i - \bar{q}_i) }{ I^{-1}\sum_{i=1}^I|T_i - \bar{q}_i| -  I^{-1}\sum_{i=1}^I(T_i - \bar{q}_i)} = \frac{\eta + \theta}{\eta - \theta}.
\end{align*}

\subsection{Proof of Theorem \ref{thm:mstat}}
As the $m$-statistics $T$  is a sum of $iid$ contributions $T_i$ under the generative model in \S\ref{sec:gen} and $\bar{q}_i=0$, the probability limits in Assumption \ref{as:plim} are simply $\theta_J = E(T_i)$ and $\eta_J = E(|T_i|)$. The form for the design sensitivity then follows immediately from Theorem \ref{thm:closed}. 

To prove that the design sensitivity is non-decreasing in $J$, define \\$T_{iJ} = \sum_{j=1}^{J+1}Z_{ij}\sum_{\ell=1}^{J+1}\psi(Y_{ij}-Y_{i\ell})$, $\bar{\theta}_J = \theta_J/J$ and $\bar{\eta}_J = \eta_J/J$. Rearrange indices so that $Z_{i1}=1$; then $T_{iJ} = \sum_{\ell=1}^{J+1}\psi(Y_{i1}-Y_{i\ell}) = \sum_{\ell=2}^{J+1}\psi(\epsilon_{i1}+\tau - \epsilon_{i\ell})$ is the sum of $J$ identically distributed random variables, each with expectation $E[\psi(\epsilon_{i1}+\tau-\epsilon_{i2})] = \bar{\theta}_J$. Further note that $\bar{\theta}_J$ does not vary with $J$. As the function $f(x) = (x+c)/(x-c)$ is decreasing in $x$ for $x > c$ and $\eta_J \geq c$, the design sensitivity $(\theta_J+\eta_J)/(\eta_J-\theta_J) = (\bar{\theta}_J + \bar{\eta}_J)/(\bar{\eta}_J-\bar{\theta}_J)$ is non-decreasing in $J$ whenever $\bar{\eta}_J$ is non-increasing in $J$.

Consider $\epsilon_{i1},...,\epsilon_{iJ}, \epsilon_{i(J+1)},\epsilon_{i(J+2)}\overset{iid}{\sim} F_\epsilon$ and condition upon $\varepsilon_{i1}$. Then, given $\varepsilon_{i1}$, $T_{iJ}/J$ is the average of $J$ $iid$ random variables, and $T_{i(J+1)}/(J+1)$ is the average of $J+1$ random variables. By Example 3.A.29 of \citet{sha07}, if $X_1,...,X_J, X_{J+1}$ are $iid$ random variables, then for any convex $g$
\begin{align*}
E\left\{g\left(J^{-1}\sum_{j=1}^JX_i\right)\right\} &\geq E\left\{g\left((J+1)^{-1}\sum_{j=1}^{J+1} X_i\right)\right\}.
\end{align*}
Applying this conditional on $\epsilon_{i1}$ yields that \begin{align*}E\left\{J^{-1}\left|\sum_{\ell=2}^{J+1}\psi(\epsilon_{i1}+\tau - \epsilon_{i\ell})\right|\mid \epsilon_{i1}\right\} &\geq E\left\{(J+1)^{-1}\left |\sum_{\ell=2}^{J+2}\psi(\epsilon_{i1}+\tau - \epsilon_{i\ell})\right|\mid \epsilon_{i1}\right\}\end{align*}

Integrating over the distribution $\varepsilon_{i1}$ yields $\bar{\eta}_J$ on the left-hand side and $\bar{\eta}_{J+1}$ on the right-hand side, completing the proof that $\bar{\eta}_J$ is non-increasing. A necessary and sufficient condition for a strict inequality $\tilde{\Gamma}_{tilt, J} < \tilde{\Gamma}_{tilt, J+1}$ is
\begin{align*}
\text{pr}\left(\min_{2\leq \ell \leq J+2}\psi(\epsilon_{i1}+\tau-\epsilon_{i\ell})< \sum_{\ell=2}^{J+2}\psi(\epsilon_{i1}+\tau-\epsilon_{i\ell}) < \max_{2\leq \ell \leq J+2}\psi(\epsilon_{i1}+\tau-\epsilon_{i\ell}) \right) > 0,
\end{align*}
This stems from the proof technique used to prove the cited inequality. It ensures that among the $J+1$ possible leave-one-out averages formed from the $J+1$ terms $\{\psi(\epsilon_{i1}+\tau-\epsilon_{i\ell}): \ell=2,...,J+2\}$, some are positive and some are negative.

\subsection{Proof of Proposition \ref{prop:compare}}

The design sensitivities $\tilde{\Gamma}_{tilt}$ and $\tilde{\Gamma}_{conv}$ are the values $\Gamma$ that solve, respectively

\begin{align*}
\text{plim}\; I^{-1}\sum_{i=1}^I (T_i - \bar{q}_i) - I^{-1}\{(\Gamma-1)/(1+\Gamma)\}\sum_{i=1}^I|T_i-\bar{q}_i| &= 0;\\
\text{plim}\; I^{-1}\sum_{i=1}^I (T_i - \bar{q}_i) - I^{-1}\{(\Gamma-1)/(1+\Gamma)\}\sum_{i=1}^IM_{\Gamma i} = 0.
\end{align*}

Assume the uniqueness of these thresholds. We prove the result for $m_{\tilde{\Gamma}_{tilt}}$; the proof for $m_{\tilde{\Gamma}_{conv}}$ is analogous. The function $\text{plim}\; I^{-1}\sum_{i=1}^I (T_i - \bar{q}_i) - I^{-1}\{(\Gamma-1)/(1+\Gamma)\}\sum_{i=1}^IM_{\Gamma i}$ is monotone nonincreasing in $\Gamma$. Letting $\tilde{\Gamma}_{tilt}$ be the design sensitivity for the tilted statistic, we see that if $\text{plim}\; I^{-1}\sum_{i=1}^IM_{\tilde{\Gamma}_{tilt}i} - I^{-1}\sum_{i=1}^I|T_i-\bar{q}_i| > 0$, then $\text{plim}\; I^{-1}\sum_{i=1}^I (T_i - \bar{q}_i) - I^{-1}\{(\tilde{\Gamma}_{tilt}-1)/(1+\tilde{\Gamma}_{tilt})\}\sum_{i=1}^IM_{\tilde{\Gamma}_{tilt} i} < 0$. As the function is monotone nonincreasing, this implies that $\tilde{\Gamma}_{conv}$, the zero of this equation, falls below $\tilde{\Gamma}_{tilt}$. Conversely, if $\text{plim}\; I^{-1}\sum_{i=1}^IM_{\tilde{\Gamma}_{tilt} i} - I^{-1}\sum_{i=1}^I|T_i-\bar{q}_i| < 0$,  then $\text{plim}\;I^{-1}\sum_{i=1}^I (T_i - \bar{q}_i) - I^{-1}\{(\tilde{\Gamma}_{tilt}-1)/(1+\tilde{\Gamma}_{tilt})\}\sum_{i=1}^IM_{\tilde{\Gamma}_{tilt} i} > 0$, implying that $\tilde{\Gamma}_{conv} > \tilde{\Gamma}_{tilt}$.

\section{Theorem \ref{thm:designsens}}

Before proceeding with the proofs, we introduce asymptotic order notation for the small-$\tau$ regime. For a random variable $R_\tau$ and a positive deterministic scale $a_\tau$, we write $R_\tau = O_{L1}(a_\tau)$ if $\lim\sup_{\tau\downarrow 0} E|R_\tau|/a_\tau < \infty$. Similarly, we write $R_\tau = o_{L1}(a_\tau)$ to describe a random variable for which $ E|R_\tau|/a_\tau\rightarrow 0$ as $\tau \downarrow 0$. We use $O(a_\tau)$ and $o(a_\tau)$ when $R_\tau$ is constant, rather than random; for instance, if $R_\tau = O_{L1}(a_\tau)$, then $E|R_\tau| = O(a_\tau)$.

Redefine indices so that the first unit in each matched set receives the treatment and the remaining $J$ receive the control. Then under the generative model of \S \ref{sec:gen} \begin{align*} Y_{i1}&=\tau+\alpha_i+\epsilon_{i1}\\
 Y_{ij} &= \alpha_i+\epsilon_{ij} \;\;\; (j=2,...,J+1),\end{align*} where $\epsilon_{i1},...,\epsilon_{i,J+1}\overset{iid}{\sim}F_\epsilon$. 

 As $T=\sum_{i=1}^I T_i$ and $T_i = \sum_{j=1}^{J+1}Z_{ij}\sum_{\ell=1}^{J+1}\psi(Y_{ij}-Y_{i \ell})$ are $iid$ (note the $\alpha_i$ is cancelled out when subtracting $Y_{ij}$ and $Y_{i\ell}$), we consider any particular $T_i$ and drop the index $i$ for ease of notation. Let $\theta(\tau) = JE\{\psi(\epsilon_{1}-\epsilon_{2} + \tau)\}$, let $\eta(\tau) = E|\sum_{\ell=2}^{J+1}\psi(\epsilon_{1}-\epsilon_{\ell}+\tau)|$. Then, for $0 < \theta(\tau) < \eta(\tau)$, we have $\tilde{\Gamma}(\tau) = \{\eta(\tau)+\theta(\tau)\}/\{\eta(\tau)-\theta(\tau)\}$ is the design sensitivity for the tilted sensitivity analysis for a given value of $\tau >0$ in (\ref{eq:gen}). Let $\kappa(\tau) = \theta(\tau)/\eta(\tau) = \{\tilde{\Gamma}(\tau)-1\}/\{\tilde{\Gamma}(\tau)+1\}$. At $\tau = 0$, we have $\tilde{\Gamma}(0) = 1$ and $\kappa(0) = 0$. 

Recall that $\psi(\cdot)$ is assumed odd, and that $\psi(0) = 0$. Define $q_j(\tau)$ as 

\begin{align}\label{eq:qj}q_j(\tau) &= \begin{cases} \sum_{\ell=2}^{J+1}\psi(\epsilon_1-\epsilon_\ell + \tau) & j=1\\
 \psi(\epsilon_j-\epsilon_1-\tau) + \sum_{\ell=2}^{J+1}\psi(\epsilon_j-\epsilon_\ell) & j=2,...,J+1.\end{cases}
\end{align}

$q_j(\tau)$ is the value for $\sum_{\ell=1}^{J+1}\psi(Y_j-Y_\ell)$ for any particular $\tau$, such that $E(q_1(\tau)) = \theta(\tau)$ and $E|q_1(\tau)| = \eta(\tau)$. Define $U_\tau(\epsilon,c)$ as
\begin{align}\label{eq:fixed}
U_\tau(\epsilon, c) &= \frac{1}{J+1}\sum_{j=1}^{J+1}|q_j(\tau)-\kappa(\tau)c| 
\end{align} The worst-case expectation for the conventional sensitivity analysis at $\tilde{\Gamma}(\tau)$ is $\kappa(\tau)$ times the fixed point of (\ref{eq:fixed}). Let $\tilde{M}(\tau)= M_{\tilde{\Gamma}(\tau)}$ be the fixed point of (\ref{eq:fixed}), such that $U_\tau(\epsilon, \tilde{M}_{\tau}) = \tilde{M}(\tau)$, and let $m(\tau) = E[\tilde{M}(\tau)]$. By Proposition \ref{prop:compare} and the fact that $T_i$ are $iid$, the comparison of the design sensitivities of the tilted and conventional sensitivity analyses amounts to a calculation of $m(\tau)-\eta(\tau)$ under the uniqueness and monotonicity assumptions used in Proposition \ref{prop:compare}: positive values imply the tilted sensitivity analysis is superior, negative values imply the conventional approach is superior.

The proofs will involve expansions taken at $\tau=0$. With that in mind, define  $Q_j := q_j(0)$, and let $\bar{A} = (J+1)^{-1}\sum_{j=1}^{J+1}|Q_j|$. Note that $Q_j$ are exchangeable random variables for $j=1,...,J+1$, and that $E(\bar{A}) = E|Q_1|$.  Sections \ref{sec:1order}-\ref{sec:2order} develop first- and second-order expansions respectively for comparing the tilted and conventional approach in the small$-\tau$ regime. The remaining sections apply these results to the statistics considered in Theorem \ref{sec:ds}.

\subsection{First-order expansions}\label{sec:1order}
Observe that $\bar{A} = U_0(\epsilon, \bar{A})$. The following proposition establishes that $\bar{A}$ itself approximates $\tilde{M}(\tau)$ to order $O_{L1}(\tau)$. If one then plugs $\bar{A}$ into $U_\tau(\epsilon, \bar{A})$ in one step of fixed-point iteration, one attains an approximation to $\tilde{M}(\tau)$ with error of order $O_{L1}(\tau^2)$.
\begin{proposition}\label{prop:fixedorder}

Suppose that $\kappa(\tau) = O(\tau)$, $0 < \kappa(\tau) < 1$, that $(J+1)^{-1}\sum_{j=1}^{J+1}|q_j(\tau)-Q_j| = O_{L1}(\tau)$, and that $E|Q_j| < \infty$. Then,

\begin{align*}
\tilde{M}(\tau) &= \bar{A} + O_{L1}(\tau)\\
&= U_\tau(\epsilon,\bar{A}) + O_{L1}(\tau^2).
\end{align*}
\end{proposition}
\begin{proof}
Since $\tilde{M}(\tau)$ is a fixed point,
\begin{align*}
|\tilde{M}(\tau) - U_\tau(\epsilon, \bar{A})| &= \left|U_\tau(\epsilon, \tilde{M}(\tau))- U_\tau(\epsilon, \bar{A})\right|\\
&=\left|\frac{1}{J+1}\sum_{j=1}^{J+1}|q_j(\tau)-\kappa(\tau)\tilde{M}(\tau)| - \frac{1}{J+1}\sum_{j=1}^{J+1}|q_j(\tau)-\kappa(\tau)\bar{A}|\right|\\
&\leq |\kappa(\tau)||\tilde{M}(\tau) - \bar{A}|,
\end{align*} with the last line using the reverse triangle inequality. To control the absolute difference $|\tilde{M}(\tau)-\bar{A}|$, we add and subtract $U_\tau(\epsilon, \bar{A})$ and use this result along with the triangle inequality (both standard and reverse).

\begin{align*}
\left|\tilde{M}(\tau) - \bar{A}\right| &\leq |\tilde{M}(\tau) - U_\tau(\epsilon, \bar{A})| + |U_{\tau}(\epsilon,\bar{A})-\bar{A}|\\
&\leq |\kappa(\tau)||\tilde{M}(\tau) - \bar{A}| + |U_{\tau}(\epsilon,\bar{A})-\bar{A}|\\
& = |\kappa(\tau)||\tilde{M}(\tau) - \bar{A}| + \frac{1}{J+1}\left|\sum_{j=1}^{J+1}|q_{j}(\tau)-\kappa(\tau)\bar{A}| - |q_{j}(0)|\right|\\
&\leq |\kappa(\tau)||\tilde{M}(\tau) - \bar{A}| + \frac{1}{J+1}\sum_{j=1}^{J+1}|q_{j}(\tau)-\kappa(\tau)\bar{A} - Q_j|\\
&\leq |\kappa(\tau)||\tilde{M}(\tau) - \bar{A}| + \frac{1}{J+1}\sum_{j=1}^{J+1}|q_{j}(\tau)- Q_j| + |\kappa(\tau)|\bar{A} 
\end{align*}

Rearranging, we obtain
\begin{align*}\left|\tilde{M}(\tau) - \bar{A}\right| \leq \frac{\frac{1}{J+1}\sum_{j=1}^{J+1}|q_{j}(\tau)- Q_j|+ |\kappa(\tau)|\bar{A}}{1-|\kappa(\tau)|}\end{align*}

As $(J+1)^{-1}\sum_{j=1}^{J+1}|q_{j}(\tau)- Q_{j}| = O_{L1}(\tau)$, $\kappa(\tau) = O(\tau)$, and $\bar{A} = O_{L1}(1)$ we then have 

\begin{align*}\left|\tilde{M}(\tau) - \bar{A}\right| = \frac{O_{L1}(\tau)+O(\tau)O_{L1}(1)}{1-O(\tau)} = O_{L1}(\tau)\end{align*}

Hence,

\begin{align*}|\tilde{M}(\tau)-U_\tau(\epsilon,\bar{A})| &\leq \kappa(\tau)\left|\tilde{M}(\tau) - \bar{A}\right|\\
&= O(\tau)O_{L1}(\tau) = O_{L1}(\tau^2)\end{align*}
\end{proof}

Before proceeding, we state an initial lemma regarding the expected absolute value which will be useful when obtaining first-order expansions.

\begin{lemma}\label{lemma:outfirst}
Consider random variables $A_0$, $A_1$ with $E|A_0| < \infty$, $E|A_1| < \infty$. Then,

\begin{align*}
|A_0 + \tau A_1+ o_{L1}(\tau)| &= |A_0| + \tau \{A_1\text{sgn}(A_0)\} + \tau |A_1| \1\{A_0=0\} + o_{L1}(\tau)\end{align*} 
Consequently,
\begin{align*}
E|A_0 + \tau A_1+ o_{L1}(\tau)| &= E|A_0| + \tau E\{A_1\text{sgn}(A_0)\} + \tau E|A_1| \1\{A_0=0\} + o(\tau)\end{align*} 
\end{lemma}
\begin{proof}
First consider $|A_0 + \tau A_1|$. Consider $D_\tau = \{|A_0+\tau A_1| - |A_0|\}/\tau$. On $A_0\neq 0$, we see that $\text{sgn}(A_0+\tau A_1) = \text{sgn}(A_0)$ for $\tau$ sufficiently small. Therefore, $D_\tau \rightarrow A_1\text{sgn}(A_0)$ as $\tau\downarrow 0$. For $A_0=0$, we have instead that $|A_0+\tau A_1| = \tau |A_1|$, so on this event $D_\tau \rightarrow |A_1|\1\{A_0=0\}$. Hence  $D_\tau \rightarrow A_1\text{sgn}(A_0) + |A_1|\1\{A_0=0\}$ almost surely as $\tau\downarrow 0$. Moreover, using the reverse triangle inequality, we have that $|D_\tau| \leq |A_1|$, so that $|D_\tau - A_1\text{sgn}(A_0) - |A_1|\1\{A_0=0\}|\leq 2|A_1|$. As we have assumed $E|A_1| < \infty$, dominated convergence gives $E|D_\tau - A_1\text{sgn}(A_0) - |A_1|\1\{A_0=0\}|\rightarrow 0$. Equivalently after multiplying by $\tau$, we have $|A_0 + \tau A_1| = |A_0| + \tau A_1\text{sgn}(A_0) + \tau |A_1|\1\{A_0=0\} + o_{L1}(\tau)$. As $||A_0 + \tau A_1 + o_{L1}(\tau)| - |A_0+\tau A_1|| = o_{L1}(\tau)$ by the reverse triangle inequality, we have $|A_0 + \tau A_1 + o_{L1}(\tau)| = |A_0| + \tau A_1\text{sgn}(A_0) + \tau |A_1|\1\{A_0=0\} + o_{L1}(\tau)$.

\end{proof}

\begin{corollary}\label{cor:firstderiv}
For any random variable with the expansion $A_\tau = A_0 + \tau A_1 + o_{L1}(\tau)$
\begin{align*}
\frac{\partial}{\partial \tau} E(A_\tau)\Bigr|_{\tau = 0^+} &= E(A_1)\\
\frac{\partial}{\partial \tau} E|A_\tau|\Bigr|_{\tau = 0^+} &= E\{A_1\text{sgn}(A_0)\} + E|A_1| 1 \{A_0=0\}
\end{align*}
\end{corollary}
\begin{corollary}\label{cor:Kpsi}
For any random variable with the expansion $A_\tau = A_0 + \tau A_1 + o_{L1}(\tau)$, let $A_\tau^-= -A_0 + \tau A_1 + o_{L1}(\tau)$. Then
\begin{align*}
\left\{\frac{\partial}{\partial \tau} E|A_\tau|\Bigr|_{\tau = 0^+} - \frac{\partial}{\partial \tau} E|A^-_\tau|\Bigr|_{\tau = 0^+} \right\}/2 &= E\{A_1\text{sgn}(A_0)\}.
\end{align*}
\end{corollary}

We are now ready to state first-order expansion theorems for comparing the tilted and conventional sensitivity analysis.

\begin{proposition}\label{prop:firstorder}
Suppose that $q_{j}(\tau)$ admits the following first-order expansions
\begin{align*}
q_j(\tau) &=  Q_j +  \tau D_j + o_{L1}(\tau).
\end{align*}

with $E(Q_j) = 0$, $E|Q_j|$, $E|D_j| <\infty$, $E|Q_j| > 0$. Then,
\begin{align*}
\kappa(\tau) &= \tau \frac{E(D_1)}{E|Q_1|} + o(\tau),
\end{align*}
and

\begin{align*}m(\tau)- \eta(\tau) &= \tau \left[\frac{1}{J+1}\sum_{j=1}^{J+1}E\{(D_j-\beta\bar{A})\text{sgn}(Q_j)\}-E\{D_1\text{sgn}(Q_1)\}\right]\\ &+ \tau \left[\frac{1}{J+1}\sum_{j=1}^{J+1}E|(D_j-\beta\bar{A}) \1\{Q_j=0\}| - E|D_1 \1\{Q_1=0\}|\right] + o(\tau),\end{align*}
 where  $\beta = E(D_1)/E|Q_1|$.
\end{proposition}

\begin{proof}
Recall $\kappa(\tau) = \theta(\tau)/\eta(\tau) = E(q_1(\tau))/E|q_1(\tau)|$. Using the assumed expansions for $q_j$, that $E(Q_1) = 0$, and applying Lemma \ref{lemma:outfirst} to $\eta(\tau) = E|q_1(\tau)|$,
\begin{align*}
\kappa(\tau) &= \frac{E(Q_1)+\tau E(D_1) + o(\tau)}{E|Q_1| + o(1)}\\
&= \tau \frac{E(D_1)}{E|Q_1|} + o(\tau),
\end{align*}
so $\kappa(\tau) = O(\tau)$. Define $\beta = E(D_1)/E|Q_1|$. The assumed first-order expansions imply $(J+1)^{-1}\sum_{j=1}^{J+1}|q_j(\tau) - Q_j| = O_{L1}(\tau)$. Then Proposition  \ref{prop:fixedorder} applies:
\begin{align*}
q_j(\tau) - \kappa(\tau)\tilde{M}(\tau) &= Q_j + \tau(D_j-\beta\bar{A}) + o_{L1}(\tau).
\end{align*}From Lemma \ref{lemma:outfirst},
\begin{align*}
m(\tau)&= \frac{1}{J+1}\sum_{j=1}^{J+1}E|q_j(\tau) - \kappa(\tau)\tilde{M}(\tau)|\\
&= \frac{1}{J+1}\sum_{j=1}^{J+1}E|Q_j + (D_j-\beta\bar{A})\tau + o_{L1}(\tau)|\\
&= E(\bar{A}) + \tau E\left[\frac{1}{J+1}\sum_{j=1}^{J+1}\{(D_j-\beta\bar{A})\text{sgn}(Q_j) + |D_j-\beta\bar{A}|1\{Q_j = 0\}\} \right]+o(\tau).
\end{align*}Again applying Lemma \ref{lemma:outfirst} to $\eta(\tau) = E|q_1(\tau)|$,
\begin{align*}
\eta(\tau) = E|q_1(\tau)| &= E|Q_1 + D_1\tau + o_{L1}(\tau)|\\
&= E|Q_1| + \tau E\{D_1\text{sgn}(Q_1)\} + \tau E|D_1 \1\{Q_1=0\}| + o(\tau)
\end{align*}

Therefore,
\begin{align*}
m(\tau) - \eta(\tau) &= E\{\tilde{M}(\tau)\} - E|q_1(\tau)|\\
&= \tau \left[\frac{1}{J+1}\sum_{j=1}^{J+1}E\{(D_j-\beta\bar{A})\text{sgn}(Q_j)\}-E\{D_1\text{sgn}(Q_1)\}\right]\\ &+ \tau \left[\frac{1}{J+1}\sum_{j=1}^{J+1}E|(D_j-\beta\bar{A}) \1\{Q_j=0\}| - E|D_1 \1\{Q_1=0\}|\right] + o(\tau)
\end{align*}
\end{proof}

\subsection{Second-order expansions}\label{sec:2order}

For linear scores and $\psi_{0,h}$, $J$ odd, we will need a second-order expansion for the absolute value of random variables which themselves admit a second-order expansion. Consider a random variable $A_0$ which has a density $f$ in a neighborhood of 0 which is continuous at 0. For a deterministic $b$, the second-order term in the expansion of $E|A_0+b|$ about $b=0$ is $E(b^2\delta_0(A_0)) = b^2f(0)$, where $\delta_0(A_0)$ is the Dirac delta at 0. For $b = \tau a_1 + \tau^2 a_2$, the terms of order $\tau^2$ become $\tau^2 E\{a_2\text{sgn}(A_0)\} + \tau^2E\{a_1^2\delta_0(A_0)\} = \tau^2 E\{a_2\text{sgn}(A_0)\} + \tau^2a_1^2f(0).$ In what follows we consider how the expansions behave when $a_1$ and $a_2$ are replaced with random variables $A_1$ and $A_2$.

We begin by defining a version of the conditional expectation which is referred to within (\ref{eq:slicesup}), which will be used when dealing with objects of the form $E\{A_1\delta_0(A_0)\}$ when $A_1$ and $A_0$ are both random. This needs to be stated explicitly, as conditional expectations are only defined up to null sets. We then state two lemmas which, as written, are more or less tautological but provide a target for subsequent primitive conditions. Primitive conditions for the case of $\psi_{lin}$ will be presented later.

\begin{definition}\label{def:slice}
For random variables $V$ and $W$, suppose that

\begin{itemize}
\item[(i)] $V$ admits a density $g_V$ in a neighborhood of 0 that is continuous at 0, and satisfies $0 < g_V(0) < \infty$.
\item[(ii)] $E|W|< \infty$, and within a neighborhood of 0 there exists a function $g_{V,W}$ that is continuous at 0 such that for any Borel set $B$ within that neighborhood, $E(W \1\{V\in B\}) = \int_{B}g_{V,W}(x)dx$.
\end{itemize}
Then, we define the conditional expectation of $W$ at $V=0$ as the density ratio version\begin{align*} E(W\mid V=0) := g_{V,W}(0)/g_V(0).
\end{align*}

\end{definition}

\begin{lemma}\label{lemma:outsecond}
Suppose that $E|A_0| < \infty$ with $\text{pr}(A_0=0) = 0$. Suppose further that conditions (i) and (ii) within Definition \ref{def:slice} hold for $V = A_0$ and $W = A_1^2$. Let $f_0 = g_{A_0}(0)$ be the density of $A_0$ evaluated at 0 and let $W_\tau = A_1+\tau A_2$ with $E|A_2|<\infty$. Suppose that $E\{|A_0+\tau W_\tau|-|A_0|-\tau W_\tau\text{sgn}(A_0)\} = f_0\tau^2E\{A_1^2\mid A_0=0\} + o(\tau^2)$. Then, \begin{align*}&E|A_0 + \tau A_1 + \tau^2 A_2 + o_{L1}(\tau^2)|\\ &= E|A_0| + \tau E\{A_1\text{sgn}(A_0)\}+ \tau^2 \left[E\{A_2\text{sgn}(A_0)\} + f_0E\{A_1^2 \mid A_0=0\}\right] + o(\tau^2) \end{align*}
\end{lemma}

\begin{proof}

First observe $E|A_0+\tau W_\tau+o_{L1}(\tau^2)| = E|A_0+\tau W_\tau| + o(\tau^2)$ by the triangle inequality. Taking $C_\tau = |A_0+\tau W_\tau| -|A_0|-\tau W_\tau\text{sgn}(A_0)$, we have $E|A_0+\tau W_\tau| = E|A_0|+\tau E\left\{W_\tau\text{sgn}(A_0)\right\} + E(C_\tau)$. Plugging in the definition of $W_\tau$ gives $\tau E\left\{W_\tau\text{sgn}(A_0)\right\} = \tau E\{A_1\text{sgn}(A_0)\} +\tau^2 E\{A_2\text{sgn}(A_0)\}$. As we have assumed $E(C_\tau) = f_0\tau^2E\{A_1^2\mid A_0=0\} + o(\tau^2)$, the proof is complete.
\end{proof}

\begin{lemma}\label{lemma:outrice}
Define $B_\tau = A_0+\tau A_1 + \tau^2 A_2$ for random variables $A_0, A_1,A_2$. Consider a random variable $a_\tau = O_{L1}(\tau^2)$. If $E|a_\tau \1\{\text{sgn}(B_\tau)\neq\text{sgn}(A_0)\}| = o(\tau^2)$ and $E|a_\tau \1\{|B_\tau|\leq |a_\tau|\}| = o(\tau^2)$, then\begin{align*}
E|B_\tau + a_\tau + o_{L1}(\tau^2)| &= E|B_\tau| + E\{\text{sgn}(A_0)a_\tau\} + o(\tau^2)
\end{align*}
\end{lemma}

\begin{proof}

First observe $E|B_\tau + a_\tau+o_{L1}(\tau^2)| = E|B_\tau+a_\tau| + o(\tau^2)$ by the triangle inequality. Consider the function $H(t, a) = |t+a| - |t| - a\;\text{sgn}(t)$, and observe that $|t|-a\;\text{sgn}(t)$ = $(t-a)\text{sgn}(t)$. Then, we see that $H(t,a)$ equals zero whenever $|t|>|a|$ as this implies $t$ and $t+a$ have the same sign. In general, we have $0\leq H(t,a)\leq 2|a|\1\{|t|\leq|a|\}$. Applying this for $t=B_\tau$ and $a = a_\tau$, we have $|E|B_\tau+a_\tau| - E|B_\tau| - E\{a_\tau\;\text{sgn}(B_\tau)\}| \leq 2E\{|a_\tau|\1\{|B_\tau|\leq|a_\tau|\}\}$, which is $o(\tau^2)$ by assumption. As $|\text{sgn}(B_\tau)-\text{sgn}(A_0)| \leq 2\1\{\text{sgn}(B_\tau)\neq \text{sgn}(A_0)\}$, we further have $|E\{\text{sgn}(B_\tau)a_\tau\} - E\{\text{sgn}(A_0)a_\tau\}|\leq 2E\{|a_\tau|\1\{\text{sgn}(B_\tau)\neq\text{sgn}(A_0)\}\} = o(\tau^2)$ by the assumptions. This completes the proof.

\end{proof}

Assuming additional terms in the expansions for $q_j(\tau)$ and $\kappa(\tau)$, we can now attain a second-order comparison between the tilted and conventional sensitivity analyses, useful when the first-order terms cancel out.
\begin{proposition}\label{prop:secorder}
Suppose that $q_{j}(\tau)$ admits the expansion
\begin{align*}
q_j(\tau) &=  Q_j +  \tau D_j + \tau^2 H_j + r_j(\tau) + o_{L1}(\tau^2)\\
\end{align*}
with $E(Q_j)=0$, $E|Q_j|, E|D_j|, E|H_j| < \infty$, and $E\{r_1(\tau)\} = o(\tau^2)$. Suppose further that the first-order coefficient in the expansion of $m(\tau)-\eta(\tau)$ displayed in Proposition \ref{prop:firstorder} evaluates to zero, and that $\text{pr}(Q_j = 0) = 0 $ so that there are no atoms at 0.

Define $\bar{R} =  (J+1)^{-1}\sum_{j=1}^{J+1}\{(D_j-\beta\bar{A})\text{sgn}(Q_j)\}$, with $\beta = E(D_1)/E|Q_1|$ and $\rho = E(H_1)/E|Q_1| - E(D_1)E\{D_1\text{sgn}(Q_1)\}/\{E|Q_1|\}^2$.  Suppose that the tuples $(Q_1,D_1,H_1, r_1(\tau))$ and $(Q_j, D_j-\beta\bar{A}, H_j-\rho\bar{A}-\beta\bar{R}, r_j(\tau))$ for $j=1,...,J+1$ satisfy the assumptions imposed on $(A_0, A_1, A_2, a_\tau)$  in Lemmas \ref{lemma:outsecond} and \ref{lemma:outrice}. Let $f_0$ denote the density of $Q_j$ evaluated at 0. Further suppose that $\lim_{\tau \downarrow 0} E[\text{sgn}(Q_j)r_j(\tau)]/\tau^2 = \mathcal{R}_j < \infty$. Then,

\begin{align*}
&m(\tau) - \eta(\tau) \\
&= \tau^2 \left(\frac{1}{J+1}\sum_{j=1}^{J+1}\left[E\{(H_j-\rho\bar{A}-\beta\bar{R})\text{sgn}(Q_j)\} + f_0E\{(D_j-\beta \bar{A})^2\mid Q_j=0\} + \mathcal{R}_j\right]\right)\\& - \tau^2\left[ E\{H_1\text{sgn}(Q_1)\} + f_0E\{D_1^2\mid Q_1=0\}+\mathcal{R}_1\right] + o(\tau^2)\\
\end{align*}
\end{proposition}
\begin{proof}
Under the stated assumptions, $E(q_1(\tau))$ admits the expansion $\tau E(D_1) + \tau^2E(H_1)+ o(\tau^2)$. From this and Corollary \ref{cor:firstderiv}, $\kappa(\tau)$ admits the second-order expansion
\begin{align*}
\kappa(\tau) &= \tau \beta + \tau^2\rho + o(\tau^2),\\
\rho &:= \frac{E(H_1)}{E|Q_1|}-\frac{E(D_1)E\{D_1\text{sgn}(Q_1)\}}{\{E|Q_1|\}^2}.
\end{align*}

Using Lemma \ref{lemma:outfirst} and remembering $\text{pr}(Q_j=0)=0$,
\begin{align*}
\tilde{M}(\tau) &= \frac{1}{J+1}\sum_{j=1}^{J+1}|q_j(\tau) - \kappa(\tau)\tilde{M}(\tau)|\\
&= \frac{1}{J+1}\sum_{j=1}^{J+1}|Q_j + (D_j-\beta\bar{A})\tau + o_{L1}(\tau)|\\
&= 
\frac{1}{J+1}\sum_{j=1}^{J+1}|Q_j| +  \tau\{(D_j-\beta\bar{A})\text{sgn}(Q_j) + |D_j-\beta\bar{A}|\1\{Q_j = 0\}\} + o_{L1}(\tau)\\
&= \bar{A} + \tau \left[\frac{1}{J+1}\sum_{j=1}^{J+1}\{(D_j-\beta\bar{A})\text{sgn}(Q_j)\} \right]+o_{L1}(\tau) 
\end{align*}

Recalling $\bar{R} =  (J+1)^{-1}\sum_{j=1}^{J+1}\{(D_j-\beta\bar{A})\text{sgn}(Q_j)\}$ we have $\tilde{M}(\tau) = \bar{A}+\tau\bar{R} + o_{L1}(\tau)$. Now, using the assumed expansions along with Lemmas \ref{lemma:outsecond} and \ref{lemma:outrice},
\begin{align*}
m(\tau) &= E\left\{\tilde{M}(\tau)\right\}\\ &= \frac{1}{J+1}\sum_{j=1}^{J+1}E|q_j(\tau) - \kappa(\tau)\tilde{M}(\tau)|\\
&= \frac{1}{J+1}\sum_{j=1}^{J+1}E|Q_j + \tau(D_j-\beta\bar{A})+  \tau^2 (H_j - \beta\bar{R}-\rho\bar{A})+ r_j(\tau) +  o_{L1}(\tau^2)|\\
&= E(\bar{A}) + \tau E(\bar{R}) + E\left[\frac{1}{J+1}\sum_{j=1}^{J+1}\left\{\text{sgn}(Q_j)r_j(\tau)\right\}\right]\\
&+ \tau^2\left(\frac{1}{J+1}\sum_{j=1}^{J+1}\left[ E\{(H_j-\rho\bar{A}-\beta\bar{R})\text{sgn}(Q_j)\} +  f_0E\{(D_j-\beta \bar{A})^2\mid Q_j=0\}\right]\right) + o(\tau^2).
\end{align*}

Similarly for $\eta(\tau)$
\begin{align*}
\eta(\tau) = E|q_1(\tau)| &= E|Q_1| + \tau[E\{D_1\text{sgn}(Q_1)\}] + E\{\text{sgn}(Q_1)r_1(\tau)\}\\
&+ \tau^2 \left[E\{H_1\text{sgn}(Q_1)\} + f_0E\{D_1^2\mid Q_1=0\}\right]+o(\tau^2).
\end{align*}

Recall again that $E|Q_1| = E(\bar{A})$. Then, under our assumption of equality of the first-order coefficients and using that $\mathcal{R}_j = \lim_{\tau \downarrow 0} E\{\text{sgn}(Q_j)r_j(\tau)\}/\tau^2 < \infty $,

\begin{align*}
&m(\tau) - \eta(\tau) \\
&= \tau^2 \left(\frac{1}{J+1}\sum_{j=1}^{J+1}\left[E\{(H_j-\rho\bar{A}-\beta\bar{R})\text{sgn}(Q_j)\} + f_0E\{(D_j-\beta \bar{A})^2\mid Q_j=0\} + \mathcal{R}_j\right]\right)\\& - \tau^2\left[ E\{H_1\text{sgn}(Q_1)\} + f_0E\{D_1^2\mid Q_1=0\}+\mathcal{R}_1\right] + o(\tau^2)\\
\end{align*}
\end{proof}

\subsection{Proof for linear scores, $\psi_{lin}$}
\begin{proposition}\label{prop:linear}
Suppose that $F_\epsilon$ admits a density that is bounded and continuous, and that $E(\epsilon_i^2) < \infty$. Then, the conditions of Proposition \ref{prop:firstorder} hold. Moreover, when $K(\psi,F_\epsilon)=0$, the conditions required for Proposition \ref{prop:secorder} hold. We further have
\begin{align*} L(\psi_{lin}, F_\epsilon) &=\left[\frac{1}{J+1}\sum_{j=1}^{J+1}E|(D_j-\beta\bar{A}) \1\{Q_j=0\}| - E|D_1 \1\{Q_1=0\}|\right] = 0.
\end{align*}
Therefore, $K(\psi_{lin},F_\epsilon) > 0$ implies the tilted approach is superior in the small-$\tau$ regime, and $K(\psi_{lin}, F_\epsilon) < 0$ implies the conventional approach is superior in the small-$\tau$ regime. For $K(\psi_{lin}, F_\epsilon) = 0$, the second-order comparison implies that the comparison is determined by the inequality
\begin{align}\label{eq:slicesup}
\frac{f_0E(\bar{A}^2\mid Q_1=0) + E(\bar{A}\bar{S}^2) - E(\bar{A})\{E(\bar{S})\}^2}{\{E(\bar{A})\}^2} > f_0\left(\frac{J-1}{J}\right),
\end{align}
where $\bar{S} = (J+1)^{-1}\sum_{j=1}^{J+1}\text{sgn}(Q_j)$.
\end{proposition}

\begin{proof}
The derivative of $\psi_{lin}(x+\tau)$ with respect to $\tau$ is 1. Therefore, $D_1 = J$ and $D_j = -1$ for $j=2,...,J+1$, $H_j=0$ for all $j$, and $r_j(\tau) = 0$ for all $j$. The assumptions imply $E|Q_1|>0$, such that $\beta = J/E|Q_1| > 0$. Therefore, Proposition \ref{prop:firstorder} applies. The assumptions and the form of $\psi_{lin}$ imply $\text{pr}(Q_1=0) = 0$, such that 
\begin{align*} L(\psi_{lin}, F_\epsilon) &=\left[\frac{1}{J+1}\sum_{j=1}^{J+1}E|(D_j-\beta\bar{A}) \1\{Q_j=0\}| - E|D_1 \1\{Q_1=0\}|\right] = 0.
\end{align*}

Therefore, if $K(\psi_{lin}, F_\epsilon)\neq 0$, the first-order comparison can be carried out through its sign: positive $K(\psi_{lin}, F_\epsilon)$ implies the tilted approach is superior in the small-$\tau$ regime, while the conventional approach is superior for negative $K(\psi_{lin}, F_\epsilon)$.

When $K(\psi_{lin}, F_\epsilon)=0$, the second-order expansion is required. Under our assumptions, $\kappa(\tau)$ admits a second-order expansion. The first order coefficient is again $\beta = J/E|Q_1|$, and the second order coefficient is $\rho = -J^2E(\text{sgn}(Q_1))/(E|Q_1|)^2$. 

We now consider a second-order expansion of $m(\tau) = E[\tilde{M}(\tau)]$. Under our assumptions, $\tilde{M}(\tau)$ can be written as the average of terms $|Q_j+\tau A_{1j} + \tau^2 A_{2j} + o_{L1}(\tau^2)|$, where $A_{1j} = (D_j-\beta \bar{A})$, and $A_{2j} = -\beta \bar{R} - \rho\bar{A}$, with $\bar{R}$ defined as in the proof of Proposition \ref{prop:secorder}. We thus require a second-order expansion of $E|Q_j+\tau A_{1j} + \tau^2 A_{2j} + o_{L1}(\tau^2)|$

We proceed for $j=1$; the proof is analogous for $j=2,...,J+1$. To begin, we consider quantities in the numerator of the conditional expectation in Definition \ref{def:slice} for $A_{11}^2 = (D_1-\beta\bar{A})^2$. As $D_1,..,D_{J+1}$ are constants, we only need to verify the requirements for $V=Q_1$, $W=\bar{A}^2$ and $W=\bar{A}$. 

We begin with the required conditions for $Q_1$. To derive the density of $Q_1$, note that $Q_1 = J\epsilon_1 - \sum_{\ell=2}^{J+1}\epsilon_\ell$, so that $\epsilon_1 = \left(Q_1+\sum_{\ell=2}^{J+1}\epsilon_\ell\right)/J$. Letting $X = (X_1,...,X_J) = (\epsilon_2,...,\epsilon_{J+1})$ and recalling that $\epsilon_j$ are $iid$, the joint density of $(Q_1, X)$ is 
\begin{align*}
p(q_1, x) &= \frac{1}{J}p_\epsilon\left(\frac{q_1 + \sum_{\ell=1}^J x_\ell}{J}\right)\prod_{\ell=1}^{J}p_\epsilon(x_\ell),
\end{align*} 
and the marginal for $Q_1$ is thus
\begin{align*}
g_{Q_1}(q_1) &= \int p(q_1, x) dx
\end{align*}
$p(q_1,x)$ is bounded and continuous in its first argument for any fixed value of the second by our assumptions. Therefore, dominated convergence shows that
$g_{Q_1}(q_1)$ is also continuous everywhere, and hence at 0. Moreover, note that since $p_\epsilon$ is a continuous density, it must be positive on a compact interval $\mathcal{I}$. If $x_1,...,x_J$ are all within $\mathcal{I}$, so too is their average. Therefore, for some $C>0$, $g_{Q_1}(0) \geq \int_{\mathcal{I}^J} p(0, x) \geq \int_{\mathcal{I}^J} C^{J+1}/J = |\mathcal{I}|^{J}C^{J+1}/J > 0$. This establishes condition (i) within Definition \ref{def:slice} applies to $Q_1$.

We now consider $\bar{A}$ and $\bar{A}^2$. Recall that $\bar{A} = (J+1)^{-1}\sum_{j=1}^{J+1}|Q_j|$, and $Q_j = J\epsilon_j - \sum_{\ell\neq j}\epsilon_\ell$. For $|Q_1| < \delta$, there exists a $C_\delta$ such that $\max |Q_\ell| \leq C_\delta(1+\sqrt{\sum_{\ell=1}^{J}X_\ell^2})$. Therefore, there is a constant $C_\delta'$ such that $|\bar{A}| \leq C_\delta'(1+\sqrt{\sum_{\ell=1}^{J}X_\ell^2})$ and $\bar{A}^2 \leq C_\delta'(1+\sum_{\ell=1}^{J}X_\ell^2)$.  

Let $\bar{a}(q_1, x)$ be the realized value of $\bar{A}$ when $Q_1=q_1$, $X = x$ and recall $(X_1,...,X_J)=(\epsilon_2,...,\epsilon_{J+1})$. We have that\begin{align*}
\{\bar{a}(q_1,x)\}^2p(q_1,x) \leq C'_\delta\left(1+\sum_{\ell=1}^{J}x_\ell^2\right)p(q_1,x),
\end{align*}
which is integrable as $E(X_\ell^2) = E(\epsilon_\ell^2) < \infty$ and $p_\epsilon$ is bounded by assumption.

For $|q_1|\leq \delta$, define 
\begin{align*} g_{Q_1,\bar{A}^2}(q_1) &:= \int \{\bar{a}(q_1,x)\}^2p(q_1,x)\;dx.\end{align*} We have $E(\bar{A}^2\1\{Q_1\in B\}) = \int_{B} g_{Q_1,\bar{A}^2}(q_1) d q_1$. By the established integrable bound, dominated convergence yields continuity of $g_{Q_1,\bar{A}^2}$ at 0. By this, in tandem with our results on the density for $Q_1$ in a neighborhood of 0, we may use the ratio form for $E\{\bar{A}^2\mid Q_1=0\}$ as defined in Definition \ref{def:slice}. A similar proof shows the conditions stated in Definition \ref{def:slice} hold for $\bar{A}$, justifying the definition of $E\{\bar{A}\mid Q_1=0\}$.

We now verify that the conditions in Lemma \ref{lemma:outsecond} hold, which will provide the required expansion. For realized values $q_1$ and $x$ of $Q_1$ and $X=(\epsilon_2,...,\epsilon_{J+1})$, let $b_{\tau}(q_1,x)$ be the realized value of $\tau a_{1}(q_1,x)+\tau^2 a_{2}(q_1,x)$, where $a_{1}(q_1,x)$ is the realized value of $D_1-\beta \bar{A}$ and $a_{2}(q_1,x)$ is the realized value of $-\beta\bar{R}-\rho\bar{A}$. The difference of interest is between $|q_1+b_{\tau}(q_1,x)|$ and the first-order expansion with respect to $b_{\tau}(q_1,x)$. It may be written as
\begin{align*}\Delta_\tau(q_1,x) := |q_1 + b_{\tau}(q_1,x)| - |q_1|-b_{\tau}(q_1, x)\text{sgn}(q_1).
\end{align*}

Before proceeding, as in Lemma \ref{lemma:outrice} we record a few properties of $H(t, a) = |t+a| - |t| - a\;\text{sgn}(t)$ that will be useful in what follows. First, it is bounded as $0\leq H(t,a)\leq 2|a|\1\{|t|\leq|a|\}$. Furthermore, it satisfies $\int H(t,a)dt = a^2$. To see this, note that for $a > 0$, $H(t,a) = 2(t+a)\1\{-a \leq t \leq 0\}$, so that $\int H(t,a)dt = 2\int_{-a}^0(t+a)\;dt = a^2$. For $a < 0$, $H(t,a) = -2(t+a)\1\{0\leq t \leq -a\}$, so again $\int H(t,a)\;dt = a^2$ by symmetry.

We proceed by partitioning into the events that $||X||_2 = \sqrt{\sum_{j=1}^JX_j^2} \leq K$ versus $\geq K$, and then control the tail contribution sending $K\rightarrow \infty$. 

The contributions to $E(\Delta_\tau(Q_1,X))$ on the event $||X||_2  \leq K$, scaled by $\tau^2$, is
\begin{align*}
\frac{1}{\tau^2}\int_{||X||_2 \leq K}\int \Delta_\tau(q_1, x)p(q_1,x)\;dq_1\;dx
\end{align*}
Changing variables as $q_1=t \tau$, we have $\Delta_\tau(\tau t, x) = \tau H(t, a_{1}(\tau t, x) + \tau a_{2}(\tau t, x))$. As $dq_1 = \tau dt$, we have
\begin{align*}
\frac{1}{\tau^2}\int_{||x||_2 \leq K}\int \Delta_\tau(q_1, x)p(q_1,x)\;dq_1\;dx &= \int_{||x||_2 \leq K}\int \left\{H(t, a_{1}(\tau t, x) + \tau a_{2}(\tau t, x))\right\}p(\tau t,x)\;dt\;dx
\end{align*}
As pointwise we have that $\left\{H(t, a_{1}(\tau t, x) + \tau a_{2}(\tau t, x))\right\}p(\tau t,x)$ converges to $\left\{H(t, a_{1}(0, x))\right\}p(0,x)$, dominated convergence and our previous result on the integral of $H(t, a)$ gives
\begin{align*}\int_{||x||_2 \leq K}\int \left\{H(t, a_{1}(\tau t, x) + \tau a_{2}(\tau t, x))\right\}p(\tau t,x)\;dt\;dx &\rightarrow \int_{||x||_2\leq K}\int \left\{H(t, a_{1}(0, x))\right\}p(0,x)\;dt\;dx\\
&=\int_{||x||_2\leq K}\left\{a_{1}(0,x)\right\}^2p(0,x) dx
\end{align*}

It now remains to control the tail $||X||_2  > K$ as $\tau \rightarrow 0$.
\begin{align*}
\Delta_\tau(q_1,x) & \leq 2|b_\tau(q_1,x)|\1\{|q_1|\leq |b_\tau(q_1,x)|\}
\end{align*}

By analogous arguments to those used to establish the conditional expectation identity, there exists a constant such that $|\bar{A}|\leq C (|Q_1|+||X||_2)$. As $|D_j|\leq J$ and $\bar{R} = (J+1)^{-1}\sum_{j=1}^{J+1}(D_j-\beta \bar{A})\text{sgn}(Q_j)$, there exists a $C_1$ such that $|a_{1}(q_1,x)|\leq C_1(1+|q_1| + ||X||_2)$ and such that $|a_{2}(q_1,x)|\leq C_1(1+|q_1|+||X||_2)$; this implies $|b_\tau(q_1,x)|\leq C\tau(1+||x||_2 +|q_1|)$. On the event that $\Delta_\tau$ is nonzero we have $|q_1| \leq |b_\tau(q_1,x)|$, which further implies that $|q_1| \leq C_1\tau(1+|q_1|+||x||_2 )$. Hence, rearranging we have for $\tau$ sufficiently small, $|q_1|\leq C_2\tau(1+||x||_2 )$, and hence that $|b_\tau(q_1,x)|\leq C_3\tau(1+||x||_2 )$ on the event that $\Delta_\tau$ is nonzero.

Recall further that $p_\epsilon$ is bounded, so that $p(q_1,x) \leq C_4\prod_{\ell=1}^{J}p_\epsilon(x_\ell)$ Therefore, for fixed $||x||$ and $\tau$ sufficiently small,

\begin{align*}
\frac{1}{\tau^2} \int \Delta_\tau(q_1, x)p(q_1,x)\;dq_1 &\leq \frac{1}{\tau}C_5(1+||x||_2 )\int_{|q_1|\leq C_2\tau(1+||x||_2 )}\prod_{\ell=1}^{J}p_\epsilon(x_\ell)\;dq_1\\
&\leq C_6 (1+||x||_2 )^2\prod_{\ell=1}^{J}p_\epsilon(x_\ell)\end{align*}

As $E(X_\ell^2)<\infty$, sending $K\rightarrow \infty$ we get
\begin{align*}
E(\Delta_{\tau}(Q_1,X)\1\{||X||_2 \geq K\})/\tau^2 \leq \int_{||x||_2 \geq K}C_6 (1+||x||_2 )^2\prod_{\ell=1}^{J}p_\epsilon(x_\ell)\rightarrow 0,
\end{align*}
so that
\begin{align*}
E(\Delta_{\tau}(Q_1,X))/\tau^2 &\rightarrow \int\left\{a_{1}(0,x)\right\}^2p(0,x) dx\\
&= D_1^2g_{Q_1}(0) -2\beta D_1g_{Q_1,\bar{A}}(0) + \beta^2g_{Q_1,\bar{A}^2}(0)\\
&= f_0\left\{D_1^2-2\beta D_1E(\bar{A}\mid Q_1=0) + \beta^2E(\bar{A}^2\mid Q_1=0)\right\}.
\end{align*}

Hence,
\begin{align*} E(\Delta_\tau(Q_1,X)) &= \tau^2f_0\left\{D_1^2-2\beta D_1E(\bar{A}\mid Q_1=0) + \beta^2E(\bar{A}^2\mid Q_1=0)\right\} + o(\tau^2).\end{align*}

The expansion for $\eta(\tau) = E|q_1(\tau)|$ is established by the same argument, with $A_1=D_1$ and $A_2=0$. We have thus justified the components of the second-order comparison in Proposition \ref{prop:secorder}. To attain the simplification (\ref{eq:slicesup}), we note first that $\mathcal{R}_j=0$, $H_j=0$, $\beta = J/E|Q_1|$, and $\rho = -\beta^2E(\bar{S})$, where $S_j = \text{sgn}(Q_j)$ and $\bar{S} = (J+1)^{-1}\sum_{j=1}^{J+1}S_j$.  We also have that $(J+1)^{-1}\sum_{j=1}^{J+1}D_j^2 = (J^2+J)/(J+1)=J$. By exchangeability, $E(\bar{A}^2\mid Q_j=0) = E(\bar{A}^2\mid Q_1=0)$ and $E(\bar{A}\mid Q_j=0) = E(\bar{A}\mid Q_1=0)$

We now consider $(J+1)^{-1}\sum_{j=1}^{J+1}(-\rho\bar{A}-\beta\bar{R})S_j = -\rho\bar{A}\bar{S} - \beta\bar{R}\bar{S}$. Recall that $\bar{R} = (J+1)^{-1}\sum_{j=1}^{J+1}D_jS_j - \beta\bar{A}\bar{S}$. Note that $(J+1)^{-1}\sum_{j=1}^{J+1}E(D_jS_j)=0$ since $D_j$ sums to 0 and $S_j$ are exchangeable. Note further that $(J+1)^{-1}\sum_{j=1}^{J+1}E(\bar{S}D_jS_j)=0$, also by the fact that $D_j$ sums to zero and $S_j$ are exchangeable. We thus have $E(\bar{R}) = -\beta E(\bar{A}\bar{S})$ and $-\beta E(\bar{R}\bar{S}) = \beta^2E(\bar{A}\bar{S}^2)$.

As we are considering $K(\psi_{lin}, F_\epsilon) = 0$, we must have that the first-order terms in Proposition \ref{prop:firstorder} cancel. As we are in the atomless case, we must have $E(\bar{R}) = JE(\bar{S})$. As $E(\bar{R}) = -\beta E(\bar{A}\bar{S})$, we must have $E(\bar{A}\bar{S}) = -JE(\bar{S})/\beta = -E|Q_1|E(\bar{S})$. Hence, $-\rho E(\bar{A}\bar{S}) = -\beta^2E|Q_1|\{E(\bar{S})\}^2$.

Noting $E(\bar{A}) = E|Q_1|$, the second-order expansion is thus
\begin{align*}  
m(\tau) - \eta(\tau) &= \tau^2 \left(\frac{1}{J+1}\sum_{j=1}^{J+1}\left[E\{(-\rho\bar{A}-\beta\bar{R})\text{sgn}(Q_j)\} + f_0E\{(D_j-\beta \bar{A})^2\mid Q_j=0\}\right]\right)\\& - \tau^2 f_0ED_1^2 + o(\tau^2)\\
&= \tau^2\left[-\beta^2E(\bar{A})\left\{E(\bar{S})\right\}^2+\beta^2E(\bar{A}\bar{S}^2) + f_0 J + \beta^2f_0E\{\bar{A}^2\mid Q_1=0\} - f_0J^2\right] + o(\tau^2)\\
\end{align*}
Recalling that $\beta = J/E(\bar{A})$, the second-order coefficient is positive if and only if
\begin{align*}
 \frac{f_0E\{\bar{A}^2\mid Q_1=0\}+E(\bar{A}\bar{S}^2)-E(\bar{A})\left\{E(\bar{S})\right\}^2}{\{E(\bar{A})\}^2} &> f_0\left(\frac{J-1}{J}\right)\\
\end{align*}

\end{proof}
\subsection{Proof for linear scores, Gaussian errors}
We now demonstrate that when $F_\epsilon$ is Gaussian and we use $\psi_{lin}$, the inequality in (\ref{eq:slicesup}) holds for all $J\geq 2$. 

In the proofs it is easier to work with $n=J+1$ to align formulae with standard results from the multivariate normal. Without loss of generality assume $E(\epsilon_j^2)=1$. Observe that as $\epsilon_j$ are $iid$ Gaussian, $(Q_1,...,Q_{n})$ are exchangeable and Gaussian with $E(Q_1) = 0$, $\text{var}(Q_1) = n(n-1)$, and $E(Q_1Q_2) = -n$, so $Corr(Q_1,Q_2) = -1/(n-1)$. They further satisfy $(Q_1,...,Q_{n})  \overset{d}{=} -(Q_1,...,Q_{n})$. This implies that $E(\bar{S}) = 0$. The remainder of the proof involves computing and combining terms of the form $E\{Q_1S_2\}$, $E\{|Q_1|S_2S_3\}$, $E|Q_1|$, $E(|Q_2|\mid Q_1=0)$, and $E(|Q_2Q_3|\mid Q_1=0)$. 

We begin with $E(\bar{A}^2\mid Q_1=0)$. As $(Q_1,...,Q_{n})$ is jointly normal and supported on the $n-1$ dimensional hyperplane $\sum_{j=1}^nQ_j = 0$, the distribution for $(Q_2,...,Q_n)$  given $Q_1=0$ remains jointly normal, supported on the $n-2$ dimensional hyperplane $\sum_{j=2}^n Q_j=0$. Note $Q_1=0$ implies that $\epsilon_{1} = \sum_{\ell\neq 1}\epsilon_\ell/(n-1)$. Therefore, $Q_2 \mid Q_1=0$ can be expressed as $(n-1)\epsilon_2-n/(n-1)\sum_{\ell \neq 1,2}\epsilon_\ell - \epsilon_2/(n-1) = \left\{n/(n-1)\right\}\left\{(n-2)\epsilon_2 - \sum_{\ell\neq 1,2}\epsilon_\ell \right\}$. This is simply $n/(n-1)$ times the unconditional distribution for $(Q_1,...,Q_{n-1})$ when there are $J-1$ rather than $J$ controls.  Let $\tilde{Q}_1,...,\tilde{Q}_{n-1}$ be distributed according to the unconditional distribution for $Q$ with $J-1$ controls. It has $E(\tilde{Q}_j)= 0$, $\text{var}(\tilde{Q}_j) = (n-1)(n-2)$, and covariance $E(\tilde{Q}_1\tilde{Q}_2)=-(n-1)$, such that the correlation is $-1/(n-2)$. As $\bar{A}$ divides by $n$ rather than $n-1$, the distribution of $\bar{A}\mid Q_1=0$ equals the distribution of  $(n-1)^{-1}\sum_{\ell=1}^{n-1}|\tilde{Q}_\ell|$. From this, we see that $E(\bar{A}^2\mid Q_1=0) = (n-1)^{-2}[(n-1)\text{var}(\tilde{Q}_1) + (n-1)(n-2)E\{|\tilde{Q}_1\tilde{Q}_2|\}]$, where
\begin{align*}
\text{var}(\tilde{Q}_1) &= (n-1)(n-2)\\
E\{|\tilde{Q}_1\tilde{Q}_2|\} &= \frac{2(n-1)}{\pi}\left\{\sqrt{(n-2)^2-1}  - \text{arcsin}(-1/(n-2))\right\}
\end{align*}
For $n > 3$, the formula $E|\tilde{Q}_1\tilde{Q}_2|$ may be found in \citet{nab51} among other places. At $n=3$ $(\tilde{Q}_1,\tilde{Q}_2)$ are degenerate as $\tilde{Q}_1$ and $\tilde{Q}_2$ are perfectly anti-correlated, and a direct calculation shows that $E|\tilde{Q}_1\tilde{Q}_2| = E(\tilde{Q}_1^2) = 2$, agreeing with the formula above $n > 3$. Combining terms and using oddness of arcsin,
\begin{align*}
E(\bar{A}^2\mid Q_1=0) &= (n-2)\left\{1+(2/\pi)\left\{\sqrt{(n-2)^2-1} + \text{arcsin}(1/(n-2))\right\}\right\}
\end{align*}

We now consider $E(\bar{A}\bar{S}^2)$. Noting $S_i^2=1$, its expectation is 
\begin{align*}
E(\bar{A}\bar{S}^2) &= n^{-3}\left\{n^2E|Q_1| + 2n(n-1)E(Q_1S_2) + n(n-1)(n-2)E(|Q_1|S_2S_3)\right\}\\
&= \frac{1}{n}E|Q_1| + \frac{2(n-1)}{n^2}E(Q_1S_2) + \frac{(n-1)(n-2)}{n^2}E(|Q_1|S_2S_3)
\end{align*}

The first two expectations can be computed in a straightforward way. Using the standard formula for absolute moments of Gaussians, $E|Q_1| = \sqrt{n(n-1)}\sqrt{2/\pi}$.  Using the tower property and that $\sum Q_j = 0$, $E(Q_1S_2) = E(E(Q_1\mid Q_2)S_2) = E(-Q_2 S_2)/(n-1) = -E|Q_2|/(n-1) = -\sqrt{n/(n-1)}\sqrt{2/\pi}$. 

Consider $E(|Q_1|S_2S_3) = E(Q_1S_1S_2S_3)$ and $n\geq 4$ so that the joint distribution of $(Q_1,Q_2,Q_3)$ is not singular; the degenerate case $n=3$ will then follow by taking limits. Set $G(Q_1,Q_2,Q_3) = S_1S_2S_3$ and apply Gaussian integration by parts using distributional derivatives:
\begin{align*}
E(Q_1S_1S_2S_3) &= \text{var}(Q_1)\;E\left\{(\partial/\partial Q_1) G(Q_1,Q_2,Q_3)\right\}\\&+ 2\text{cov}(Q_1,Q_2)\;E\left\{(\partial/\partial Q_2)G(Q_1,Q_2,Q_3)\right\}
\end{align*}   

The distributional derivative of $\text{sgn}(Q_1)$ is $2\delta_0(Q_1)$, where $\delta_0$ is the Dirac delta at 0. Hence, $E\left\{(\partial/\partial Q_1) G(Q_1,Q_2,Q_3)\right\} =2 E[S_2S_3\delta_0(Q_1)] = 2f_0E(S_2S_3\mid Q_1=0)$, where $f_0$ is the density of $Q_1$. The conditional expectation $E(S_2S_3\mid Q_1=0)$ equals $4p_{\geq 0, \geq 0} - 1$, where $p_{\geq 0, \geq 0}$ is the conditional probability that $(Q_2, Q_3)$ lie in the nonnegative orthant given $Q_1=0$. Orthant probabilities are widely used in order constrained statistical literature, and the needed orthant probability is $p_{\geq 0, \geq 0} = 1/4 + 1/(2\pi)\text{arcsin}(-1/(n-2))$, as $-1/(n-2)$ is the conditional correlation between $Q_2$ and $Q_3$ given $Q_1$; see, for instance, \citet[\S 3.5]{sil05}. This yields $E(S_2S_3\mid Q_1=0) = (2/\pi)\text{arcsin}(-1/(n-2))$. Recalling $\text{var}(Q_j) = n(n-1)$ and $\text{cov}(Q_1,Q_2)=-n$ and using exchangeability

\begin{align*}
E(Q_1S_1S_2S_3) &= \text{var}(Q_1)\;\;E\left\{(\partial/\partial Q_1) G(Q_1,Q_2,Q_3)\right\} + 2\text{cov}(Q_1,Q_2)\;\;E\left\{(\partial/\partial Q_2)G(Q_1,Q_2,Q_3)\right\}\\
&= -2f_0n(n-3)(2/\pi)\text{arcsin}(1/(n-2))\\
&= -E|Q_1|\{(n-3)/(n-1)\}(2/\pi)\text{arcsin}(1/(n-2))
\end{align*} 
where the last line uses that $2f_0n(n-1) = E|Q_1|$. 

For the case $n=3$, we proceed by adding $iid$ Gaussian perturbations to each coordinate and then taking limits. Define $Q_{\lambda j} = Q_j + \lambda X_j$, where $X_j$ are $iid$ standard Gaussian for $j=1,...,3$ and independent of $(Q_1,Q_2,Q_3)$. This inflates the variance $v_\lambda=\text{var}(Q_{\lambda j})$ to $n(n-1) + \lambda^2$ but leaves the covariance as $c_\lambda = \text{cov}(Q_{\lambda 1},{Q}_{\lambda 2}) =  -n$. The conditional correlation between $Q_{\lambda 3}$ and $Q_{\lambda 2}$ given $Q_{\lambda 1}=0$ simplifies to $c_\lambda/(v_\lambda+c_\lambda)$. Applying the previous integration by parts argument to $(Q_{\lambda 1},...,Q_{\lambda 3})$ and $S_{\lambda j} = \text{sgn}(Q_{\lambda j})$ for any fixed $\lambda > 0$ returns
\begin{align*}
E(Q_{\lambda 1}S_{\lambda 1}S_{\lambda 2}S_{\lambda 3}) &=(v_\lambda+2c_\lambda)2f_{0\lambda}(2/\pi)\text{arcsin}(c_\lambda/(v_\lambda+c_\lambda)),
\end{align*}where $f_{0\lambda}$ is the density of $Q_{1\lambda}$ evaluated at 0. Sending $\lambda$ to 0, we have $(Q_{\lambda 1},Q_{\lambda 2},Q_{\lambda 3}) \rightarrow (Q_1,Q_2,Q_3)$ a.s. Dominated convergence then recovers the result for $n=3$ that $E(Q_1S_1S_2S_3)=0$ as $v_\lambda+2c_\lambda =\lambda^2\rightarrow 0$.

For $n\geq 3$, we thus have
\begin{align*}
E(\bar{A}\bar{S}^2)&= E|Q_1|/n + 2(n-1)E(Q_1S_2)/n^2 + \frac{(n-1)(n-2)}{n^2}E(|Q_1|S_2S_3)\\
&= E|Q_1|/n -2E|Q_1|/n^2- E|Q_1|\{(n-3)(n-2)/n^2\}(2/\pi)\text{arcsin}(1/(n-2))\\
&= E|Q_1|\left(\frac{n-2}{n^2}\right)\{1-(n-3)(2/\pi)\text{arcsin}(1/(n-2))\}.
\end{align*}

We are now ready to plug terms into (\ref{eq:slicesup}). Recall $n = J+1$ and that $E|Q_1|$ = $E(\bar{A})$. Note further that $f_0E|Q_1| = 1/\pi$ and that  $2f_0n(n-1) = E|Q_1|$. For convenience, let $\alpha_n = \text{arcsin}(1/(n-2))$ and $s_n = \sqrt{(n-2)^2-1}$.

\begin{align*} &\frac{f_0E(\bar{A}^2\mid Q_1=0) + E(\bar{A}\bar{S}^2) - E(\bar{A})\{E(\bar{S})\}^2}{\{E(\bar{A})\}^2} - f_0\left(\frac{n-2}{n-1}\right)\\
&= \frac{E|Q_1|(n-2)/(2n(n-1))\left\{1+(2/\pi)\left\{s_n + \alpha_n\right\}\right\}}{\{E|Q_1|\}^2}+ \frac{ E|Q_1|\left((n-2)/n^2\right)\{1-(n-3)(2/\pi)\alpha_n\}}{\{E|Q_1|\}^2}\\ &- 1/\{E|Q_1|\pi\}\left(\frac{n-2}{n-1}\right)\\
\end{align*}

Multiplying by the factor $E|Q_1|n^2(n-1)\pi/(n-2)$, we obtain 
\begin{align*}V_n&= \{n/2\}\left\{\pi+2\left\{s_n + \alpha_n\right\}\right\}+ \left(n-1\right)\{\pi-(n-3)(2)\alpha_n\}-n^2.
\end{align*}
We want to show $V_n > 0$.
For $n=3$ a direct computation yields $s_n=0$, $\alpha_n=\pi/2$, and
\begin{align*}
V_3 &= \{3/2\}\left\{\pi+(2)\left\{\pi/2\right\}\right\}+ \left(2\right)\{\pi\}-9\\
&= 3\pi + 2\pi-9 > 0.
\end{align*}

For $n \geq 4$, we now establish bounds on $\alpha_n$ and $s_n$. The coefficient on $s_n$ is positive, while for $\alpha_n$ it is $n - (n-1)(n-3)2$, which is negative for $n\geq 4$. Hence to lower bound $V_n$, we need to lower bound $s_n$ and upper bound $\alpha_n$. For all $x > 1$ we have that $(x^2-1)\geq (x-1/x)^2$. Taking the square root we have $s_n \geq (n-2)-1/(n-2)$. We further have for all $t\leq 1/2$ that $1/\sqrt{1-t^2}\leq (1+3t^2)$.  Using the integral representation $\text{arcsin}(x) = \int_{0}^x 1/\sqrt{1-t^2}\; dt$, integrating the bound then yields $\text{arcsin}(x)\leq x+x^3$ for $x\leq 1/2$, implying that for $n\geq 4$ $\alpha_n \leq 1/(n-2) + 1/(n-2)^3$.

Therefore, for $n\geq 4$ and setting $x=n-2$.

\begin{align*}V_n&= (x+2)s_{x+2} - (x+2)^2 + \pi(3x+4)/2 + (-2x^2+x+4)\alpha_{x+2}\\
&\geq (x+2)(x-1/x)-(x+2)^2+\pi(3x+4)/2+(-2x^2+x+4)(1/x+1/x^3)\\
&=(3\pi/2-4)x+(2\pi-4)+1/x^2+4/x^3.
\end{align*}

As $x\geq 2$, observe that $(3\pi/2-4)x$, $(2\pi-4)$, $1/x^2$, and $4/x^3$ are all positive. Therefore the desired result holds for $x\geq 2$, or equivalently $n\geq 4$. Combining this with the direct calculation at $n=3$ completes the proof.

\subsection{Intuition for the skewness measure $K(\psi, F_\epsilon)$}

Here we provide intuition for the skewness functional $K(\psi, F_\epsilon)$ by stating an alternate form using the first-order expansion from Proposition \ref{prop:firstorder}. We then describe qualitatively why the measure tends to be positive for right-skewed distributions through an illustration with $\psi_{lin}$.

We first re-express the definition of $K(\psi, F_\epsilon)$ using the notation in the proofs. Let $K(\psi, F_\epsilon) = \{B(\psi, F_\epsilon) - B(\psi, F_{-\epsilon})\}/2$, where for $\epsilon_j \overset{iid}{\sim} G$
\begin{align*}
B(\psi,G) &=   \frac{\partial}{\partial \tau}\left\{\frac{1}{J+1}\sum_{j=1}^{J+1}E|q_{j}(\tau) - \tau \beta\bar{A} | -  E_G|q_1(\tau)|\right\}\Bigr|_{\tau=0^+}\end{align*}

Define $q^{-}_{j}(\tau)$ by replacing $\epsilon_j$ with $-\epsilon_j$ in (\ref{eq:qj}), and define $\tilde{M}^{-}(\tau)$ as the fixed point when $q^{-}_j(\tau)$ replaces $q_j(\tau)$.

\begin{proposition}\label{prop:Kpsi} Under the generative model of \S \ref{sec:gen} and under the assumptions of Proposition \ref{prop:fixedorder} and \ref{prop:firstorder}, \begin{align*} K(\psi, F_\epsilon) = \frac{1}{J+1}\sum_{j=1}^{J+1}E\{(D_j-\beta\bar{A})\text{sgn}(Q_j)\} - E\{D_1\text{sgn}(Q_1)\} \end{align*}
\end{proposition}
\begin{proof}
If $q_j(\tau)$ admits the expansion $Q_j + D_j\tau + o_{L1}(\tau)$, $q^{-}_j(\tau)$ admits the expansion $-Q_j + D_j\tau + o_{L1}(\tau)$ by oddness of $\psi$. Moreover, by Propositions \ref{prop:fixedorder} and \ref{prop:firstorder} $\kappa(\tau)\tilde{M}^{-}(\tau)$ admits the expansion $\tau \beta \bar{A} + o(\tau)$. We then consider

\begin{align*} A_{\tau,j} &= Q_j +\tau (D_j - \beta\bar{A})+o_{L1}(\tau)\\
A^-_{\tau,j} &= -Q_j +\tau (D_j - \beta\bar{A})+o_{L1}(\tau) \end{align*}
and
\begin{align*}
\tilde{A}_{\tau,1} &= Q_1+\tau D_1 + o_{L1}(\tau)\\
\tilde{A}^-_{\tau,1} &= -Q_1+\tau D_1 + o_{L1}(\tau)
\end{align*}
By Corollary \ref{cor:Kpsi}
\begin{align*}
\frac{1}{2}\left( \frac{\partial}{\partial \tau} E|A_{\tau,j}| - \frac{\partial}{\partial \tau} E|A^-_{\tau,j}|\right) &= E\left\{\left(D_j-\beta\bar{A}\right)\text{sgn}(Q_j)\right\}\\
\frac{1}{2}\left( \frac{\partial}{\partial \tau} E|\tilde{A}_{\tau,1}| - \frac{\partial}{\partial \tau} E|\tilde{A}^-_{\tau,1}|\right) &= E\left\{D_1\text{sgn}(Q_1)\right\} 
\end{align*}
The result then follows.
\end{proof}

We now describe qualitatively why the skewness measure tends to be positive for right-skewed distributions, with the discussion catered to linear $\psi$. Recall that for $\psi_{lin}$, $D_1=J$, $D_j=-1$ for $j=2,...,J+1$, and $\beta = J/E|Q_1|$. As the $D_j$ sum to zero and $Q_j$ are exchangeable, for linear scores the skewness measure equals
\begin{align*}
K(\psi, F_\epsilon) &= -JE(\bar{S}\bar{A})/E|Q_1| - J E(S_1),\end{align*}
where $S_i = \text{sgn}(Q_i)$.

First consider $E(S_1)$. Recall that $Q_1 = J\epsilon_1-\sum_{\ell=2}^{J+1}\epsilon_\ell = (J+1)(\epsilon_1-\bar{\epsilon})$. For a right-skewed distribution, $S_i = \text{sgn}(\epsilon_1-\bar{\epsilon})$ tends to be negative, as $\bar{\epsilon}$ is pulled up by outlying observations. This provides a positive contribution to the skewness measure through $-JE(S_1)$. Now consider $E(\bar{S}\bar{A})$. By the same argument as before, $E(\bar{S})$ tends to be negative for right-skewed data. Moreover, it tends to be more negative precisely when there are larger outlying observations, which also result in larger values for $\bar{A}$. This suggests a negative value for $\text{Cov}(\bar{A},\bar{S})$, and hence for $E(\bar{S}\bar{A}) = E(\bar{S})E(\bar{A}) + \text{Cov}(\bar{A},\bar{S})$ since $E(\bar{S})$ tends to be negative. Hence, the contribution $-JE(\bar{S}\bar{A})/E|Q_1|$ is also typically positive.

\subsection{Proof for inner trimming: $\psi_{\iota,h}$, $0<\iota<h$}
\begin{proposition}\label{prop:inner} Suppose that $F_\epsilon$ is absolutely continuous and admits a density $p_\epsilon$ such that $p_\epsilon(x) > 0$ for all $x$ in an open interval of length greater than $\iota$. Then, the conditions of Proposition \ref{prop:firstorder} are satisfied.

Define the contribution from the atom event

\begin{align*} L(\psi, F_\epsilon) &=\left[\frac{1}{J+1}\sum_{j=1}^{J+1}E|(D_j-\beta\bar{A}) \1\{Q_j=0\}| - E|D_1 \1\{Q_1=0\}|\right]
\end{align*}

Then, $L(\psi_{\iota,h}, F_\epsilon) > 0$. Therefore, the tilted sensitivity analysis is superior in the small-$\tau$ regime whenever $K(\psi_{\iota,h}, F_\epsilon) > -L(\psi_{\iota,h}, F_\epsilon)$. If the inequality is reversed, the conventional sensitivity analysis is superior in the small-$\tau$ regime.
\end{proposition}
\begin{proof}
The almost-everywhere derivative of $\psi_{\iota,h}(x+\tau)$ with respect to $\tau$ is $\{h/(h-\iota)\}\1\{\iota < |x| < h\}$. Recall that the first unit receives the treatment and the remaining $J$ receive the control. Define $D_j$ as

\begin{align*}
D_1 &= \left(\frac{h}{h-\iota}\right)\sum_{\ell=2}^{J+1}\1\{\iota< |\epsilon_1-\epsilon_\ell|<h\}\\
D_j &= - \left(\frac{h}{h-\iota}\right)\1\{\iota< |\epsilon_j-\epsilon_1|< h\}\;\;\;(j=2,...,J+1).
\end{align*}

As $\epsilon_{j}-\epsilon_\ell \in \{\pm \iota, \pm h\}$ is an event of measure 0, we have the representation
\begin{align*}
q_j(\tau) &= Q_j + \tau D_j + o_{L1}(\tau),
\end{align*}

with $E|D_j| < \infty$, and $E|Q_j| < \infty$ by boundedness of $\psi$. Moreover, by the conditions on $p_\epsilon$ we have that $E|Q_j| > 0$, as the event that $|\psi(\epsilon_j-\epsilon_\ell)| > 0$ occurs with positive probability. This establishes that Proposition \ref{prop:firstorder} may be applied.

We now establish that $L(\psi, F_\epsilon) > 0$. Under the conditions on $p_\epsilon$, the event $Q_j = 0$ occurs with positive probability. In particular, $Q_j$ is equal to 0 on the event that all $\ell = 1,...,J+1$ $|\epsilon_{j}-\epsilon_\ell|$ are either below $\iota$ or above $h$ and that the number of differences above $h$ equals the number of differences below $-h$. On this event, $D_j=0$. On the event $D_j\neq 0$, $Q_j$ could only equal zero if the values for $\psi(\epsilon_j-\epsilon_1)$ with $\iota < |\epsilon_j-\epsilon_1| < h$ cancel out perfectly when summed together; this occurs with probability 0 since $\epsilon_1$ has a density. Therefore, $E(|D_j|\1\{Q_j=0\}) = 0$. 

Consider now $E(\bar{A}1\{Q_j=0\})$. Under the conditions on $p_\epsilon$, one can construct an event of positive probability on which for some $\delta > 0$ $\iota + \delta \leq |\epsilon_1-\epsilon_2|$, but where $|\epsilon_1-\epsilon_j| \leq \iota$, $|\epsilon_2-\epsilon_j| \leq \iota$, and $|\epsilon_\ell-\epsilon_k| \leq \iota$ for all $k,\ell = 3,...,J+1$. On this event, we have that $Q_1 = -Q_2$ with $|Q_1| = |Q_2| > 0$, but that $|Q_3|=...=|Q_{J+1}| = 0$. Therefore, on this event $\bar{A}\1\{Q_j=0\} > 0$ for $j=3,...,J+1$, implying $E(\bar{A}\1\{Q_j=0\}) > 0$. We further have that $E(D_1) > 0$ under the assumptions on $p_\epsilon$, implying $\beta > 0$, as the event that $\iota < |\epsilon_1-\epsilon_j| < h$ occurs with positive probability. 

As $D_j=0$ on the event $\{Q_j=0\}$, we thus have that $|D_1\1\{Q_1=0\}| = 0$ and $|D_j-\beta \bar{A}|\1\{Q_j=0\}  = \beta \bar{A}\1\{Q_j=0\}$ a.e. for all $j$. Therefore, as $Q_j$ are exchangeable,

\begin{align*} L(\psi, F_\epsilon) &=\left[\frac{1}{J+1}\sum_{j=1}^{J+1}E|(D_j-\beta\bar{A}) \1\{Q_j=0\}| - E|D_1 \1\{Q_1=0\}|\right]\\
&= \beta E(\bar{A}1\{Q_j=0\}) > 0. 
\end{align*}

\end{proof}

\subsection{Proof for outer trimming, $J$ even: $\psi_{0,h}$}

The proof for outer trimming with $J$ even closely mirrors that of inner trimming.
\begin{proposition}\label{prop:outereven}
Suppose that $F_\epsilon$ is absolutely continuous and admits a density $p_\epsilon$ such that $p_\epsilon(x) > 0$ for all $x$ in an open interval of length greater than $2h$. Then, the conditions of Proposition \ref{prop:firstorder} are satisfied.

Define the contribution from the atom event

\begin{align*} L(\psi, F_\epsilon) &=\left[\frac{1}{J+1}\sum_{j=1}^{J+1}E|(D_j-\beta\bar{A}) \1\{Q_j=0\}| - E|D_1 \1\{Q_1=0\}|\right]
\end{align*}

Then, $L(\psi_{0,h}, F_\epsilon) > 0$. Therefore, the tilted sensitivity analysis is superior in the small-$\tau$ regime whenever $K(\psi_{0,h}, F_\epsilon) > -L(\psi_{0,h}, F_\epsilon)$. If the inequality is reversed, the conventional sensitivity analysis is superior in the small-$\tau$ regime.
\end{proposition}
\begin{proof}
The almost-everywhere derivative of $\psi_{0,h}(x+\tau)$ with respect to $\tau$ is $\1\{|x| < h\}$. Recall that the first unit receives the treatment and the remaining $J$ receive the control. Define $D_j$ as

\begin{align*}
D_1 &= \sum_{\ell=2}^{J+1}\1\{|\epsilon_1-\epsilon_\ell|< h\}\\
D_j &= - \1\{|\epsilon_j-\epsilon_1|< h\}\;\;\;(j=2,...,J+1).
\end{align*}

As $\epsilon_{j}-\epsilon_\ell \in \{\pm h\}$ is an event of measure 0, we have the representation
\begin{align*}
q_j(\tau) &= Q_j + \tau D_j + o_{L1}(\tau),
\end{align*}with $E|D_j| < \infty$, and $E|Q_j| < \infty$ by boundedness of $\psi$. Moreover, by the conditions on $p_\epsilon$ we have that $E|Q_j| > 0$, as the event that $|\psi(\epsilon_j-\epsilon_\ell)| > 0$ occurs with positive probability. This establishes that Proposition \ref{prop:firstorder} may be applied.

We now establish that $L(\psi, F_\epsilon) > 0$. Under the conditions on $p_\epsilon$, the event $Q_j = 0$ occurs with positive probability. In particular, $Q_j$ is equal to 0 on the event that for all $\ell\neq j$, $|\epsilon_{j}-\epsilon_\ell|$ are all above $h$ and that the number of differences above $h$ equals the number of differences below $-h$; this cancellation is only possible for $J$ even. On this event, $D_j=0$. On the event $D_j\neq 0$, $Q_j$ could only equal zero if the values for $\psi(\epsilon_j-\epsilon_1)$ with $| \epsilon_j-\epsilon_1| < h$ cancel out perfectly when summed together; this occurs with probability 0 since $\epsilon_1$ has a density. Therefore, $E(|D_j|\1\{Q_j=0\}) = 0$. 

Consider now $E(\bar{A}1\{Q_j=0\})$. Consider the event where for sufficiently small $0<\delta_1<\delta_2$, $\epsilon_1 \in [-\delta_1, \delta_1]$, $h+\delta_1 \leq \epsilon_{2},...,\epsilon_{J/2+1}\leq h + \delta_1 + \delta_2$, $-h-\delta_1-\delta_2 \leq \epsilon_{J/2+2},...,\epsilon_{J+1} \leq -h-\delta_1$. Then, $Q_1=0$. For $j>2$, $|Q_j| \geq 2h > 0$. Therefore, $\bar{A}\1\{Q_1=0\} > 0$, implying $E(\bar{A}\1\{Q_1=0\}) > 0$. We further have that $E(D_1) > 0$ under the assumptions on $p_\epsilon$ implying $\beta > 0$, as the event that $|\epsilon_1-\epsilon_j| < h$ occurs with positive probability.

As $D_j=0$ on the event $\{Q_j=0\}$, we thus have that $|D_1\1\{Q_1=0\}| = 0$ and $|D_j-\beta \bar{A}|\1\{Q_j=0\}  = \beta \bar{A}\1\{Q_j=0\}$ a.e. for all $j$. Therefore, as $Q_j$ are exchangeable,

\begin{align*} L(\psi, F_\epsilon) &=\left[\frac{1}{J+1}\sum_{j=1}^{J+1}E|(D_j-\beta\bar{A}) \1\{Q_j=0\}| - E|D_1 \1\{Q_1=0\}|\right]\\
&= \beta E(\bar{A}1\{Q_j=0\}) > 0. 
\end{align*}

\end{proof}

\subsection{Discussion of outer trimming, $J$ odd}
When $J$ is odd, under the regularity conditions on $F_\epsilon$ given in Proposition \ref{prop:outereven} we have that $L(\psi_{0,h}, F_\epsilon) = 0$ as the atom occurs with probability zero. For $J$ even, an atom only occurs on the event that all $|\epsilon_{j}-\epsilon_\ell|$,  $\ell\neq j$, are above $h$ and the number of differences above $h$ equals the number of differences below $-h$. For $J$ odd, the number below $-h$ cannot equal the number above $h$; these numbers must differ by at least 1. Therefore, a second-order comparison is required when $K(\psi_{0,h},F_\epsilon) = 0$. 

The conditions within Proposition \ref{prop:outereven} provide sufficient conditions for the existence of a first-order expansion.  The almost-everywhere derivative for $\psi_{0,h}(x+\tau)$ is still $\1\{|x|<h\}$, and we still define  $D_j$ as
\begin{align*}
D_1 &= \sum_{\ell=2}^{J+1}\1\{|\epsilon_1-\epsilon_\ell|\leq h\}\\
D_j &= - \1\{|\epsilon_j-\epsilon_1|\leq h\}\;\;\;(j=2,...,J+1).
\end{align*}

Because there are no first-order atom contributions, $L(\psi_{0,h}, F_\epsilon) = 0$, we are only able to state based on first-order expansions that the tilted approach is superior in the small-$\tau$ regime when $K(\psi_{0,h},F_\epsilon) > 0$, and that the conventional approach is superior when $K(\psi_{0,h},F_\epsilon) < 0$.  

For the second-order expansion, $H_j=0$ as the almost-everywhere derivative is flat outside of the kinks in $\psi_{0,h}$; however, unlike with linear scores there is a non-negligible contribution $r_j(\tau)$ to the second-order expansion in Proposition \ref{prop:secorder}. The term $r_j(\tau)$ comes from the error in approximating the differences $\psi_{0,h}(\epsilon_j-\epsilon_\ell + \tau) - \psi_{0,h}(\epsilon_j-\epsilon_\ell)$ by $\tau$ times the almost-everywhere derivative, $\tau\1\{|\epsilon_j-\epsilon_\ell| < h\}$,  when deploying the first-order expansion $\tau D_j$. Error is incurred when $|\epsilon_j-\epsilon_\ell|$ and $|\epsilon_j-\epsilon_\ell-\tau|$ fall on different sides of the kink points $\pm h$, and this error provides a non-negligible contribution to the expansion at second order.

We now characterize the contributions from $r_j(\tau)$ to the second-order expansion and argue that ultimately they are favorable to the tilted sensitivity analysis in that their aggregate contribution within Proposition \ref{prop:secorder} is nonnegative under the assumed conditions. We will not do so under primitive conditions, but instead assume that the requisite assumptions for Lemmas \ref{lemma:outsecond} and \ref{lemma:outrice} hold.

Let $v_h(x,\tau) = \psi_{0,h}(x + \tau) - \psi_{0,h}(x) - \tau\1\{|x| < h\} = (x+\tau+h)\1\{-h-\tau < x < -h\} + (h-x-\tau)\1\{h-\tau < x < h\}$. Then, we have
\begin{align*}
r_1(\tau) &= \sum_{j=2}^{J+1}v_h(\epsilon_1-\epsilon_j,\tau)\\
r_j(\tau) &= v_h(\epsilon_j-\epsilon_1,-\tau) = -v_h(\epsilon_1-\epsilon_j,\tau) ;\;\;\;(j=2,...,J+1),
\end{align*}
the equality in the display for $r_j(\tau)$ coming from oddness of $\psi$.

Set $X_{1\ell} = \epsilon_1-\epsilon_\ell$. We now study the limiting quantities $\lim_{\tau \downarrow 0} E\{\text{sgn}(Q_1)v_h(X_{1\ell}, \tau)\}/\tau^2$ and, for $j>1$,  $\lim_{\tau \downarrow 0} E\{\text{sgn}(Q_j)v_h(X_{1j}, \tau)\}/\tau^2$. This will then provide the forms for $\mathcal{R}_j$ appearing within Proposition \ref{prop:secorder}. 

Under suitable regularity conditions, there exists a function $g$ such that
\begin{align*}
E\{\text{sgn}(Q_1)v_{h}(X_{1j},\tau)\} &= \int g(x)v_h(x,\tau)\;dx\\
&= \int_{-h-\tau}^{-h}(x+\tau+h)g(x)\;dx + \int_{h-\tau}^{h}(h-\tau-x)g(x)\;dx\\
\end{align*}

To simplify further, in the first integral set $v=-h-x$ so that $x+\tau+h = (\tau-v)$. In the second, define $v=h-x$ so that $(h-\tau-x) = (v-\tau)$. Changing the limits of integration accordingly, we obtain

\begin{align*}
E\{\text{sgn}(Q_1)v_{h}(X_{1j},\tau)\}   &= \int g(x)v_h(x,\tau)\;dx\\
&= \int_{-h-\tau}^{-h}(x+\tau+h)g(x)\;dx + \int_{h-\tau}^{h}(h-\tau-x)g(x)\;dx\\
&= \int_0^\tau(\tau-v)g(-h-v)\;dv -  \int_0^\tau(\tau-v)g(h-v)\;dv\\
&= \int_0^\tau(\tau-v)\left\{g(-h-v)-g(h-v)\right\}\;dv
\end{align*}
Changing variables  $v = \tau t$ and dividing by $\tau^2$
\begin{align*}\frac{1}{\tau^2}\int_0^\tau(\tau-v)\left\{g(-h-v)-g(h-v)\right\}\;dv &= \frac{1}{\tau^2}\int_0^1\tau^2(1-t)\left\{g(-h-\tau t) - g(h-\tau t)\right\}\;dt\\
&= \int_0^1(1-t)\left\{g(-h-\tau t) - g(h-\tau t)\right\}\;dt\\
\end{align*}
Taking limits as $\tau \downarrow 0$ and assuming continuity of  $g$ at $\pm h$
\begin{align*}
\lim_{\tau\downarrow 0}\frac{1}{\tau^2}E\{\text{sgn}(Q_1)v_{h}(X_{1j},\tau)\} &= \int_0^1(1-t)\left\{g(-h) - g(h)\right\}\;dt\\
&= \left\{g(-h) - g(h)\right\}/2
\end{align*}

An analogous argument provides a form for $\lim_{\tau \downarrow 0}\frac{1}{\tau^2}E\{\text{sgn}(Q_j)v_{h}(X_{1j},\tau)\}$ for another function $\tilde{g}$.

By definition of $r_j(\tau)$ and using exchangeability, this yields

\begin{align*}
\mathcal{R}_1 &= J\left\{g(-h) - g(h)\right\}/2\\
\mathcal{R}_j &= - \left\{\tilde{g}(-h) - \tilde{g}(h)\right\}/2
\end{align*}

Using exchangeability, the aggregate contribution of these terms in Proposition \ref{prop:secorder} is
\begin{align*}
\left\{\frac{1}{J+1}\sum_{\ell=1}^{J+1}\mathcal{R}_\ell\right\} - \mathcal{R}_1&= \left(\frac{J}{2(J+1)}\right)\left\{\tilde{g}(h) +Jg(h) - \tilde{g}(-h)-Jg(-h)\right\}
\end{align*}

Now, observe that $X_{j\ell} = \epsilon_j-\epsilon_\ell$ is symmetric regardless of the underlying law of $F_\epsilon$. Therefore, letting $f$ be the density of $X_{j\ell}$, we have $f(x) = f(-x)$, and by exchangeability
\begin{align*}
g(h) &= f(h)E\{\text{sgn}(Q_1)\mid X_{1j} = h\}\\
\tilde{g}(h) &= f(h)E\{\text{sgn}(Q_j)\mid X_{1j} = h\}\\
g(-h) &= f(-h)E\{\text{sgn}(Q_1)\mid X_{1j} = -h\} = f(h)E\{\text{sgn}(Q_j)\mid X_{1j} = h\}\\
\tilde{g}(-h) &= f(-h)E\{\text{sgn}(Q_j)\mid X_{1j} = -h\} = f(h)E\{\text{sgn}(Q_1)\mid X_{1j} = h\}
\end{align*}

So
\begin{align*}
&\left(\frac{J}{2(J+1)}\right)\left\{\tilde{g}(h) +Jg(h) - \tilde{g}(-h)-Jg(-h)\right\}\\&=f(h)\left(\frac{J(J-1)}{2(J+1)}\right)E\{\text{sgn}(Q_1)-\text{sgn}(Q_j) \mid X_{1j}=h\}.
\end{align*}

Now, we argue that $E\{\text{sgn}(Q_1)-\text{sgn}(Q_j) \mid X_{1j}=h\} \geq 0$. Recall that $X_{1j} = \epsilon_1-\epsilon_j$, such that on this event $\epsilon_1 = \epsilon_j+h$. Consider now $Q_1$ and recall that $\psi$ is monotone nondecreasing.

\begin{align*}Q_1 &= \sum_{\ell=1}^{J+1}\psi(\epsilon_1-\epsilon_\ell) = \sum_{\ell=1}^{J+1}\psi(\epsilon_j-\epsilon_\ell + h)\\
&\geq \sum_{\ell=1}^{J+1}\psi(\epsilon_j-\epsilon_\ell) = Q_j,\\
\end{align*}
so $\text{sgn}(Q_1)\geq \text{sgn}(Q_j)$. Therefore,
\begin{align*}
\left\{\frac{1}{J+1}\sum_{\ell=1}^{J+1}\mathcal{R}_\ell\right\} - \mathcal{R}_1&= f(h)\left(\frac{J(J-1)}{2(J+1)}\right)E\{\text{sgn}(Q_1)-\text{sgn}(Q_j) \mid X_{1j}=h\} \geq 0.
\end{align*}
Hence the terms provide a nonnegative contribution to the second-order expansion of $m(\tau)-\eta(\tau)$. This does not establish that the tilted sensitivity analysis is superior in the small-$\tau$ regime, as one would need to evaluate the remaining terms in the second-order expansion.

\bibliographystyle{apalike}
\bibliography{../bibliography.bib}
\end{document}